\documentclass[11pt]{article}

\usepackage{amsmath}
\usepackage{amsfonts}
\usepackage{amssymb}
\usepackage{latexsym}
\usepackage{graphicx}
\usepackage{psfrag}
\usepackage{stmaryrd}

\newcommand{\opn}[1]{\operatorname{#1}}
\newcommand{\veps}{\varepsilon}
\newcommand{\real}{\opn{Re}}
\newcommand{\imag}{\opn{Im}}
\newcommand{\phm}{\phantom{-}}

\newcommand{\der}[2]{\frac{\partial #1}{\partial #2}}
\newcommand{\varder}[2]{\frac{\delta #1}{\delta #2}}
\newcommand{\pr}{\partial}
\newcommand{\e}[1]{{(#1)}}
\newcommand{\mc}[1]{\mathcal{#1}}
\newcommand{\jt}{\textstyle}
\newcommand{\jd}{\displaystyle}
\newcommand{\js}{\scriptstyle}

\usepackage{epsfig}
\newcommand{\mypsdraft}{\psdraft}
\renewcommand{\mypsdraft}{\psfull}
\newcommand{\mypsfull}{\psfull}

\newcommand{\ignore}[1]{}

\usepackage{theorem}
\theoremstyle{plain}
\theoremheaderfont{\scshape}
\newtheorem{theorem}{Theorem}

\newtheorem{conjecture}[theorem]{Conjecture}

\theorembodyfont{\rmfamily}
\newtheorem{remark}[theorem]{Remark}
\newtheorem{example}[theorem]{Example}

\title{Global Paths of Time-Periodic Solutions of the Benjamin-Ono
Equation Connecting Pairs of Traveling Waves}

\author{
David M. Ambrose
\thanks{Department of Mathematics, Drexel University,
  Philadelphia, PA 19104 ({\tt ambrose@math.drexel.edu}).
  This work was supported in part by the National Science Foundation
  through grant DMS-0926378.}
\and
Jon Wilkening
\thanks{Department of Mathematics and Lawrence Berkeley National
  Laboratory, University of California, Berkeley, CA 94720 ({\tt
    wilken@math.berkeley.edu}).  This work was supported in part by the
  Director, Office of Science, Computational and Technology Research,
  U.S. Department of Energy under Contract No. DE-AC02-05CH11231.}
}

\begin{document}

\maketitle

\begin{abstract}
  We classify all bifurcations from traveling waves to non-trivial
  time-periodic solutions of the Benjamin-Ono equation that are
  predicted by linearization.  We use a spectrally accurate numerical
  continuation method to study several paths of non-trivial solutions
  beyond the realm of linear theory.  These paths are found to either
  re-connect with a different traveling wave or to blow up.  In the
  latter case, as the bifurcation parameter approaches a critical
  value, the amplitude of the initial condition grows without bound
  and the period approaches zero.  We then prove a theorem that gives
  the mapping from one bifurcation to its counterpart on the other
  side of the path and exhibits exact formulas for the time-periodic
  solutions on this path.  The Fourier coefficients of these solutions
  are power sums of a finite number of particle positions whose
  elementary symmetric functions execute simple orbits (circles or
  epicycles) in the unit disk of the complex plane.  We also find
  examples of interior bifurcations from these paths of already
  non-trivial solutions, but we do not attempt to analyze their
  analytic structure.
\end{abstract}

\vspace*{5pt}
{\bf Key words.} Periodic solutions, Benjamin-Ono equation,
non-linear waves, solitons,
bifurcation, continuation, exact solution, adjoint equation,
spectral method

\vspace*{5pt}
{\bf AMS subject classifications.} 65K10, 37M20, 35Q53, 37G15

\thispagestyle{empty}

\section{Introduction}

The Benjamin-Ono equation is a non-local, non-linear dispersive
equation intended to describe the propagation of internal waves in a
deep, stratified fluid \cite{benjamin:67, davis:67, ono:75}.  In spite
of non-locality, it is an integrable Hamiltonian
system with meromorphic particle
solutions \cite{case:mero,case:remarkable}, $N$-soliton solutions
\cite{matsuno:BO:sol}, and $N$-phase multi-periodic solutions
\cite{satsuma:ishimori:79, dobro:91, matsuno:04}.  A bilinear
formalism \cite{satsuma:ishimori:79} and a B\"acklund transformation
\cite{nakamura:backlund:79, bock:kruskal:79, matsuno:backland:85} have
been found to generate special solutions of the equation, and,
in the non-periodic setting of rapidly decaying initial conditions, an
inverse scattering transform has been developed
\cite{fokas:ablowitz:83, kaup:matsuno:98} that exploits an interesting
Lax pair structure in which the solution plays the role of a
compatibility condition in a Riemann-Hilbert problem.

It is common practice in numerical analysis to test a numerical method
using a problem for which exact solutions can be found.  Our initial
interest in Benjamin-Ono was to serve as such a test problem.
Although many of the tools mentioned above can be used to study
time-periodic solutions, they do not generalize to problems such as
the vortex sheet with surface tension
\cite{ambrose:03,ambrose:wilkening:vtx} or the true water wave
\cite{tolandPlotnikov, tolandPlotnikovIooss}, which are not known to
be integrable.  Our goal in this paper is to develop tools that
\emph{will} generalize to these harder problems and use them to study
bifurcation and global reconnection in the space of time-periodic
solutions of \mbox{B-O}.  Specifically, we employ a variant of the
numerical continuation method we introduced in \cite{benj1} for this
purpose, which yields solutions that are accurate enough that we are
able to recognize their analytic form.

Because we approached the problem from a completely different
viewpoint, our description of these exact solutions is very different
from previously known representations of multi-periodic solutions.
Rather than solve a system of non-linear algebraic equations at each
$x$ to find $u(x,t)$ as was done in \cite{matsuno:04}, we represent
$u(x,t)$ in terms of its Fourier coefficients $c_k(t)$, which turn out
to be power sums $c_k=2[\beta_1^k+\cdots+\beta_N^k]$ of a collection
of $N$ particles $\beta_j(t)$ evolving in the unit disk of the complex
plane as the zeros of a polynomial $z\mapsto P(z,t)$ whose
coefficients execute simple orbits (circles or epicycles in
$\mathbb{C}$).  The connection between the new representation and
previous representations will be explored elsewhere \cite{benj3}.

Many of our findings on the structure of bifurcations and
reconnections in the manifold of time-periodic solutions of the
Benjamin-Ono equation are likely to hold for other systems as well.
One interesting pitfall we have identified by applying our method to
an integrable problem is that degenerate bifurcations can exist that
are not predicted by counting linearly independent, periodic solutions
of the linearization about traveling waves.  Although it is possible
that such degeneracy is a consequence of the symmetries that make this
problem integrable, it is also possible that other problems such as
the water wave will also possess degenerate bifurcations that are
invisible to a linearized analysis.  We have also found that one
cannot achieve a complete understanding of these manifolds of
time-periodic solutions by holding e.g.~the mean constant and varying
only one parameter.  In some of the simulations where we hold the
mean fixed, the solution (i.e.~the $L^2$ norm of the initial condition)
blows up as the parameter approaches a critical value rather than
reconnecting with another traveling wave.  However, if the mean is
simultaneously varied, it is always possible to reconnect.
Thus, although numerical continuation with more than one parameter is
difficult, it will likely be necessary to explore multi-dimensional
parameter spaces to achieve a thorough understanding of time-periodic
solutions of other problems.

On the numerical side, we believe our use of certain Fourier modes of
the initial conditions as bifurcation parameters will prove useful in
many other problems beyond Benjamin-Ono.  We also wish to advocate the
use of variational calculus and optimal control for the purpose of
finding time-periodic solutions (or solving other two point boundary
value problems).  For ODE, a competing method known as orthogonal
collocation (e.g.~as implemented in AUTO \cite{doedel91b}) has proved
to be a very powerful technique for solving boundary value problems.
This approach becomes quite expensive when the dimension of the system
increases, and is therefore less competitive for PDE than it is for
ODE.  For PDE, many authors do not attempt to find exact periodic
solutions, and instead point out that typical solutions of certain
equations do tend to pass near their initial states at a later time
\cite{camassa:lee:08}. If true periodic solutions are sought, a more
common approach has been to either iterate on a Poincar\'e map and use
stability of the orbit to find time-periodic solutions
\cite{cabral:rosa}, or use a shooting method \cite{stoer:bulirsch,
  DV:lorentz:04} to find a fixed point of the Poincar\'e map.

In a shooting method, we define a functional
$F(u_0,T)=[u(\cdot,T)-u_0]$ that maps initial conditions and a
supposed period to the deviation from periodicity.  The equation $F=0$
is then solved by Newton's method, where the Jacobian $J=DF$ is either
computed using finite differences \cite{DV:couette:07} or by solving
the variational equation
repeatedly to compute each column of $J$.  We have found that it is
much more efficient (by a factor of the number of columns of $J$) to
instead minimize the scalar functional $G=\frac{1}{2}\|F\|^2$ via a
quasi-Newton method in which the gradient $DG$ is computed by solving
an adjoint PDE.

Bristeau et.~al.~\cite{glowinski1} developed a
similar approach for linear (but two- or three-dimensional) scattering
problems.  Three dimensional problems are intractable by the standard
shooting approach as $J$ could easily have $10^5$ columns.  However, the
gradient of $G$ can be computed by solving a single adjoint PDE.  The
success of the method then boils down to a question of the number of
iterations required for the minimization algorithm to converge.  For
linear problems, Bristeau et.~al.~have had success using conjugate
gradients to minimize $G$.  We find that BFGS \cite{bfgs} works very
well for nonlinear problems like the Benjamin-Ono equation and the
vortex sheet with surface tension \cite{ambrose:wilkening:vtx}.

To find non-trivial time-periodic solutions in the present work, we
use a symmetric variant of the algorithm described in \cite{benj1}.
Although the original method works well, we use the symmetric variant
for the simulations in this paper because evolving to $T/2$ requires
half the time-steps and yields more accurate answers (as there is less
time for numerical roundoff error to corrupt the calculation).
Moreover, the number of degrees of freedom in the search space of
initial conditions is also cut in half and the condition number of the
problem improves when we eliminate phase shift degrees of freedom via
symmetry rather than including them in the penalty function described
in Section~\ref{method}.  Although we do not make use of it, there is
a procedure known as the Meyer-Marsden-Weinstein reduction
\cite{meyer,marsden:weinstein} that allows one to reduce the dimension
of a symplectic manifold on which a group acts symplectically.  This
allows one to eliminate actions of the group (e.g.~translations) from
the phase space.  Equilibria and periodic solutions of the reduced
Hamiltonian system correspond to (families of) relative equilibria and
relative periodic solutions \cite{wulff:bif:rel} of the original
system.

This paper is organized as follows.  In Section~\ref{sec:lin}, we
discuss stationary, traveling and particle solutions of B-O, linearize
about traveling waves, and classify all bifurcations predicted by
linear theory from traveling waves to non-trivial time-periodic
solutions.  Some of the more technical material from this section
is given in Appendix~\ref{sec:T:a0:a}.  In Section~\ref{sec:num}, we
present a collection of numerical experiments using our continuation
method to follow several paths of non-trivial solutions beyond the
realm of linear theory in order to formulate a theorem that gives the
global mapping from one traveling wave bifurcation to its counterpart
on the other side of the path.  In Section~\ref{sec:exact}, we study
the behavior of the Fourier modes of the time-periodic solutions found
in Section~\ref{sec:num} and state a theorem about the exact form of
these solutions, which is proved in Appendix~\ref{sec:proof}.
Finally, in Section~\ref{sec:interior}, we discuss interior
bifurcations from these paths of already non-trivial solutions to
still more complicated solutions.  Although the existence of such a
hierarchy of solutions was already known \cite{satsuma:ishimori:79},
bifurcation between various levels of the hierarchy has not previously
been discussed.

\section{Bifurcation from Traveling Waves}\label{sec:lin}

In this section, we study the linearization of the Benjamin-Ono
equation about stationary solutions and traveling waves by solving an
infinite dimensional eigenvalue problem in closed form.  Each
eigenvector corresponds to a time-periodic solution of the linearized
equation.  The traveling case is reduced to the stationary case by
requiring that the period of the perturbation (with a suitable spatial
phase shift) coincide with the period of the traveling wave.  The main
goal of this section is to devise a classification scheme of the
bifurcations from traveling waves so that in later sections we can
describe which (local) bifurcations are connected together by a global
path of non-trivial time-periodic solutions.

\subsection{Stationary, Traveling and Particle Solutions}
\label{sec:stat:trav}

We consider the Benjamin-Ono equation on the periodic interval
$\mathbb{R}\big/2\pi\mathbb{Z}$, namely
\begin{equation} \label{eqn:BO}
u_{t}=Hu_{xx}-uu_{x}.
\end{equation}
Here $H$ is the Hilbert transform, which has the symbol
$\hat{H}(k)=-i\operatorname{sgn}(k).$ The Benjamin-Ono equation
possesses solutions \cite{case:mero,benj1} of the form
\begin{equation} \label{eqn:beta:rep}
  u(x,t) = \alpha_0 + \sum_{l=1}^N \phi(x;\beta_l(t)),
\end{equation}
where $\alpha_0$ is the mean, $\beta_1(t)$, \dots, $\beta_N(t)$ are
the trajectories of $N$ particles evolving in the unit disk
$\Delta$ of the complex plane and governed by the ODE
\begin{equation} \label{eqn:beta:ode}
  \dot{\beta}_l = \sum_{\parbox{.3in}{$\js m=1\\[-8pt]m\ne l$}}^N
  \frac{-2i\beta_l^2}{\beta_l-\beta_m} +
  \sum_{m=1}^N \frac{2i\beta_l^2}{\beta_l - \bar{\beta}_m^{-1}}
  + i(2N-1-\alpha_0)\beta_l, \qquad (1\le l\le N),
\end{equation}
and $\phi(x;\beta)$ is the function with Fourier representation
\begin{equation}\label{eqn:ubeta:hat}
  \hat{\phi}(k;\beta) = \left\{
  \begin{array}{cl}
    0, & k=0 \\
    2\beta^k, & k>0 \\
    2\bar{\beta}^{|k|}, & k<0
  \end{array}
  \right\}, \qquad
  \beta\in\Delta=\{z:|z|<1\}.
\end{equation}
The function $\phi(x;\beta)$ has a peak centered at $x=\arg(\bar{\beta})$
with amplitude growing to infinity as $|\beta|$ approaches~1.  The
$N$-hump traveling waves (with a spatial period of $2\pi/N$) are a
special case of the particle solutions given by (\ref{eqn:beta:rep})
and (\ref{eqn:beta:ode}):
\begin{equation}\label{eqn:N:trav}
  u_\text{trav}(x,t;\alpha_0,N,\beta) =
  \alpha_0 + \sum_{l=1}^N \phi(x;\beta_l(t)),
  \quad \beta_l(t) = \sqrt[N]{\beta} e^{-ict}, \quad
  c = \alpha_0 - N\alpha(\beta).
\end{equation}
Each $\beta_l$ is assigned a distinct $N$th root of $\beta$ and
$\alpha(\beta)$ is the mean of the one-hump stationary solution, namely
\begin{equation}\label{eqn:alpha:beta}
  \alpha(\beta) = \frac{1 - 3|\beta|^2}{1 - |\beta|^2}, \qquad\qquad
  |\beta|^{2} = \frac{1 - \alpha(\beta)}{3 - \alpha(\beta)}.
\end{equation}
The solution (\ref{eqn:N:trav}) moves to the right when $c>0$.  Indeed,
it may also be written
\begin{equation}\label{eqn:N:trav:stat}
  u_\text{trav}(x,t;\alpha_0,N,\beta)=u_\text{stat}(x-ct;N,\beta)+c,
\end{equation}
where $u_\text{stat}$ is the $N$-hump stationary solution
\begin{equation}\label{eqn:N:stat}
  u_\text{stat}(x;N,\beta) =
  N\alpha(\beta) + \!\!\!\!\!\sum_{\{\gamma\,:\,\gamma^N=\beta\}}
  \!\!\!\phi(x;\gamma)
  \;=\; N\alpha(\beta) + N\phi(Nx;\beta).
\end{equation}
The Fourier representation of $u_\text{stat}$ is
\begin{equation} \label{eqn:uhat:k}
  \hat{u}_\text{stat}(k;N,\beta) = \begin{cases}
    N\alpha(\beta), & k=0, \\
    2N\beta^{k/N}, & k\in N\mathbb{Z}, \; k>0, \\
    2N\bar{\beta}^{|k|/N}, & k\in N\mathbb{Z}, \; k<0, \\
    0 & \text{otherwise.}
  \end{cases}
\end{equation}
Amick and Toland have shown \cite{amick:toland:BO} that all traveling
waves of the Benjamin-Ono equation have the form
(\ref{eqn:N:trav:stat}); see also \cite{benj3}.

\subsection{Linearization about Stationary Solutions}
\label{sec:lin:stat}

Let $u(x)=u_\text{stat}(x;N,\beta)$ be an $N$-hump stationary solution.  In
\cite{benj1}, we solved the linearization of (\ref{eqn:BO}) about $u$,
namely
\begin{equation} \label{eqn:BO:lin}
  v_t = Hv_{xx}-(uv)_x=iBAv, \qquad A=H\partial_x-u, \qquad
  B = \frac{1}{i}\partial_x,
\end{equation}
by substituting $v(x,t)=\real\{Cz(x)e^{i\omega t}\}$ into (\ref{eqn:BO:lin})
and solving the eigenvalue problem
\begin{equation}
  BAz=\omega z
\end{equation}
in closed form.  Specifically, we showed that the eigenvalues
$\omega_{N,n}$ are given by
\begin{equation} \label{eqn:omega:Nn}
  \omega_{N,n} = \left\{\begin{array}{lr}
    -\omega_{N,-n}, & n<0 \\
    \phm 0 & n=0 \\
    (n)(N-n),
      & 1\le n\le N-1 \\
      (n+1-N)\big(n+1+N(1-\alpha(\beta))\big),\hspace*{-.5in} &
      n\ge N\end{array}\right\} \quad
  \parbox[c][1in][b]{1.5in}{
\includegraphics[height=1in]{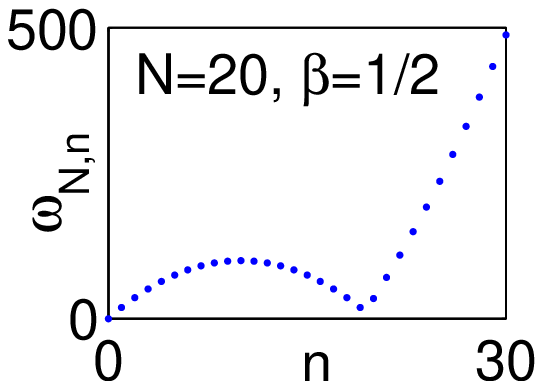}}
\end{equation}
The zero eigenvalue $\omega_{N,0}=0$ has geometric multiplicity two
and algebraic multiplicity three.  The eigenfunctions in the kernel
of $BA$ are
\begin{equation} \label{eqn:kernel:BA}
  z_{N,0}^\e{1,0}(x) = -\der{}{x} u_\text{stat}(x;N,\beta),
  \qquad z_{N,0}^\e2(x) = \der{}{|\beta|} u_\text{stat}(x;N,\beta),
\end{equation}
which correspond to changing the phase or amplitude of $\beta$ in the
underlying stationary solution.  There is also a
Jordan chain \cite{Achain} of length two associated with
$z_{N,0}^\e{1,0}(x)$, namely
\begin{equation}\label{eqn:Jordan}
  z_{N,0}^\e{1,1}(x) = 1, \qquad \left(iBA z_{N,0}^\e{1,1} =
    z_{N,0}^\e{1,0}\right),
\end{equation}
which corresponds to the fact that adding a constant to a stationary
solution causes it to travel.  The fact that all the
eigenvalues $i\omega_{N,n}$ in the linearization (\ref{eqn:BO:lin})
are purely imaginary is a consequence of the Hamiltonian structure
\cite{case:remarkable} of the Benjamin-Ono equation.  For
non-Hamiltonian systems, one does not generally expect to 
find time-periodic perturbations of traveling waves (as periodic
solutions of the linearized problem may not even exist).

The eigenfunctions $z_{N,n}(x)$ corresponding to positive eigenvalues
$\omega_{N,n}$ (with $n\ge1$) have the Fourier representation
\begin{align}
  \notag
  &\hat{z}_{N,n}(k)\Big\vert_{k=n+jN} = \left\{\begin{array}{cc}
      \left(1+\frac{N(|j|-1)}{N-n}\right)\bar\beta^{|j|-1} & j<0 \\[5pt]
      C\left(1 + \frac{Nj}{n}\right)\beta^{j+1} & j\ge0
      \end{array}\right\},  \qquad
    \Bigg(\parbox{1.56in}{\begin{center}$1\le n\le N-1$ \\
          $C = \frac{-n N}{(N-n)\big[n+(N-n)|\beta|^2\big]}$
        \end{center}}\Bigg), \\[5pt]
\label{eqn:evec:formulas}
 &\hat{z}_{N,n}(k)\Big\vert_{k=n+1-N+jN} = \left\{\begin{array}{cc}
    0 & j<0 \\[5pt]
    \frac{-\bar\beta}{(1-|\beta|^2)^2}\left[
      1 - \left(1-\frac{N}{n+1}\right)|\beta|^2\right] & j=0 \\[5pt]
    \left(1 + \frac{N(j-1)}{n+1}\right)\beta^{j-1} & j>0
  \end{array}\right\}, \qquad(n\ge N),
\end{align}
with all other Fourier coefficients equal to zero.  The eigenfunctions
corresponding to negative eigenvalues $\omega_{N,n}$ (with $n\le-1$)
satisfy $z_{N,n}(x)=\overline{z_{N,-n}(x)}$, so the Fourier
coefficients appear in reverse order, conjugated.  For $1\le n\le
N-1$, any linear combination of $z_{N,n}(x)$ and $z_{N,N-n}(x)$ is
also an eigenfunction; however, the choices here seem most natural as
they simultaneously diagonalize the shift operator (discussed below)
and yield directions along which non-trivial solutions exist beyond
the linearization.  Said differently, we have listed the first $N-1$
positive eigenvalues $\omega_{N,n}$ in an unusual order (rather than
enumerating them monotonically and coalescing multiple eigenvalues)
because this is the order that leads to the simplest description of
the global paths of non-trivial solutions connecting these traveling
waves.

\subsection{Classification of bifurcations from traveling waves}
\label{sec:classify}

Time-periodic solutions of the Benjamin-Ono equation with period $T$
have initial conditions that satisfy $F(u_0,T)=0$, where
$F:H^1\times\mathbb{R}\rightarrow H^1$ is given by
\begin{equation} \label{eqn:F:def}
  F(u_0,T) = u(\cdot,T)-u_{0}, \qquad u_t = Hu_{xx} - uu_x, \qquad
  u(\cdot,0)=u_0.
\end{equation}
We begin by linearizing $F$ about an $N$-hump stationary soution
$u_0(x)=u_\text{stat}(x;N,\beta)$.  The Fr\'echet derivative
$DF=(D_1F,D_2F):H^1\times\mathbb{R}\rightarrow H^1$ yields
directional derivatives
\begin{equation}
  \begin{aligned}
    D_1F(u_0,T)v_0 &= \der{}{\veps}\Big\vert_{\veps=0}F(u_0+\veps v_0,T) =
    v(\cdot,T)-v_0 = \left[e^{iBAT}-I\right]v_0, \\[2pt]
    D_2F(u_0,T)\tau &=
    \der{}{\veps}\Big\vert_{\veps=0}F(u_0,T+\veps\tau) = 0.
  \end{aligned}
\end{equation}
Note that $v_0\in\ker D_1F(u,T)$ if and only if the solution $v(x,t)$ of the
linearized problem is periodic with period $T$.  As a result, a basis
for the kernel $\mc{N}=\ker DF(u_0,T)$ consists of $(0;1)$ together
with all pairs $(v_0;0)$ of the form
\begin{equation} \label{eqn:v0:re:im}
  v_0(x) = \real\{z_{N,n}(x)\} \qquad \text{ or } \qquad
  v_0(x) = \imag\{z_{N,n}(x)\},
\end{equation}
where $n$ ranges over all integers such that
\begin{equation}\label{eqn:omega:condition}
  \omega_{N,n}T\in \, 2\pi\mathbb{Z}
\end{equation}
with $N$ and $\beta$ (in the formula (\ref{eqn:omega:Nn}) for
$\omega_{N,n}$) held fixed.  The corresponding periodic solutions of
the linearized problem are
\begin{equation} \label{eqn:v:re:im}
  v(x,t)=\real\{z_{N,n}(x)e^{i\omega_{N,n} t}\} \qquad\text{or}\qquad
  v(x,t)=\imag\{z_{N,n}(x)e^{i\omega_{N,n} t}\}.
\end{equation}
Negative values of $n$ have already been accounted for
in (\ref{eqn:v0:re:im}) and (\ref{eqn:v:re:im}) using $z_{N,-n}(x)=
\overline{z_{N,n}(x)}$, and the $n=0$ case always yields two vectors
in the kernel, namely those in (\ref{eqn:kernel:BA}).  These
directions do not cause bifurcations as they lead to other stationary
solutions.

Next we wish to linearize $F$ about an arbitrary traveling wave.
Suppose $u(x)=u_\text{stat}(x;N,\beta)$ is an $N$-hump stationary
solution and $U(x,t)=u(x-ct)+c$ is a traveling wave.  Then the
solutions $v$ and $V$ of the linearizations about $u$ and $U$,
respectively, satisfy $V(x,t)=v(x-ct,t)$.  Note also that
\begin{equation} \label{eqn:trav:F:0}
  F(U_0,T)=0 \quad \text{iff} \quad
  cT = \frac{2\pi \nu}{N} \quad \text{for some }
  \nu\in\mathbb{Z},
\end{equation}
where $U_0(x)=U(x,0)=u(x)+c$.  Note that $\nu$ is the number of times
the traveling wave turns over itself in one period.  Assuming
(\ref{eqn:trav:F:0}) holds, we set $\theta=2\pi\nu/N$ and compute
\begin{equation} \label{eqn:DF:trav}
\begin{aligned}[c]
  [D_1F(U_0,T)v_0](x) &= v(x-cT,T) - v_0(x) =
  [(S_\theta e^{iBAT} - I)v_0](x), \\
  [D_2F(U_0,T)\tau](x) &= U_t(x,T)\tau = -c u_x(x-cT)\tau = -cu_x(x)\tau,
\end{aligned}
\end{equation}
where $v$ solves (\ref{eqn:BO:lin}) and the shift operator $S_\theta$ 
is defined via
\begin{equation} \label{eqn:S:def}
  S_\theta z(x) = z(x-\theta), \qquad\qquad
  \hat{S}_{\theta,kl} = e^{-ik\theta}\delta_{kl}.
\end{equation}
One element of $\mc{N}=\ker DF(U_0,T)$ arises from (\ref{eqn:Jordan}),
which gives
$$e^{iBAt}1=1-tu_x \qquad \Rightarrow \qquad
D_1F(U_0,T)(-c/T)+D_2F(U_0,T)1=0,$$
and implies~$(-c/T;1)\in\mc{N}$.  This just means that we can change
the period $T$ by a small amount $\tau$ by adding the constant
$-(c/T)\tau$ to $U_0$; (this also follows from the condition
(\ref{eqn:trav:F:0}) that $cT=\theta=\text{const}$). If we wish to
change the period without changing the mean, we need to simultaneously
adjust $|\beta|$ in the underlying stationary solution
$u(x)=u_\text{stat}(x;N,\beta)$.  The other elements of $\mc{N}$ are
of the form $(v_0;0)$ with
\begin{equation} \label{eqn:v0:re:im2}
  v_0(x) = \real\{z_{N,n}(x)\} \qquad \text{ or } \qquad
  v_0(x) = \imag\{z_{N,n}(x)\}.
\end{equation}
The admissible values of $n$ here are found using (\ref{eqn:DF:trav})
together with
\begin{equation}
  S_\theta e^{iBAT}z_{N,n} = e^{i(\omega_{N,n} T - \theta k_{N,n})}z_{N,n},
  \qquad \theta = \frac{2\pi\nu}{N},
\end{equation}
where $k_{N,n}$ is the stride offset of the
non-zero Fourier coefficients of $z_{N,n}$, i.e.
\begin{equation}
  \hat{z}_{N,n}(k)\ne0 \quad \Rightarrow \quad k-k_{N,n}\in N\mathbb{Z}.
\end{equation}
Thus, instead of (\ref{eqn:omega:condition}), $n$ ranges over all
integers such that
\begin{equation} \label{eqn:omega:condition:trav}
  \omega_{N,n}T \in \,
  2\pi\left(\frac{\nu k_{N,n}}{N}+\mathbb{Z}\right), \qquad
  k_{N,n} = \begin{cases}
    -k_{N,-n}, & n<0, \\
    0 & n=0, \\
    n & 1\le n\le N-1, \\
    \opn{mod}(n+1,N) & n\ge N.
  \end{cases}
\end{equation}
As before, negative values of $n$ need not be considered once we take
real and imaginary parts in (\ref{eqn:v0:re:im2}), and the $n=0$ case
always gives the two vectors $(z_{N,0}^\e{1,0};0),
(z_{N,0}^\e{2};0)\in\mc{N}$, which lead to other traveling waves rather
than bifurcations to non-trivial solutions.

Our numerical experiments have led us to the following conjecture,
which we prove as part of Theorem~\ref{thm:bif:trav} in
Section~\ref{sec:exact}:

\begin{conjecture} For every $\beta\in\Delta$ and $(N,\nu,n,m)\in
\mathbb{Z}^4$ satisfying
\begin{equation}\label{eqn:rules}
  N\ge1, \qquad \nu\in\mathbb{Z}, \qquad n\ge1, \qquad m\ge1, \qquad
  m\in \nu k_{N,n}+N\mathbb{Z},
\end{equation}
there is a four parameter sheet of non-trivial time-periodic solutions
bifurcating from the $N$-hump traveling wave with speed index $\nu$,
($cT=2\pi\nu/N$), bifurcation index $n$, and oscillation index $m$,
($\omega_{N,n}T=2\pi m/N$).  The phase and amplitude of the traveling
wave are determined by $\beta$.
\end{conjecture}

The main content of this conjecture is that we do not have to consider
linear combinations of the $z_{N,n}$ with different values of $n$ to
find periodic solutions of the non-linear problem --- this basis is
already ``diagonal'' with respect to these bifurcations.  This is true
in spite of a small divisor problem preventing $DF(U_0,T)$ from being
Fredholm.
The decision to number the first
$N-1$ eigenvalues $\omega_{N,n}$ non-monotonically in
(\ref{eqn:omega:Nn}) and to simultaneously diagonalize the shift
operator $S_\theta$ when choosing eigenvectors $z_{N,n}$ in
(\ref{eqn:evec:formulas}) was essential to make this work.  Formulas
relating the period, $T$, the mean, $\alpha_0$, and the decay
parameter, $|\beta|$, for each of these bifurcations are given in
Appendix~\ref{sec:T:a0:a} along with a list of bifurcation rules
governing ``legal'' values of the mean.

A canonical way to
generate one of these bifurcations is to take $\beta$ real and perturb
the initial condition in the direction $v_0(x)=\real\{z_{N,n}(x)\}$.
This leads to non-trivial solutions with even symmetry at $t=0$.
Perturbation in the $\imag\{z_{N,n}(x)\}$ direction yields the same
set of non-trivial solutions, but with a spatial and temporal phase
shift:
\begin{equation}\label{eqn:phase:shift}
  \imag\{z_{N,n}(x-ct)e^{i\omega t}\} =
  \real\left\{z_{N,n}\left(\left(x-\frac{c\pi}{2\omega}\right) -
  c\left(t-\frac{\pi}{2\omega}\right)\right)
e^{i\omega\left(t-\frac{\pi}{2\omega}\right)}\right\},
\end{equation}
where $\omega=\omega_{N,n}$.
The manifold of non-trivial solutions is four dimensional with two
essential parameters (e.g.~the mean $\alpha_0$ and a parameter
governing the distance from the traveling wave) and two inessential
parameters (the spatial and temporal phase).  In our numerical
studies, we use the real part of a Fourier coefficient $c_k$ of the
initial condition (with $k$ such that $\hat{z}_{N,n}(k)\ne0$) for the
second essential bifurcation parameter.  When we discuss exact
solutions in Section~\ref{sec:exact}, a different parameter will be
used.

We remark that this enumeration of bifurcations accounts for all
time-periodic solutions of the linearization about traveling waves;
therefore, the heuristic that each bifurcation of the non-linear
problem gives rise to a linearly independent vector in the kernel
$\mc{N}$ of the linearized problem suggests that we have found all
bifurcations from traveling waves.  Interestingly, this turns out not
to be the case; the interior bifurcations we discuss in
Section~\ref{sec:interior} can occur at the endpoints of the path,
allowing for degenerate bifurcations directly from traveling waves to
higher levels in the infinite hierarchy of time-periodic solutions.
Only the transition from the first level of the hierarchy to the
second is ``visible'' to a linearized analysis about traveling waves.
The other transitions become linearly dependent on these in the
limit as the traveling wave is approached; they will be analyzed
in \cite{benj3}.

\section{Numerical Experiments} \label{sec:num}

In this section we present a collection of numerical experiments in
which we start with a given bifurcation $(N,\nu,n,m,\beta)$ and use a
symmetric variant of the method we described in \cite{benj1} for finding
periodic solutions of non-linear PDE to continue these solutions until
another traveling wave is found, or until the solution blows up as the
bifurcation parameter approaches a critical value.  We determine the
bifurcation indices $(N',\nu',n',m')$ at the other end of
the path of non-trivial solutions by fitting the data to the formulas
of the previous section.  By trial and error, we are then able to
guess a formula relating $(N',\nu',n',m')$ to $(N,\nu,n,m)$ that we
use in Section~\ref{sec:exact} to construct exact solutions.

\subsection{Numerical Method}
\label{method}

As mentioned in Section~\ref{sec:classify}, a natural choice of
spatial and temporal phase can be achieved by choosing the parameter
$\beta$ of the traveling wave to be real and perturbing the initial
condition in the direction $v_0(x) = \real\{z_{N,n}(x)\}$.  For
reasons of efficiency and accuracy (explained in the introduction), we
now restrict our search for time-periodic solutions of (\ref{eqn:BO})
to functions $u(x,t)$ that possess even spatial symmetry at $t=0$.
If we succeed in finding solutions with this symmetry, then they
(together with their phase-shifted counterparts analogous to
(\ref{eqn:phase:shift})) span the nullspace $\mc{N}=\ker DF(U_0,T)$
in the limit that the perturbation goes to zero.  Thus, we do not
expect symmetry breaking bifurcations from traveling waves that
cannot be phase shifted to have even symmetry at $t=0$.

The Benjamin-Ono equation has the property that if $u(x,t)$ is a
solution of (\ref{eqn:BO}), then so is $U(x,t)=u(-x,-t)$.  As a
result, if $u$ is a solution such that $u(x,T/2)=U(x,-T/2)$, then
$u(x,T)=U(x,0)$, i.e.~$u$ is time-periodic if the initial condition
has even symmetry.  Thus, we seek initial conditions $u_0$ with even
symmetry and a period $T$ to minimize the functional
\begin{equation} \label{eqn:Gtot:def}
  G_\text{tot}(u_0,T) = G(u_0,T) + G_\text{penalty}(u_0,T),
\end{equation}
where
\begin{equation} \label{eqn:Gsym:def}
  G(u_0,T) = \frac{1}{2}\int_0^{2\pi} [u(x,T/2)-u(2\pi-x,T/2)]^2\,dx
\end{equation}
and $G_\text{penalty}(u_0,T)$ is a non-negative penalty function to
impose the mean and set the bifurcation parameter.  To compute the
gradient of $G$ with respect to variation of the initial conditions,
we use
\begin{equation}
  \frac{d}{d\veps}\biggr|_{\veps=0}G(u_0+\veps v_0,T) = \int_0^{2\pi}
  \varder{G}{u_0}(x)v_0(x)\,dx,
\end{equation}
where the variational derivative
\begin{equation}\label{eqn:G:var1}
  \varder{G}{u_0}(x) = 2w(x,T/2), \qquad w_0(x) = u(x,T/2)-u(2\pi-x,T/2)
\end{equation}
is found by solving the following adjoint equation from $s=0$ to $s=T/2$:
\begin{equation}
  w_s(x,s) = -Hw_{xx}(x,s) + u(x,\jt\frac{T}{2}-s)w_x(x,s), \qquad
  w(\cdot,0) = w_0.
\end{equation}
Since $v_0$ is assumed symmetric in this formulation, (\ref{eqn:G:var1})
is equivalent to
\begin{equation}\label{eqn:G:var2}
  \varder{G}{u_0}(x)=w(x,T/2)+w(2\pi-x,T/2).
\end{equation}
The Benjamin-Ono and adjoint equations are solved using a
pseudo-spectral collocation method employing a fourth order
semi-implicit additive Runge-Kutta method \cite{cooper,carpenter,
  wilk228A} to advance the solution in time.  The BFGS method
\cite{bfgs,nocedal} is then used to minimize $G_\text{tot}$ (varying
the period and the Fourier coefficients of the initial conditions).
We use the penalty function
\begin{equation} \label{eqn:phi:def}
  G_\text{penalty}(u_0,T) = \frac{1}{2}\Bigl(
    [a_0(0)-\alpha_0]^2 + [a_K(0)-\rho]^2\Bigr)
\end{equation}
to specify the mean $\alpha_0$ and the real
part $\rho$ of the $K$th Fourier coefficient of the initial
condition
\begin{equation}
  u_0(x) = \sum_{k=-M/2+1}^{M/2} c_k(0)e^{ikx}, \qquad
  c_k(t) = a_k(t) + ib_k(t).
\end{equation}
The parameters $\alpha_0$ and $\rho$ serve as the bifurcation
parameters while the phases are determined by requiring that the
solution have even symmetry at $t=0$.  We generally choose $K$ to be
the first $k\ge1$ such that $\hat{z}_{N,n}(k)\ne0$.

Our continuation method consists of three stages.  First, we choose a
traveling wave and a set of bifurcation indices to begin the path of
non-trivial solutions.  We also choose a direction in which to vary
the bifurcation parameter $\rho$ and the mean $\alpha_0$.  In most of
our numerical experiments, we hold $\alpha_0$ fixed; however, in the
example of Figure~\ref{fig:football} below, we vary $\rho$ and
$\alpha_0$ simultaneously.  The traveling wave serves as the zeroth
point on the path.  The initial guess for the first point on the path
is obtained by perturbing the initial condition of the traveling wave
in the direction $\real\{z_{N,n}(x)\}$.  We use the period $T$ given
in (\ref{eqn:T:alpha}) in Appendix~\ref{sec:T:a0:a} as a starting
guess.  We then use the minimization algorithm to descend from the
starting guess predicted by linear theory to an actual time-periodic
solution.  The second stage of the continuation algorithm consists of
varying $\rho$ (and possibly $\alpha_0$), using linear extrapolation
for the starting guess (for $u_0$ and $T$) of the next solution, and
then minimizing $G_\text{tot}$ to find an actual time-periodic
solution with these values of $\rho$ and $\alpha_0$.  If the initial
value of $G_\text{tot}$ from the extrapolation step is too large, we
discard the step and try again with a smaller change in $\rho$ and
$\alpha_0$.  The final stage of the algorithm consists of identifying
the reconnection on the other side of the path.  We do this by blindly
overshooting the target values of $\rho$ and $\alpha_0$ (which we do
not know in advance).  Invariably, the algorithm will lock onto a
family of traveling waves once we reach the end of the path of
non-trivial solutions.  We look at the Fourier coefficients of the
last non-trivial solution before the traveling waves are reached and
match them with the formulas for $\hat{z}_{N',n'}(k)$ to determine the
correct bifurcation indices on this side of the path.  (A prime indicates
indices for the bifurcation at the other end of the path.)  We then
recompute the last several solutions on the path of non-trivial
solutions with appropriate values of $\rho$ and $\alpha_0$ to arrive
exactly at the traveling wave on the last iteration.  We sometimes
change $K$ in (\ref{eqn:phi:def}) to compute this reconnection to
avoid $\hat{z}_{N',n'}(K)=0$.

The running time of our algorithm (on a 2.4 GHz desktop machine)
varies from a few hours to compute one of the paths labeled $a$--$l$
in (\ref{eqn:one:hump:1})--(\ref{eqn:two:hump:234}) below, to a few
days to compute a path in which the solution blows up, such as the one
shown in Figure~\ref{fig:evolBe} below.  We always refine the mesh and
timestep enough so that the solutions are essentially exact (with
$G_\text{tot}\le 10^{-26}$ in the easy cases and $10^{-20}$ in the
hard cases).

\subsection{Global paths of non-trivial solutions}
\label{sec:global}

We now investigate the global behavior of non-trivial solutions that
bifurcate from arbitrary stationary or traveling waves.  We find that
these non-trivial solutions act as rungs in a ladder, connecting
stationary and traveling solutions with different speeds and
wavelengths by creating or annihilating oscillatory humps that grow or
shrink in amplitude until they become part of the stationary or
traveling wave on the other side of the rung.  In some cases, rather
than re-connecting with another traveling wave, the
solution blows up (i.e.~the $L^2$-norm of the initial condition
grows without bound) as the bifurcation parameter $\rho$ approaches a
critical value.  However, even in these cases a re-connection with
another traveling wave does occur if, in addition to $\rho$,
we vary the mean, $\alpha_0$, in an appropriate way.

\begin{figure}[t]
\begin{center}
\includegraphics[width=.8\linewidth,trim=10 10 0 5,clip]{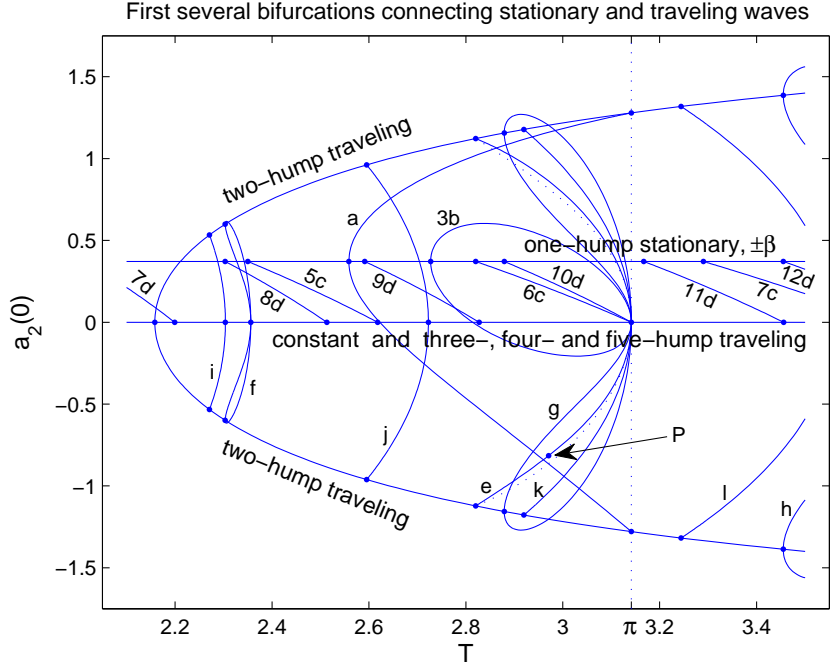}
\end{center}
\caption{
Paths of non-trivial solutions listed in equations
(\ref{eqn:one:hump:1})--(\ref{eqn:two:hump:234}).  The second Fourier
mode of the eigenvector $z_{N,n}(x)$ in the linearization is non-zero
for the pitchfork bifurcations and is zero for the one-sided,
oblique-angle bifurcations. The point labeled P corresponds to the
solution in Fig.~\ref{fig:evol2} below. }
\label{fig:bifur4}
\end{figure}

\begin{figure}[t]
\begin{center}
\includegraphics[width=.8\linewidth,trim=0 20 0 0]{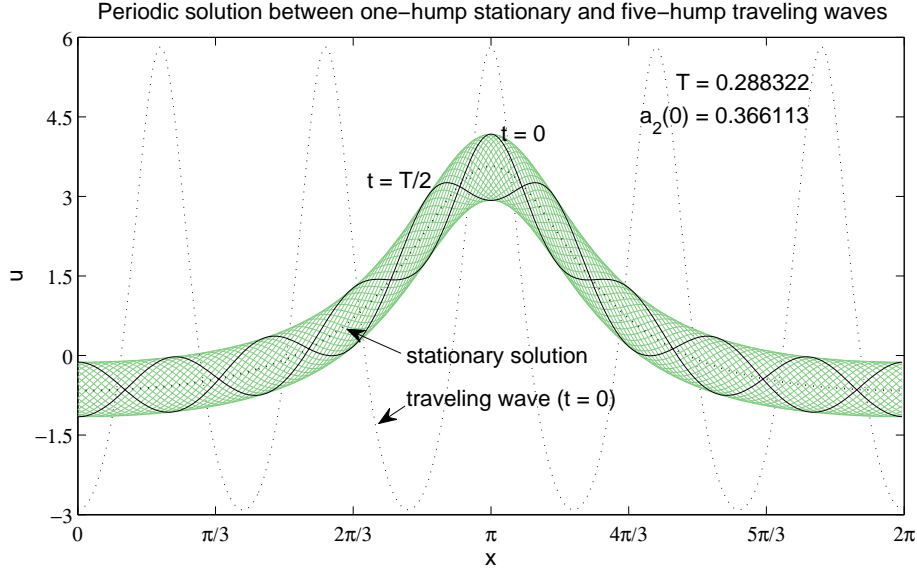}
\end{center}
\caption{Periodic solution on path $d$ connecting the one-hump
  stationary solution to the five-hump left-traveling wave
  ($\alpha_0=0.544375$).  The second Fourier mode of $z_{1,4}(x)$ is
  zero, which explains why $a_2(0)=0.366113$ for this solution is only
  $1.35\%$ of the way between the stationary solution
  $a_2(0)=0.371087$ and the five-hump traveling wave $a_2(0)=0$.  }
\label{fig:evolAa}
\end{figure}

Recall from Section~\ref{sec:classify} that we can enumerate all such
bifurcations by specifying a complex parameter $\beta$ in the unit
disk $\Delta$ along with four integers $(N,\nu,n,m)$ satisfying
(\ref{eqn:rules}), and in most cases we can solve for $|\beta|$
in terms of the mean, $\alpha_0$, using (\ref{eqn:alpha:alpha0})
in Appendix~\ref{sec:T:a0:a}.
In \cite{benj1}, we presented a detailed study of the solutions
on the path connecting a one-hump stationary solution to a two-hump
traveling wave moving left.  We denote this path by
\begin{equation} \label{eqn:one:hump:1}
a: \quad  (1,0,1,1) \qquad \longleftrightarrow \qquad (2,-1,1,1),
\end{equation}
where the label $a$ refers to the bifurcation diagram in
Figure~\ref{fig:bifur4}.  We have also computed the next several
bifurcations ($n=2,3,4$) from the one-hump stationary solution and
found that they connect up with a traveling wave with $N'=n+1$ humps
moving left with speed index $\nu'=-1$, where we denote the
bifurcation on the other side of the path by $(N',\nu',n',m')$.  By
comparing the Fourier coefficients of the last few non-trivial
solutions on these paths to those of the linearization about the
$N'$-hump traveling wave, we determined that the bifurcation and
oscillation indices satisfy $n'=n$ and $m'=1$, respectively.  Studying
these reconnections revealed that the correct way to number the
eigenvalues $\omega_{N',n'}$ was to split the double eigenvalues with
$n'<N'$ apart as we did in (\ref{eqn:omega:Nn}) by simultaneously
diagonalizing the shift operator and ordering the $\omega_{N',n'}$ via
the stride offset of the corresponding eigenvectors (rather than
monotonically).  Using this ordering, the non-trivial solutions
connect up with the $N'$-hump traveling wave along the $z_{N',n'}$
direction (without involving $z_{N',N'-n'}$).  These results are
summarized as
\begin{equation} \label{eqn:one:hump:234}
\begin{aligned}
 b: \quad  (1,0,2,1) &\qquad \longleftrightarrow \qquad (3,-1,2,1), \\
 c: \quad  (1,0,3,1) &\qquad \longleftrightarrow \qquad (4,-1,3,1), \\
 d: \quad  (1,0,4,1) &\qquad \longleftrightarrow \qquad (5,-1,4,1).
\end{aligned}
\end{equation}
The labels $a$, $b$,
$c$, $d$ in (\ref{eqn:one:hump:1}) and (\ref{eqn:one:hump:234})
correspond to the paths labeled $7d$, $8d$, $5c$, $a$, etc.~in
the bifurcation diagram.
When an integer
$p$ precedes a label, it means that the period $T$ that is plotted
is $p$ times larger than the fundamental period of the solution
represented. Thus, curve $7d$ is the image of
curve $d$ (not shown) under the linear transformation $(T,a_2)\mapsto
(7T,a_2)$.  In our labeling scheme, we just need to multiply
$\nu$, $m$, $\nu'$, $m'$ by $p$ to obtain the new path, e.g.
\begin{equation}
  7d: \quad  (1,0,4,7) \qquad \longleftrightarrow \qquad (5,-7,4,7).
\end{equation}
In this diagram, we plot $a_2(0)$ vs.~$T$ with the spatial and
temporal phases chosen so the solution is even at $t=0$.  For example,
on path $d$, as we decrease $\rho=a_2(0)$ from
$0.371087$ to 0, the solution transitions from the one-hump
stationary solution to the five-hump left-traveling wave as
shown in Figure~\ref{fig:evolAa}.

It is interesting that the paths labeled $a$ and $3b$ in
Figure~\ref{fig:bifur4} meet the one-hump stationary solutions in a
pitchfork while the other paths (such as $5c$ and $8d$) meet at an
oblique angle from one side only.  This is because the second Fourier
mode of the eigenvector $z_{1,n}(x)$ in the linearization about the
stationary solution is zero in these latter cases, so the change in
$a_2(0)$ from that of the stationary solution (namely $0.371087$) is a
higher order effect, (as is the change in $T$).  This explains the
oblique angle.  We now explain why these bifurcations occur from one
side only. When we go beyond the linearization as we have here, we
find that $c_2(t)=a_2(t)+ib_2(t)$ has a nearly circular
(epitrochoidal) orbit in case $a$, a circular orbit in case $b$, and
remains constant in time in cases $c$ and $d$ (see
Section~\ref{sec:exact}).  If one branch of the pitchfork corresponds
to $a_2(0)$, the other is $a_2(T/2)$ since the function $u(\cdot,T/2)$
also has even symmetry.  But in cases $c$ and $d$, $a_2(0)$ is equal
to $a_2(T/2)$ even though the functions $u(\cdot,0)$ and
$u(\cdot,T/2)$ are different.  These cases also become pitchforks when
a different Fourier coefficient $a_K(0)$ is used as the bifurcation
parameter.

Next we compute the first several bifurcations from the two-hump
traveling waves with mean $\alpha_0=0.544375$ and speed index
$\nu=-1$.  We set $N=2$, $\nu=-1$, $n\in\{1,2,3,4\}$ and choose the
first several legal $m$ values, i.e.~values of $m$ that satisfy the
bifurcation rules of Figure~\ref{fig:rules} in Appendix~\ref{sec:T:a0:a}.
For example, the curves labeled $i$, $j$,
$k$ and $l$ in Figure~\ref{fig:bifur4} correspond to the
bifurcations $(2,-1,4,m)$ with $m=11,13,15,17$; smaller values (and
even values) of $m$ are not allowed.  In addition to the path $a$ in
(\ref{eqn:one:hump:1}) above, we obtain the paths
\begin{equation} \label{eqn:two:hump:234}
\begin{aligned}
 e: &  & (2,-1,2,3) &\; \longleftrightarrow \; (3,-3,1,3), \\
 f: &  & (2,-1,3,6) &\; \longleftrightarrow \; (4,-5,2,6), \\
 g: &  & (2,-1,3,8) &\; \longleftrightarrow \; (4,-6,2,8), \\
 h: &  & (2,-1,3,10) &\; \longleftrightarrow \; (4,-7,2,10),
\end{aligned} \qquad
\begin{aligned}
 i: &  & (2,-1,4,11) &\; \longleftrightarrow \; (5,-8,3,11), \\
 j: &  & (2,-1,4,13) &\; \longleftrightarrow \; (5,-9,3,13), \\
 k: &  & (2,-1,4,15) &\; \longleftrightarrow \; (5,-10,3,15), \\
 l: &  & (2,-1,4,17) &\; \longleftrightarrow \; (5,-11,3,17).
\end{aligned}
\end{equation}
The paths $f$, $g$ and $h$ meet the curve representing the two-hump
traveling waves in a pitchfork bifurcation while the others meet
obliquely from one side.  This, again, is an anomaly of having chosen
the second Fourier mode for the bifurcation parameter.  The dotted
line near the path $e$ is the curve obtained when $e$ is reflected
across the $T$-axis.  Solutions on this dotted line correspond to
solutions on path $e$ shifted by $\pi/2$ in space, which changes the
sign of $\rho=a_2(0)$ but also breaks the even symmetry of the
solution at $t=0$.  The paths labeled $i$, $j$, $k$ and $l$ are
exactly symmetric when reflected about the $T$-axis because $c_2(t)$
has a circular orbit centered at zero in these cases.  It is
interesting that so many of the paths in this bifurcation diagram
terminate when $T=\pi$ (or a simple rational multiple of $\pi$).  This
is due to the fact that $T$ in (\ref{eqn:T:alpha}) in
Appendix~\ref{sec:T:a0:a} is independent of $\alpha$ when $n<N$.

The solutions $u(x,t)$ corresponding to points along the paths $b$,
$c$ and $d$ are qualitatively similar to each other.  As shown in
Figure~\ref{fig:evolAa}, these solutions look like $N'$-hump waves
traveling over a stationary one-hump carrier signal.  At one end of
the path the high frequency wave may be viewed as a perturbation of
the one-hump stationary solution, while at the other end of the path
it is more appropriate to regard the stationary solution as the
perturbation, causing the traveling wave to bulge upward as it passes
near $x=\pi$ and downward near $x=0$ and $x=2\pi$.  In all these
cases, the solution repeats itself when one of the high frequency
waves has moved left one slot to assume the shape of its left neighbor
at $t=0$.

\begin{figure}[t]
\begin{center}
\mypsdraft
\includegraphics[width=.78\linewidth,trim=0 10 0 0]{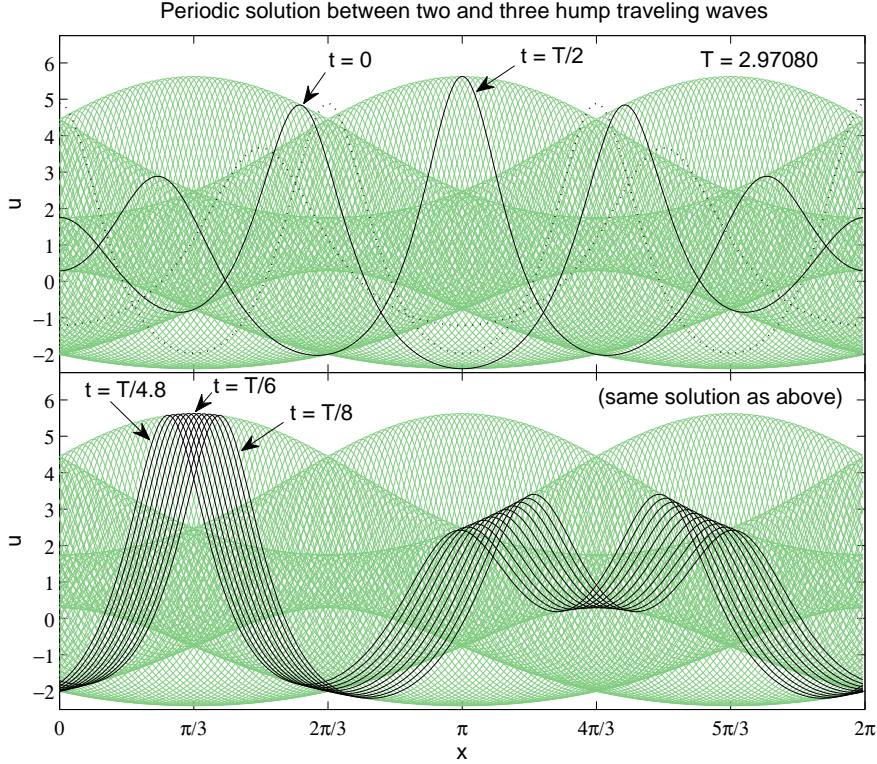}
\mypsfull
\end{center}
\caption{
Time-periodic solution (labeled P in Fig.~\ref{fig:bifur4}) 
on path $e$ connecting two- and three-hump
traveling waves.  The amplitude of each hump oscillates as it
travels left.  The dotted curves in the top panel represent the
traveling waves at each end of the path at $t=0$.
}
\label{fig:evol2}
\end{figure}

\ignore{
\begin{figure}[p]
\begin{center}
\includegraphics[width=.65\linewidth,trim=0 10 0 20,clip]{figs/ripplesEai}
\end{center}
\caption{
Three dimensional plot of the solution in figure~\ref{fig:evol2} highlighting
the braided pattern swept out by the oscillating waves.
}
\label{fig:ripples}
\end{figure}
}

By contrast, the solutions that bifurcate from the two-hump traveling
waves, i.e.~those on the paths listed in (\ref{eqn:two:hump:234}),
have the property that when a wave has moved left one slot to the
location that its neighbor occupied at $t=0$, it has acquired a
different shape and must keep progressing a number of slots before it
finally lines up with one of the initial waves.  This is illustrated
in Figure~\ref{fig:evol2} for the solution labeled P in
Figure~\ref{fig:bifur4} on the path
\begin{equation} \label{eqn:2:hump:2}
  e: \quad (2,-1,2,3) \qquad \longleftrightarrow \qquad (3,-3,1,3).
\end{equation}
This solution is qualitatively similar to the linearized solution
$(3,-3,1,3)$. There are $N'=3$ humps oscillating with the same
amplitude but with different phases as they travel left.  They do not
line up with the initial condition again until they have traveled
three slots ($\nu'=-3$) and progressed through one cycle
($m'/N'=3/3$), which leads to a braided effect when the time history
of the solution is plotted on one graph.  All the solutions on path
$e$ are \emph{irreducible} in the sense that there is no smaller time
$T$ in which they are periodic (unlike the cases labeled $3b$, $5c$,
$7d$ etc.~in Figure~\ref{fig:bifur4}, which are reducible to $b$, $c$
and $d$, respectively).  Note that although $\nu'=-3$ and $m'=3$ are
both divisible by 3, we cannot reduce $(3,-3,1,3)$ to $(3,-1,1,1)$ as
the latter indices violate the bifurcation rules of
Figure~\ref{fig:rules} in Appendix~\ref{sec:T:a0:a}.  We also mention
that at the beginning of the path, near $(2,-1,2,3)$, the braiding
effect is not present; instead, the solution can be described as two
humps bouncing out of phase as they travel left.  In one period, they
each travel left one slot ($\nu=-1$) and bounce 1.5 times ($m/N=3/2$)
to assume the shape of the other hump at $t=0$.  The transition from
this behavior to the braided behavior occurs at the point on path $e$
that a third hump becomes recognizable in the wave profile.  The
solutions on the paths $f$, $g$, $h$, $i$, $j$, $k$ and $l$ are
similar to those on path $e$, but the braiding patterns are more
complicated near the right end-points of these paths.

\begin{figure}[p]
\begin{center}
\mypsdraft
\includegraphics[width=.7\linewidth,trim=0 20 0 0]{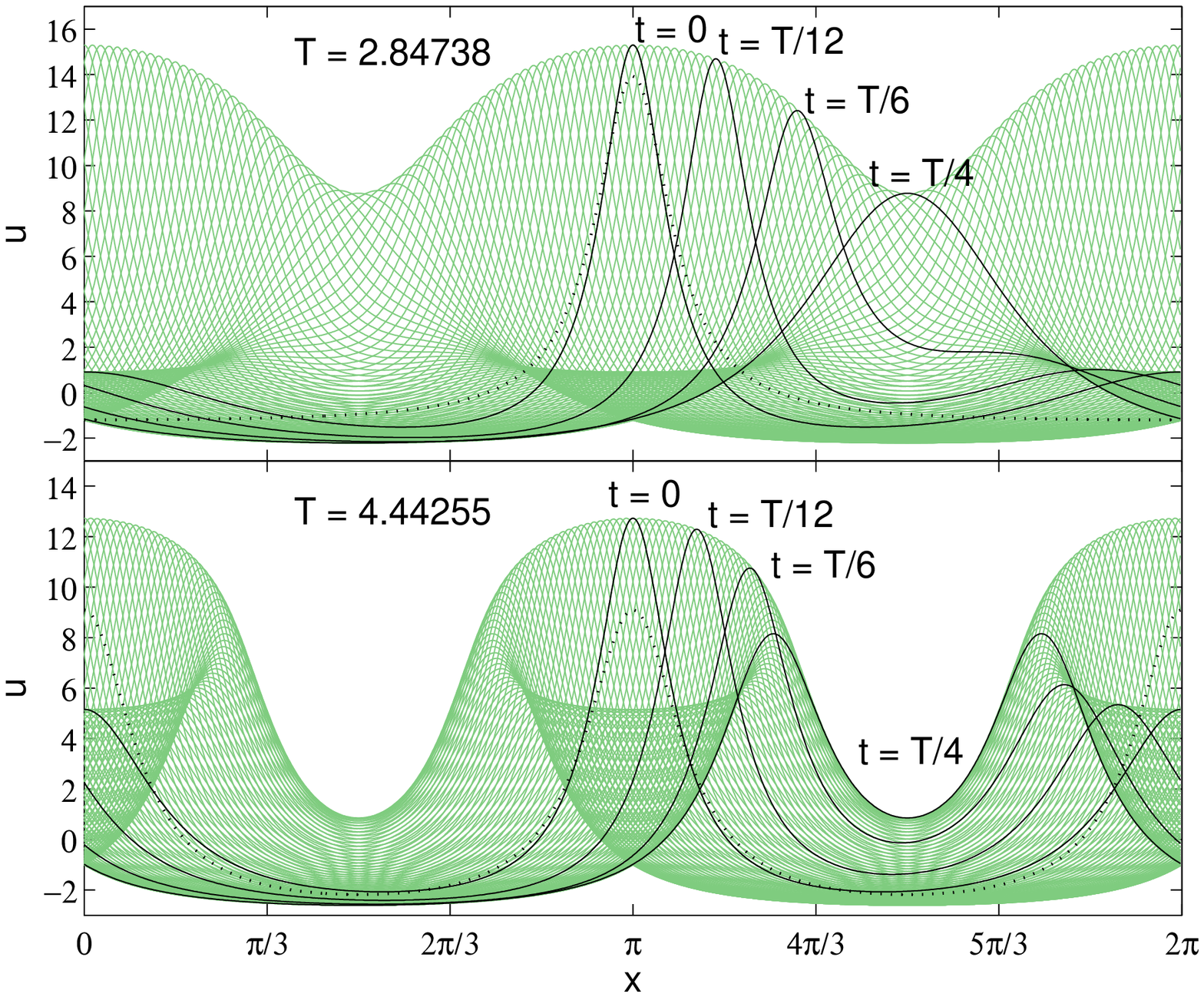}
\mypsfull
\end{center}
\caption{ Periodic solutions with mean $\alpha_0=0.544375$ between the
  one-hump right-traveling wave (dotted curve, top panel) and the
  two-hump stationary solution (dotted curve, bottom panel).
  \emph{Top:} a large, right-traveling soliton temporarily merges with
  a small, left traveling soliton at $t=\frac{T}{4}$ and
  $t=\frac{3}{4}T$.  \emph{Bottom:} two solitons traveling in opposite
  directions bounce off each other at $\frac{T}{4}$ and $\frac{3}{4}T$
  and change direction.  }
\label{fig:evolBd}
\end{figure}

All the traveling waves we have described until now move left.
To see what happens to a right-moving wave, we computed the first
bifurcation from the simplest such case and obtained the path
\begin{equation} \label{eqn:one:hump:112}
  (1,1,1,2) \qquad \longleftrightarrow \qquad (2,0,1,2).
\end{equation}
Thus, the one-hump right-traveling wave is connected to the two-hump
stationary solution.  Solutions near the left end of this path consist
of a large-amplitude, right-moving soliton traveling over a
small-amplitude, left-moving soliton.  As we progress along the path,
the amplitude of the left-moving soliton increases until the solitons
cease to fully merge at $t=T/4$ and $t=3T/4$.  Instead, a dimple forms
in the wave profile at these times and the solitons begin to bounce
off each other, trading amplitude so the right-moving wave is larger
than the left-moving wave.  This type of behavior has also been
observed by Leveque \cite{leveque} for the KdV equation for solitons
of nearly equal amplitude.  Both types of behavior (merging and
bouncing off one another) are illustrated in Figure~\ref{fig:evolBd}.
As we proceed further along this path, the solitons settle into a
synchronized dancing motion without changing their shape or deviating
far from their initial positions.
Eventually the ``dancing amplitude'' becomes
small and the non-trivial solution turns into a stationary
two-hump solution.

\begin{figure}[p]
\begin{center}
\includegraphics[height=1.7in,trim=0 20 0 0]{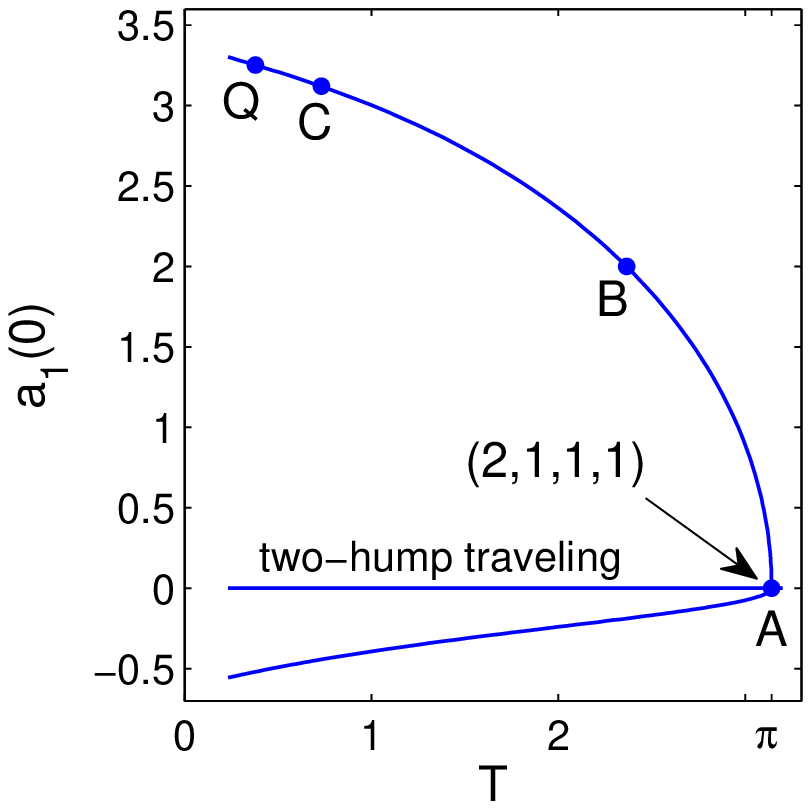}
\mypsdraft
\includegraphics[height=1.7in,trim=0 20 0 0]{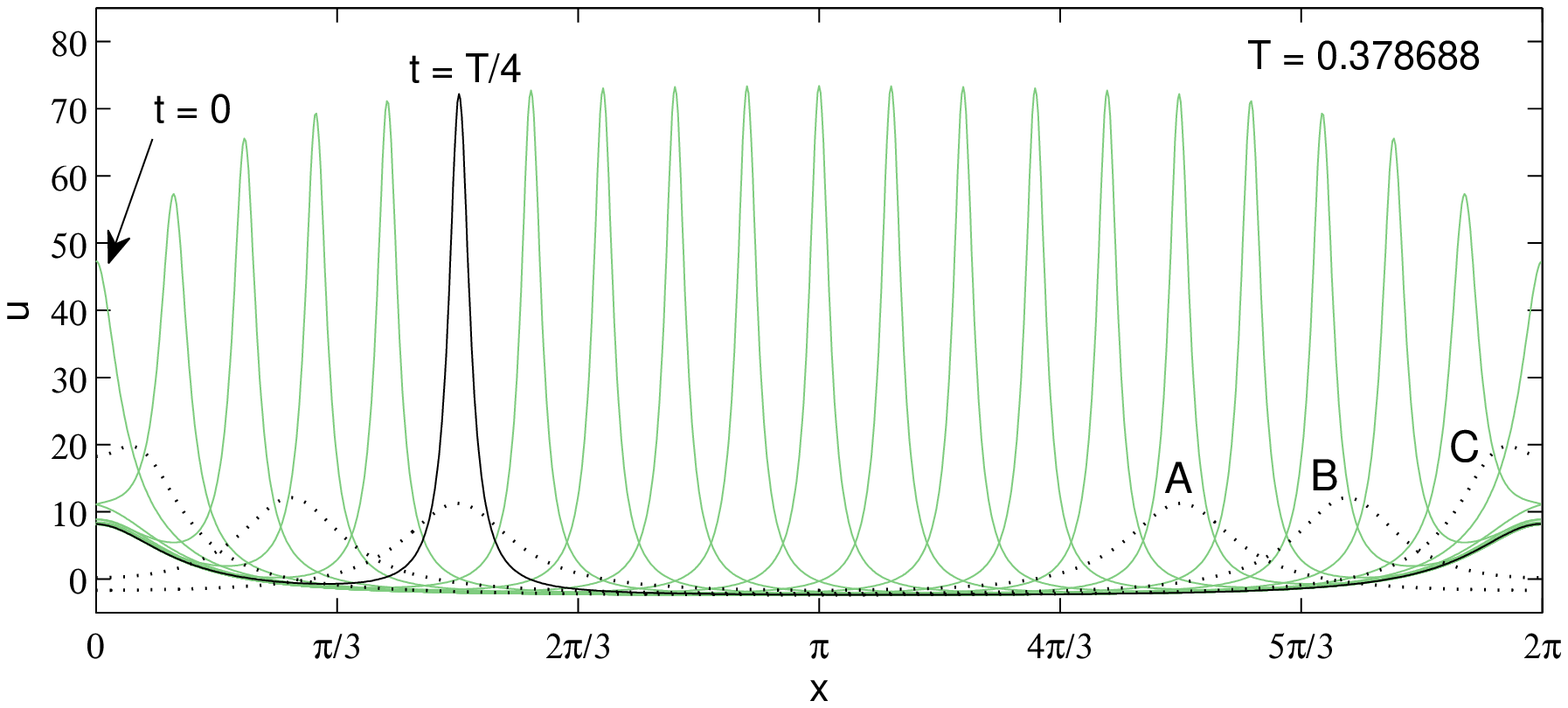}
\mypsfull
\end{center}
\caption{\emph{Left:} path of non-trivial solutions with mean
  $\alpha_0=1.2$ that bifurcates with indices $(2,1,1,1)$ from the
  two-hump traveling wave.  These solutions do not re-connect with
  another traveling wave, but instead blow up as $T\rightarrow0$.  The
  solution Q is shown at right, where a large, right-moving
  soliton travels rapidly over a small, stationary hump. The dotted
  curves are initial conditions for the points labeled A, B, C at
  left. }
\label{fig:evolBe}
\end{figure}

In order to guess a general formula for the relationship between
two traveling waves that are connected by a path of non-trivial
solutions, we generated two additional paths, namely
\begin{equation} \label{eqn:23hump:23}
\begin{aligned}
  (2,0,2,2) &\qquad \longleftrightarrow \qquad (3,-1,1,2), \\
  (3,0,3,3) &\qquad \longleftrightarrow \qquad (4,-1,1,3).
\end{aligned}
\end{equation}
After studying all the paths listed in
(\ref{eqn:one:hump:1})--(\ref{eqn:23hump:23}),
we propose the following conjecture, which we prove as part
of Theorem~\ref{thm:bif:trav} in Section~\ref{sec:exact}:

\begin{conjecture} \label{conj2}
The four-parameter sheet of non-trivial solutions with bifurcation
parameters $(N,\nu,n,m)$ coincides with the sheet with parameters
$(N',\nu',n',m')$ if and only if
\begin{alignat}{5}
  \label{eqn:conj1}
  \text{if } n &< N: &\qquad
  N' &= N-n, &\quad
  \nu' &= \frac{(N-n)\nu+m}{N}, &\quad
  n' &= N-1, &\quad
  m' &= m, \\
  \label{eqn:conj2}
  \text{if } n & \ge N: &
  N' &= n+1, &
  \nu' &= \frac{(n+1)\nu - m}{N}, &
  n' &= n+1-N, &
  m' &= m.
\end{alignat}
\end{conjecture}

By symmetry, we may interchange the primed and unprimed indices in
either formula; thus, $N'>N \;\Leftrightarrow\; n<N
\;\Leftrightarrow\; n'\ge N'$.  In most of our numerical calculations,
$N'$ turned out to be larger than $N$.  In the exact formulas of
Section~\ref{sec:exact}, we find it more convenient to adopt the
convention that $N'<N$ since, in that case, all the solutions on the
path connecting these traveling waves turn out to be $N$-particle
solutions as described in Section~\ref{sec:stat:trav}.

Equations (\ref{eqn:conj1}) and
(\ref{eqn:conj2}) are consistent with the bifurcation rules of
Appendix~\ref{sec:T:a0:a} in that
\begin{align}
  n<N,\;\;  m\in n\nu+N\mathbb{Z} \quad &\Rightarrow \quad
  \nu'\in\mathbb{Z}, \;\; m'\in(n'+1)\nu'+N'\mathbb{Z}, \\
  n\ge N,\;\; m\in (n+1)\nu+N\mathbb{Z} \quad &\Rightarrow \quad
  \nu'\in\mathbb{Z},\;\;  m'\in n'\nu'+N'\mathbb{Z}.
\end{align}
However, if the mean is held constant, they do not necessarily respect
the requirements on $\alpha_0$ listed in Figure~\ref{fig:rules} in
Appendix~\ref{sec:T:a0:a}.  For example, if $\alpha_0\le3$, then
$(2,1,1,1)$ is a valid bifurcation, but the re-connection $(1,1,1,1)$
predicted by (\ref{eqn:conj1}) is legal only if $\alpha_0=3$.
Interestingly, when we use our numerical method to follow the path of
non-trivial solutions that bifurcates from $(2,1,1,1)$ with the mean
$\alpha_0=1.2$ held constant, it does not connect up with another
traveling wave.  Instead, as illustrated in Figure~\ref{fig:evolBe},
as we vary the bifurcation parameter, the two humps (of the solutions
labeled A,B,C) grow in amplitude and merge together until they become
a single soliton traveling very rapidly on top of a small amplitude
stationary hump.  As the bifurcation parameter $\rho=a_1(0)$
approaches a critical value, the period $T$ approaches zero and the
solution blows up in $L^2(0,2\pi)$ with the Fourier coefficients of
any time-slice decaying more and more slowly.

\begin{figure}[t]
\begin{center}
\includegraphics[height=1.7in,trim=0 20 0 0]{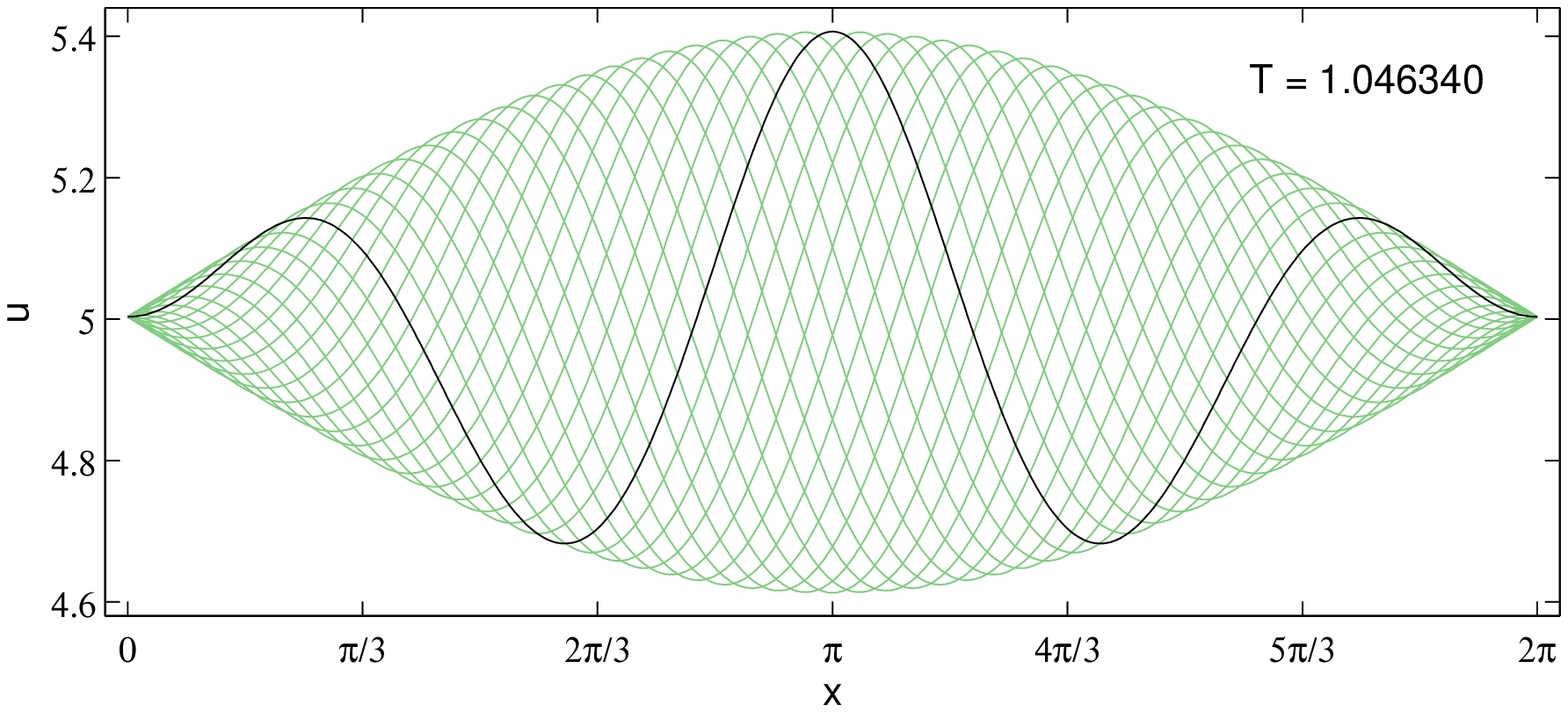}
\includegraphics[height=1.7in,trim=0 20 0 0]{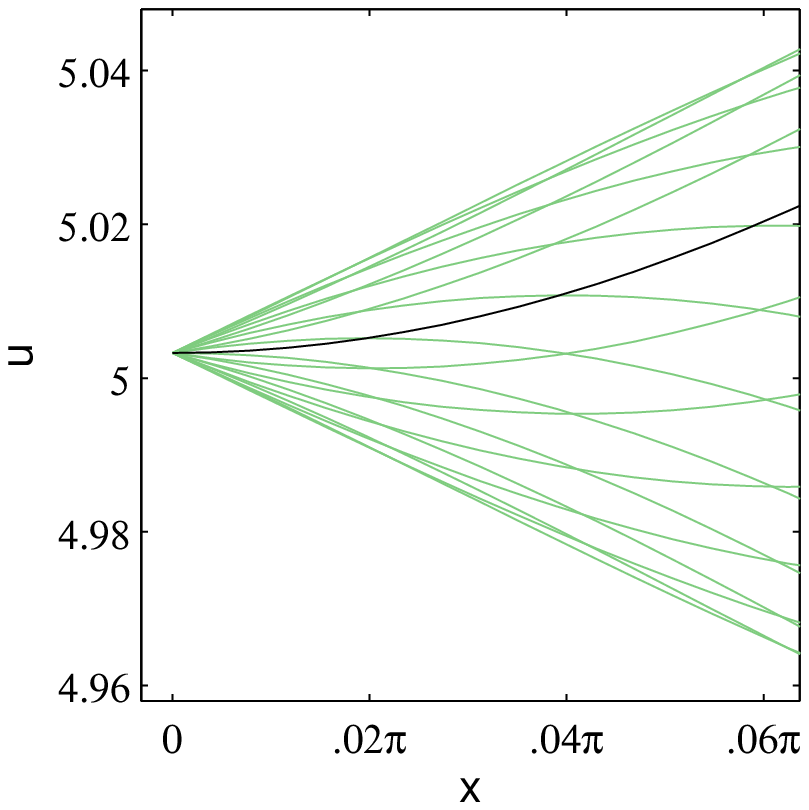}
\end{center}
\caption{\emph{Left:} One of the solutions on the path from
$\Big\{(3,1,1,1),\, \beta=-\sqrt{\frac{1}{31}}\Big\}$ to
$\Big\{(2,1,2,1),\, \beta=\frac{1}{40}\Big\}$ consists of a traveling
wave inside a football-shaped envelope.  The exact solution appears to be
of the form $u(x,t)=A+B\left(\sin \frac{x}{2}\right)
\sin\left(\frac{5}{2}x-\frac{2\pi}{T}t\right)$.
}
\label{fig:football}
\end{figure}

As another example, the bifurcation $(3,1,1,1)$ is valid when
$\alpha_0\le5$ but the reconnection $(2,1,2,1)$ is only valid if
$\alpha_0=5$.  If we hold $\alpha_0<5$ constant, the solution blows up
as we vary $\rho=a_2(0)$ from 0 to a critical value.
However, \emph{if we
simultaneously vary the mean} so that it approaches 5, we do
indeed reach a traveling wave with bifurcation indices $(2,1,2,1)$.
To check this numerically, we started at $(3,1,1,1)$ with
$\alpha_0=4.8$ (which has $\alpha=\frac{14}{15}$,
$|\beta|=1/\sqrt{31}$) and computed 40 solutions varying $\rho$
from 0 to $0.1$ and setting $\alpha_0 = 4.8+2\rho$.  The
bifurcation at the other end turned out to be $(2,1,2,1)$ with
$\alpha_0=5$, $\beta=\frac{1}{4}\rho=0.025$,
$\alpha=(1-3\beta^2)/(1-\beta^2)$, $T=\pi/(5-2\alpha)$, as predicted
by Conjecture~\ref{conj2}.
The solutions on this path have the interesting property that the
envelope of the solution pinches off into a football shape at one
point in the transition from the three-hump traveling wave to the
two-hump traveling wave.  Using a bracketing technique, we were able
to find a solution such that the value of $u(0,t)$ remained constant
in time to 8 digits of accuracy.  The result is shown in
Figure~\ref{fig:football}.

In summary, it appears that the family of bifurcations with indices
$(N,\nu,n,m)$ is always connected to the family with indices
$(N',\nu',n',m')$ given by (\ref{eqn:conj1}) and (\ref{eqn:conj2})
by a sheet of non-trivial solutions,
but we often have to vary both the mean and a Fourier coefficient of
the initial condition to achieve a re-connection.  Thus, the manifold
of non-trivial solutions is genuinely two-dimensional (or
four dimensional if phase shifts are included).  Some of its
important properties cannot be seen if we hold the mean $\alpha_0$
constant.

\section{Exact Solutions} \label{sec:exact}

In this section we use data fitting techniques to determine the
analytic form of the numerical solutions of Section~\ref{sec:num}.  We
then state a theorem that confirms our numerical predictions and
explains why some paths of solutions reconnect with traveling waves
when the mean is held fixed while others lead to blow-up.  The theorem
is proved in Appendix~\ref{sec:proof}.

\subsection{Fourier Coefficients and Lattice Sums}

\begin{figure}[p]
\begin{center}
\includegraphics[height=3in,trim=5 0 38 0, clip]{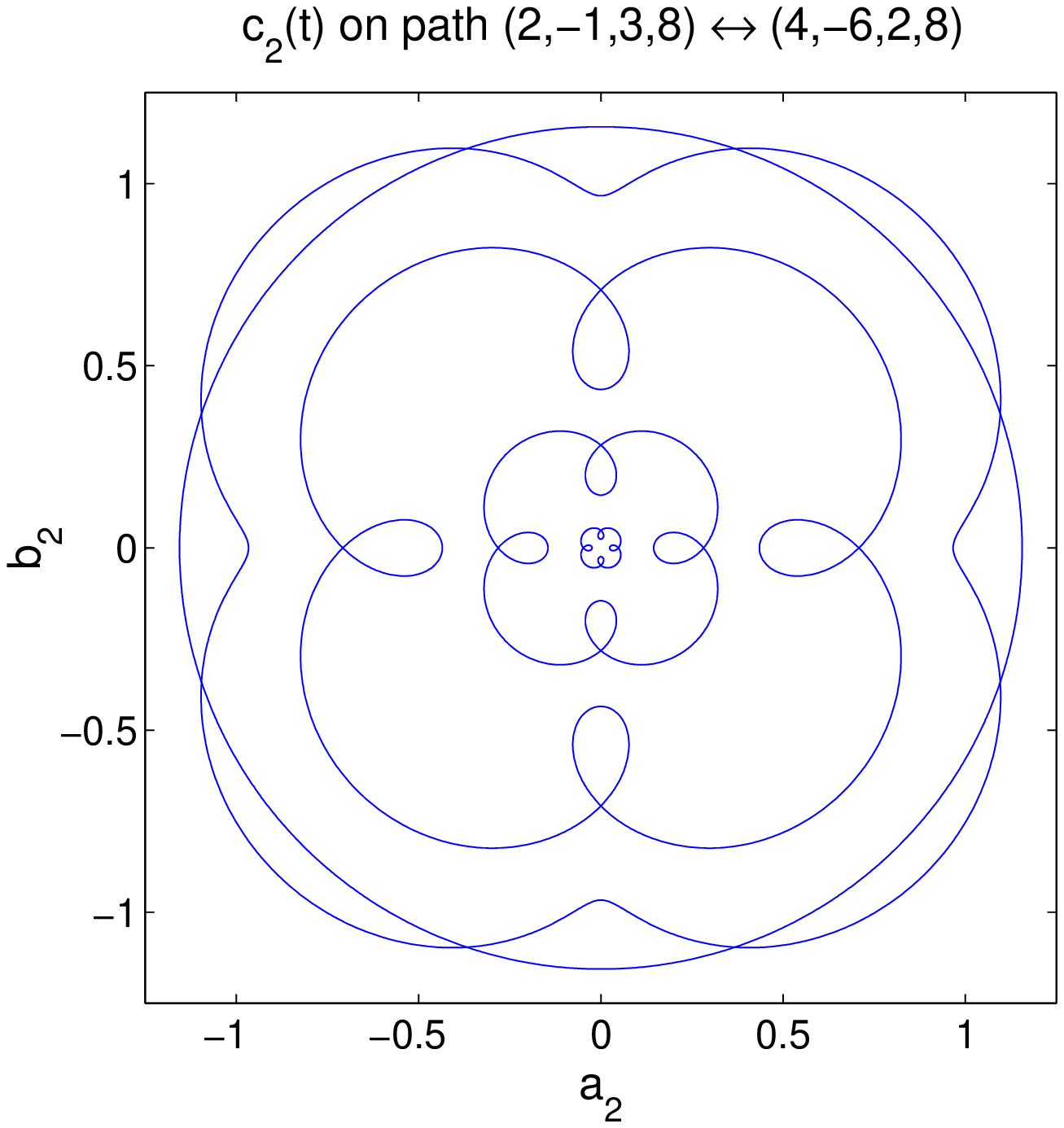}
\hfill
\includegraphics[height=3in,trim=20 0 35 0, clip]{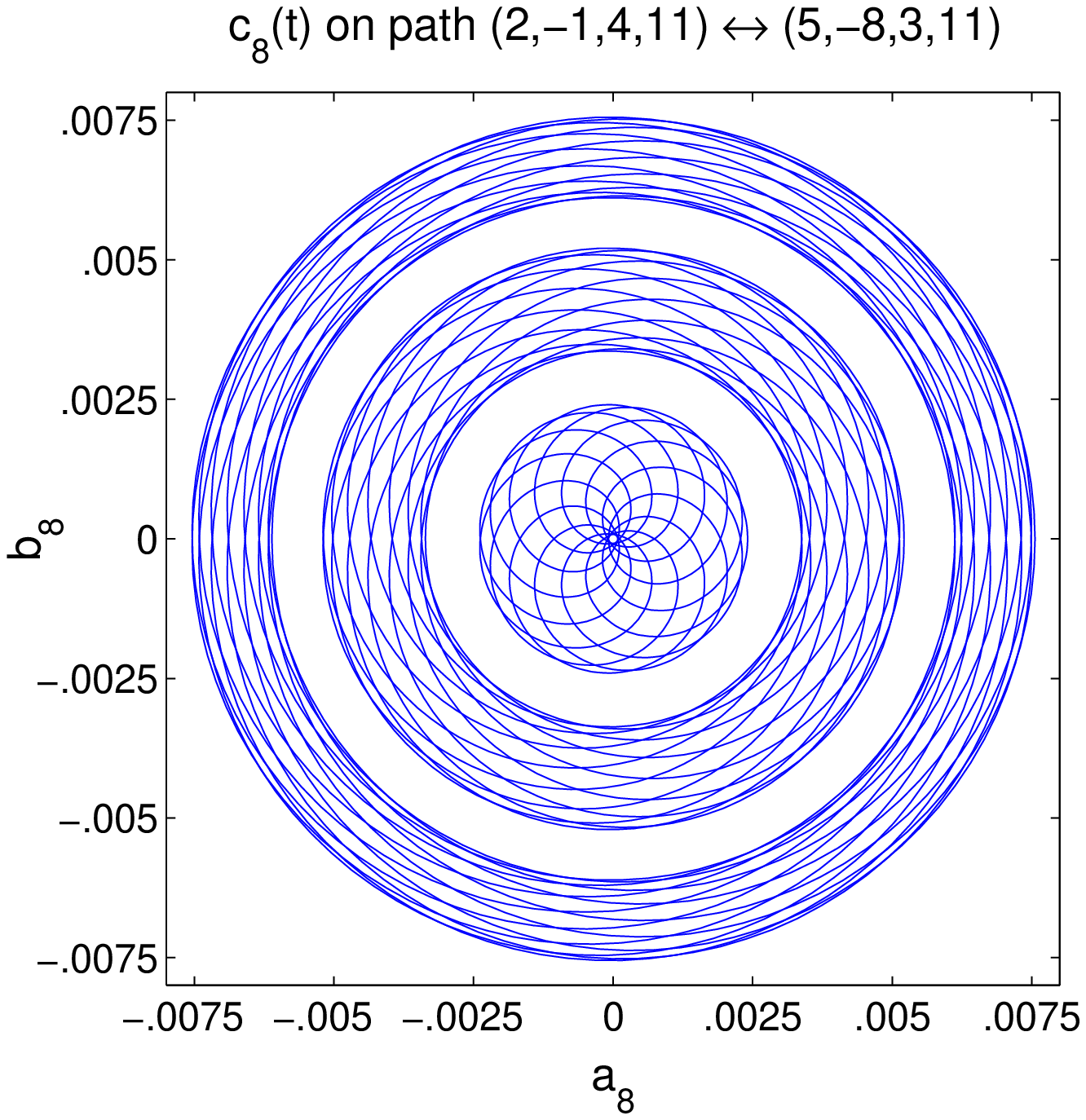}
\end{center}
\caption{
\emph{Left:} Trajectories $c_2(t)$ for
five solutions on path $g$ in (\ref{eqn:two:hump:234}).  The evolution
of $c_2(t)$ on paths $f$ and $h$ in (\ref{eqn:two:hump:234}) are
similar, but with three- and five-fold symmetry rather than four.
\emph{Right:} Trajectories $c_8(t)$ for three solutions on path $i$
in (\ref{eqn:two:hump:234}).
}
\label{fig:spiro}
\end{figure}

One striking feature of the time-periodic solutions we have found
numerically is that the trajectories of the Fourier modes $c_k(t)$ are
often circular or nearly circular.  Other Fourier modes have more
complicated trajectories resembling cartioids, flowers and many other
familiar ``spirograph'' patterns (see Figure~\ref{fig:spiro}).  This
led us to experiment with data fitting to try to guess the analytic
form of these solutions.  The first thing we noticed was that the
trajectories of the spatial Fourier coefficients are band-limited in
time, with the width of the band growing linearly with the wave number:
\begin{equation} \label{eqn:exact:form}
  u(x,t) = \sum_{k=-\infty}^\infty c_k(t)e^{ikx}
  , \qquad
  c_k(t) = \sum_{j=-\infty}^\infty c_{kj}e^{-ij\frac{2\pi}{T}t},
  \qquad 
  c_{kj}=0 \text{\,\, if\,\, } |j|>r|k|.
\end{equation}
Here $r$ is a fixed positive integer (depending on which path of
non-trivial solutions $u$ belongs to) and the $c_{kj}$ are real
numbers when a suitable choice of spatial and temporal phase is made.
Since $u$ is real, these coefficients satisfy $c_{-k,-j} = c_{kj}$.

\begin{figure}[p]
\begin{center}
\includegraphics[height=1.42in]{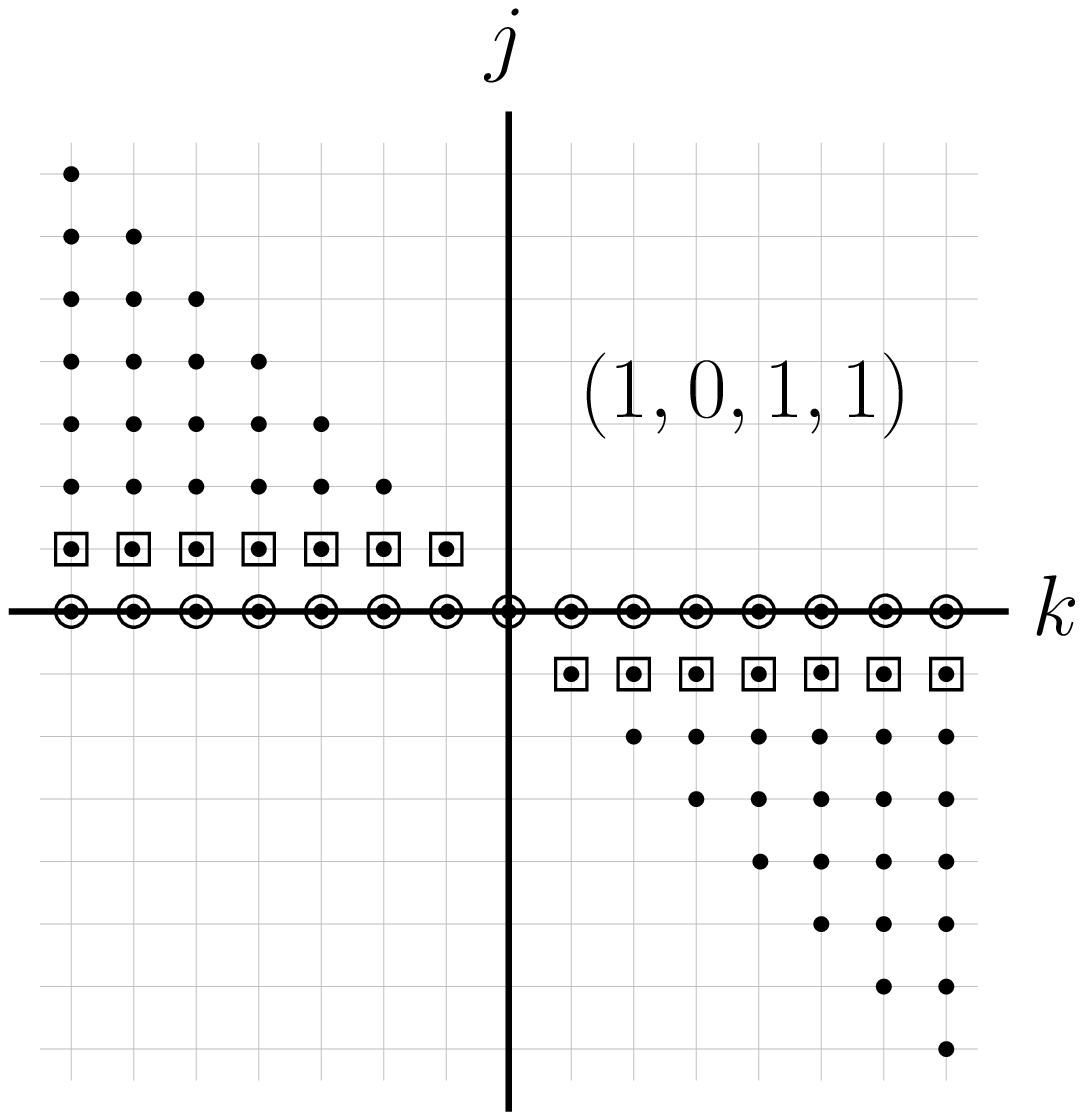}
\includegraphics[height=1.42in]{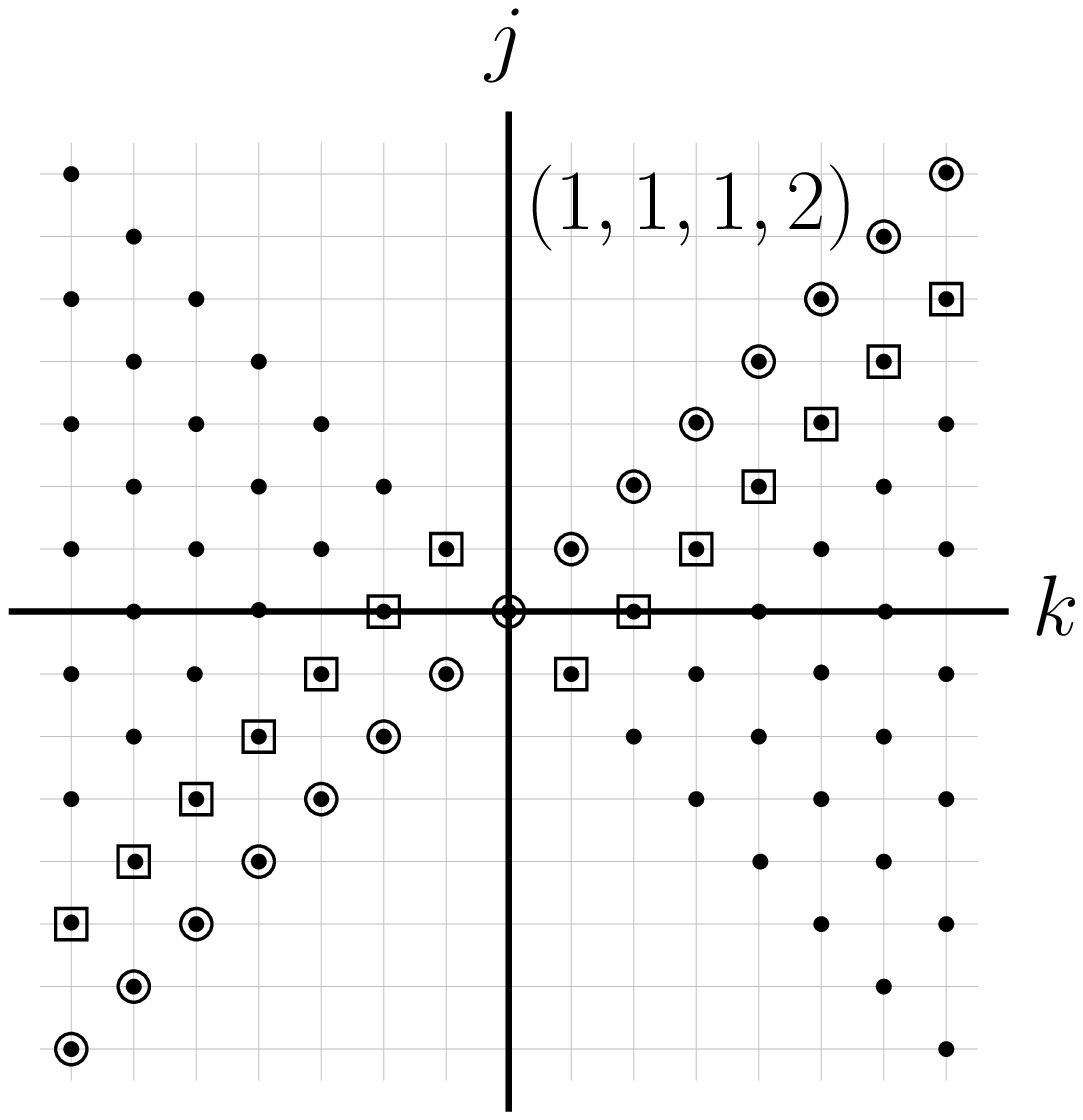}
\includegraphics[height=1.42in]{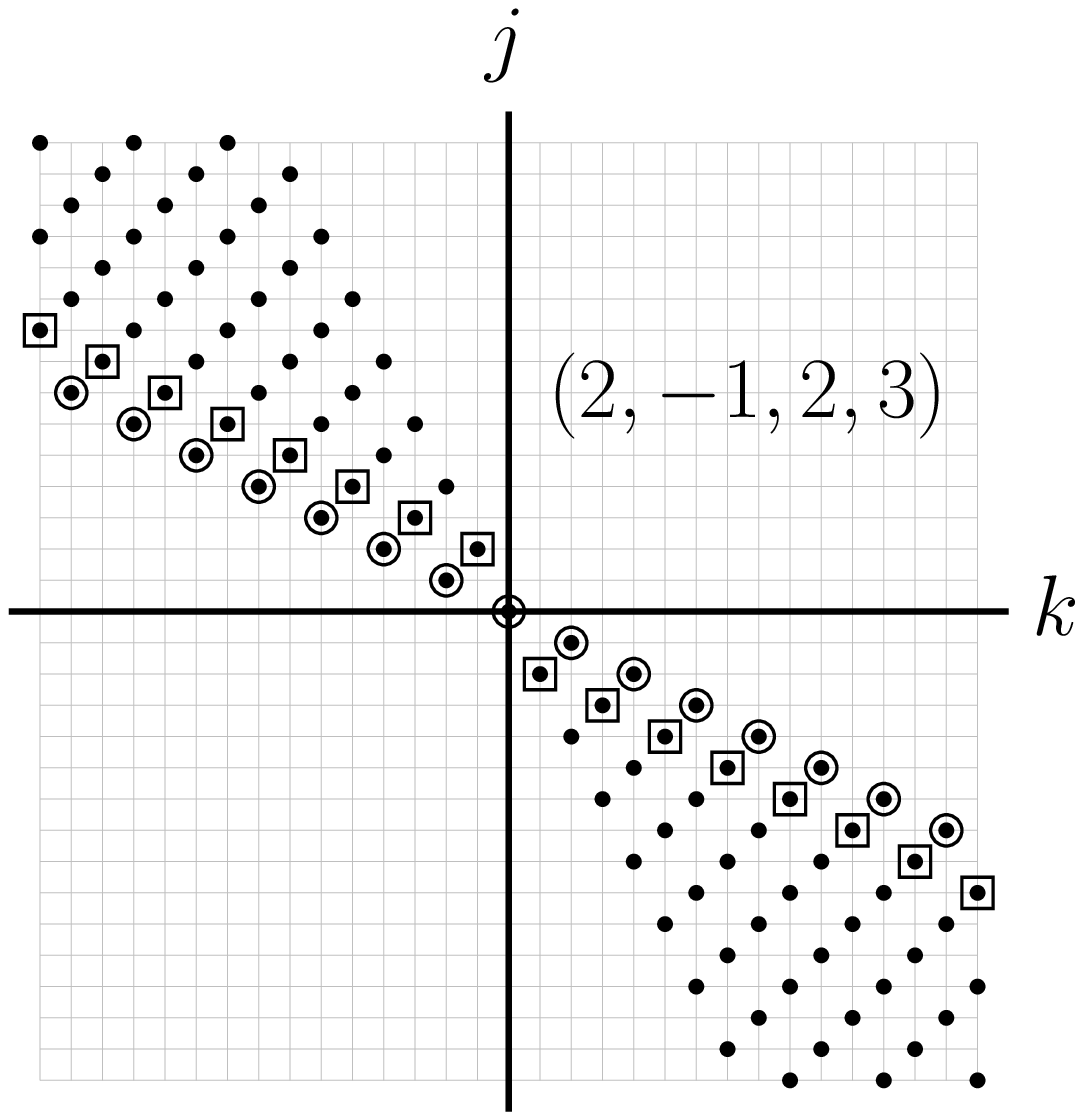}
\includegraphics[height=1.42in]{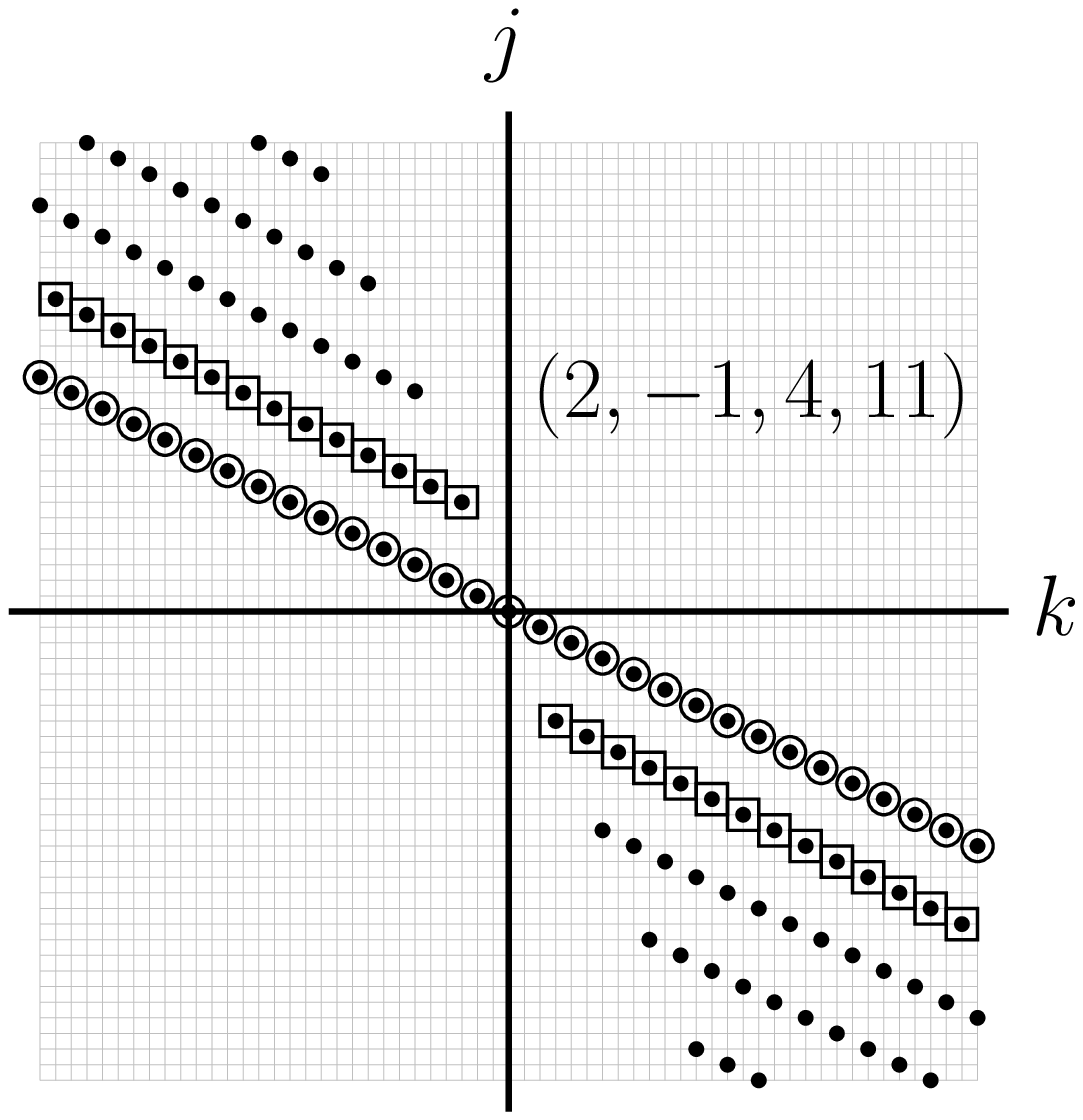} \\
\includegraphics[height=1.42in]{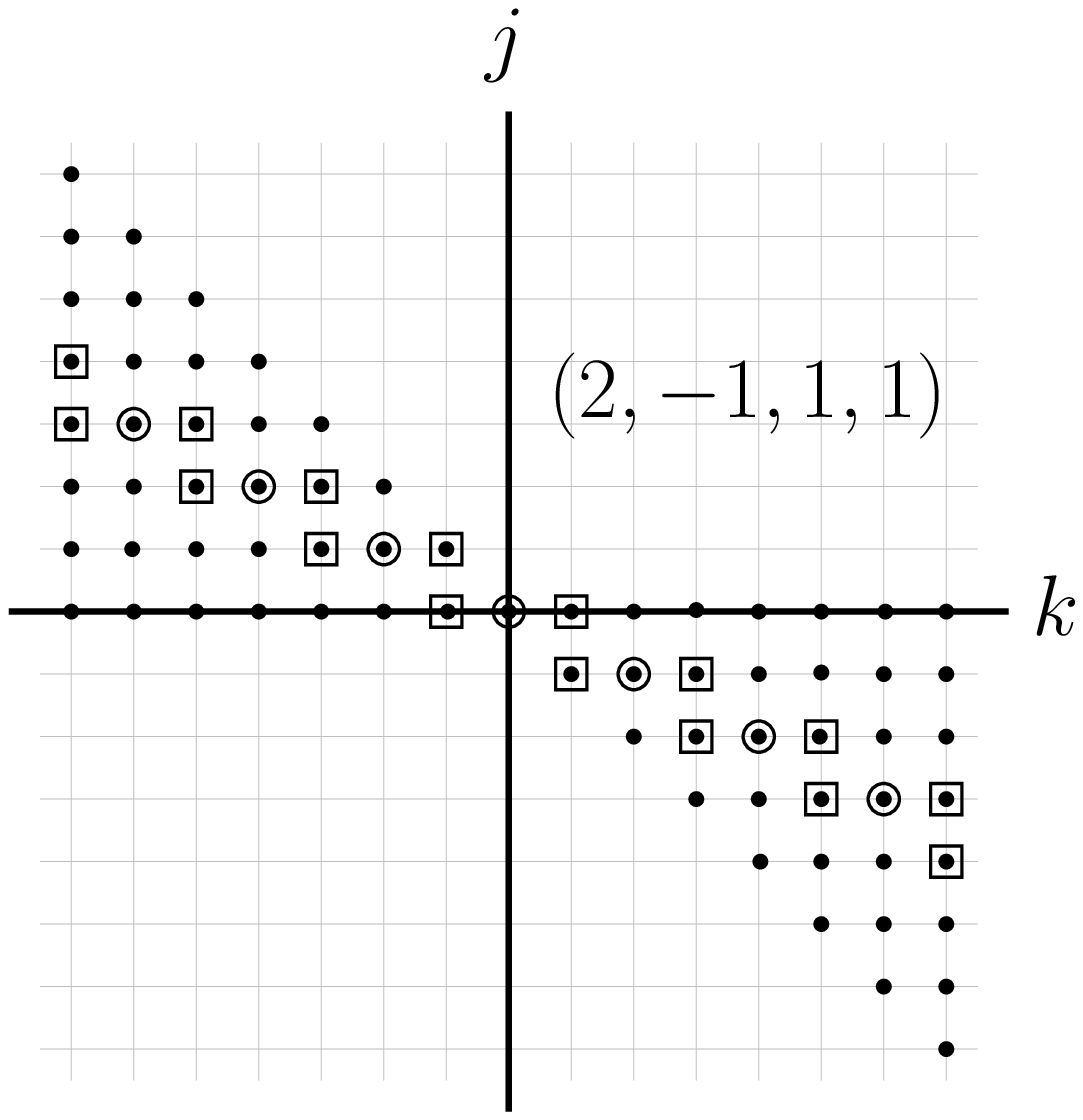}
\includegraphics[height=1.42in]{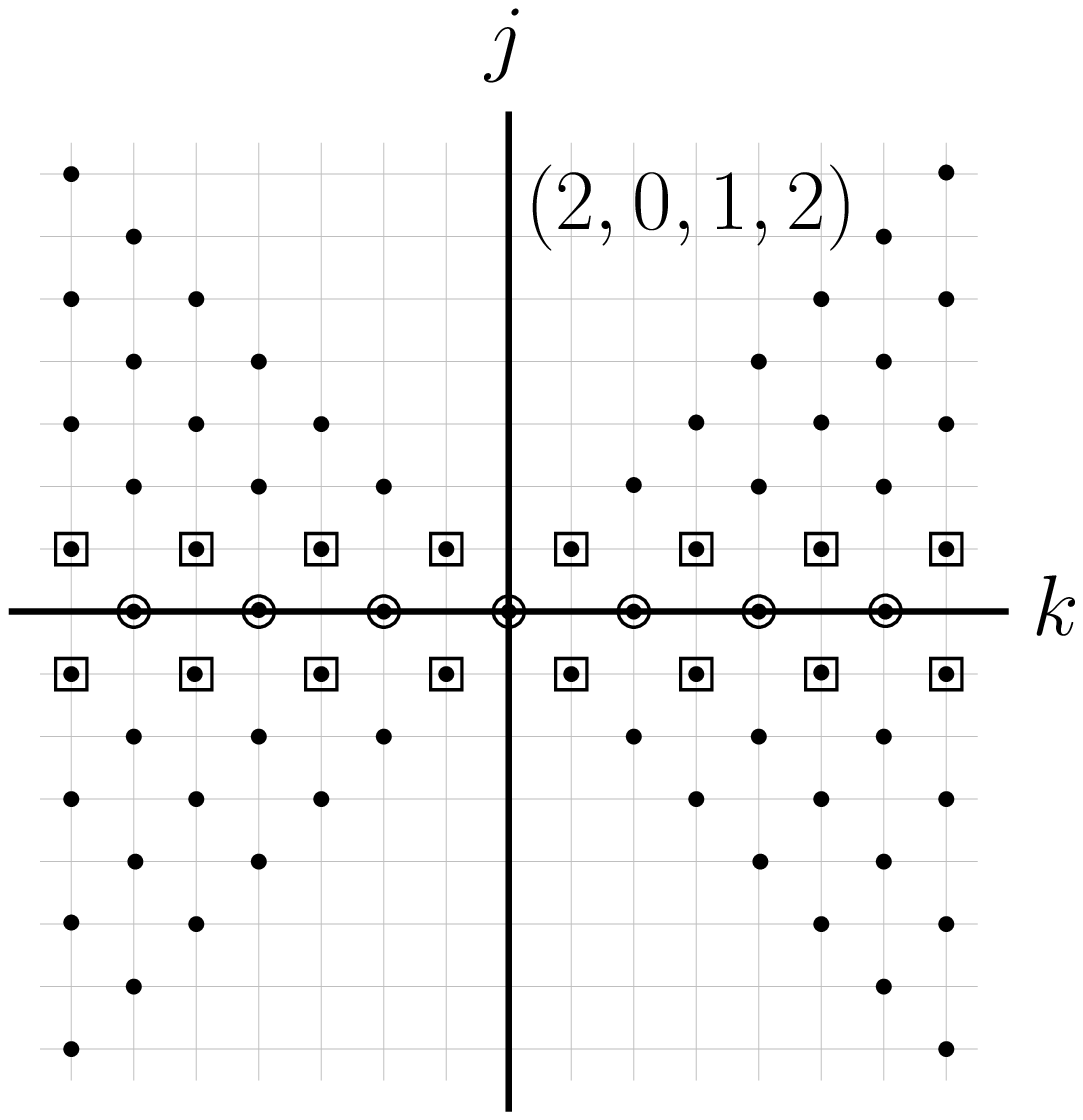}
\includegraphics[height=1.42in]{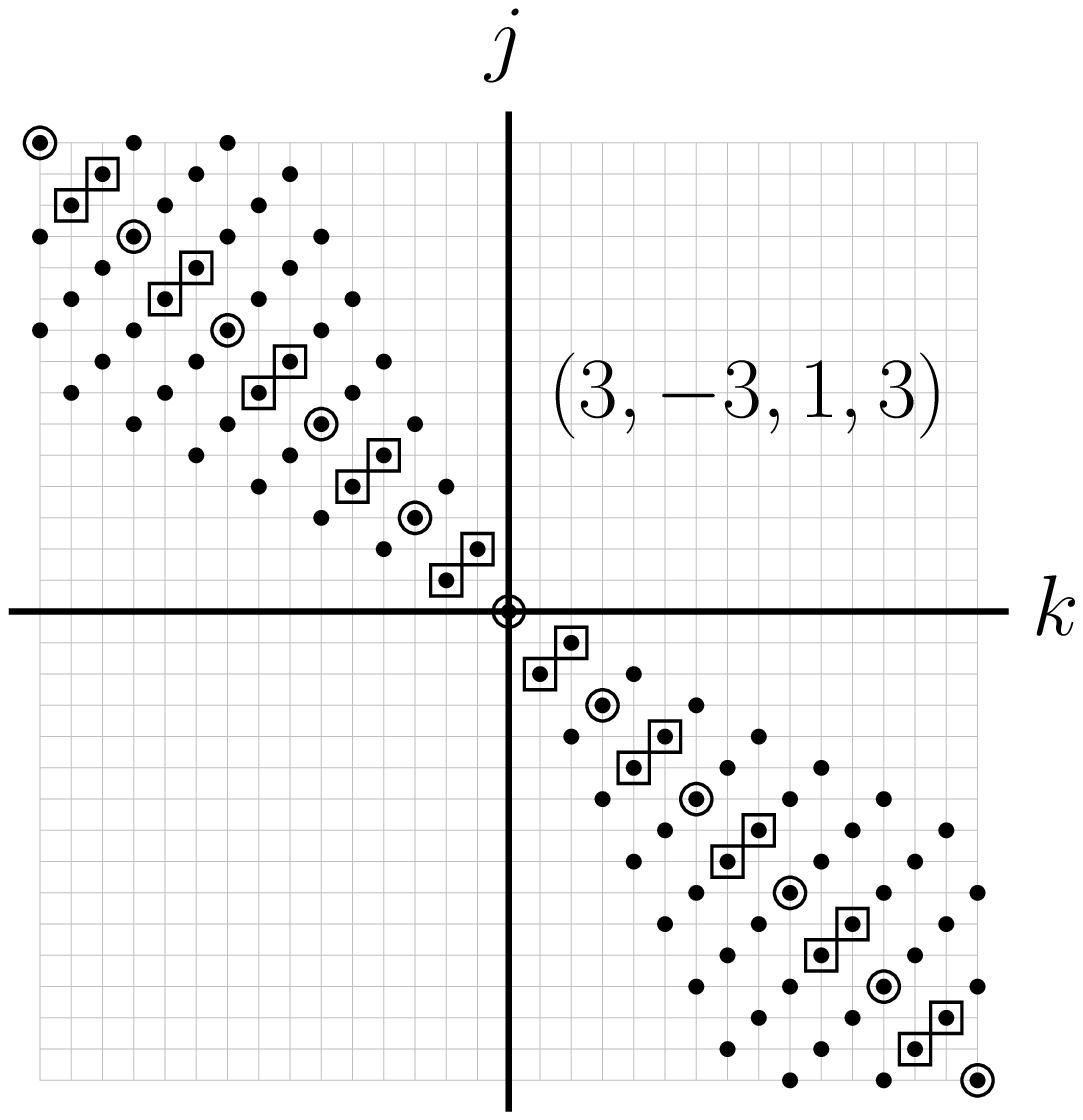}
\includegraphics[height=1.42in]{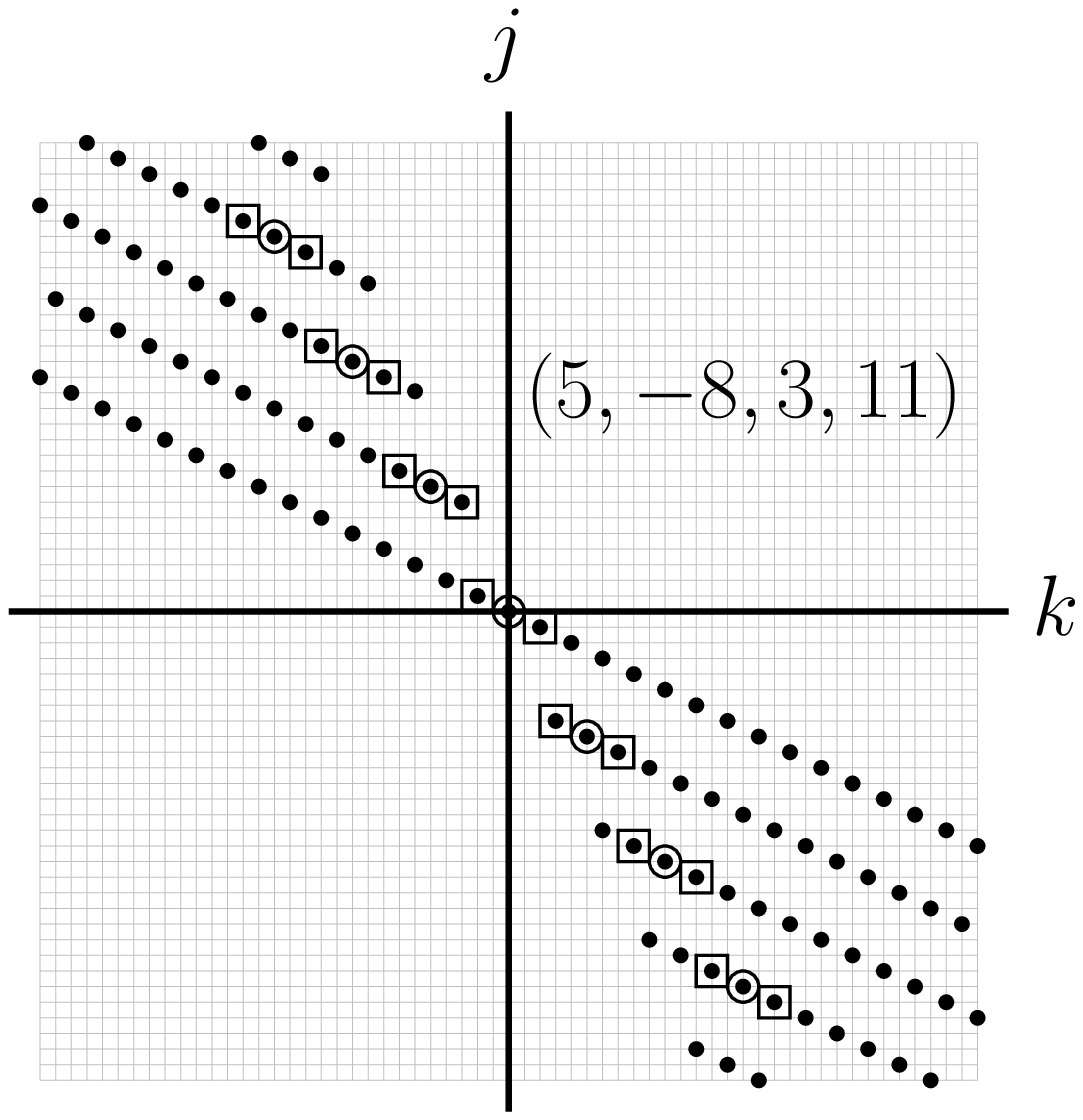}
\end{center}
\caption{
Each pair (aligned vertically) corresponds to a path of
non-trivial solutions connecting two traveling waves.  Solid dots
represent the non-zero entries $c_{kj}$ in (\ref{eqn:exact:form}) of
the exact solutions along this path; open circles represent a
traveling wave; and open squares represent the non-zero entries
$d_{kj}$ in the linearization about the traveling wave. }
\label{fig:lattice}
\end{figure}

Each path of non-trivial time-periodic solutions has a lattice pattern
of non-zero Fourier coefficients $c_{kj}$ associated with it.  In
Figure~\ref{fig:lattice}, we show the lattice of integers $(k,j)$ such
that $c_{kj}\ne0$ for solutions on the paths
\begin{equation}
\begin{aligned}
  (1,0,1,1) &\; \longleftrightarrow \; (2,-1,1,1), \\
  (1,1,1,2) &\; \longleftrightarrow \; (2,0,1,2), \\
\end{aligned} \qquad
\begin{aligned}
  (2,-1,2,3) &\; \longleftrightarrow \; (3,-3,1,3), \\
  (2,-1,4,11) &\; \longleftrightarrow \; (5,-8,3,11).
\end{aligned}
\end{equation}
All solutions on a given path have the same lattice pattern (of solid
dots), but different paths have different patterns.  One may show that
if $u(x,t)$ is of the form (\ref{eqn:exact:form}) and
\begin{equation} \label{eqn:lattice:formulation}
  \frac{k}{2}\sum_{l,p} c_{lp}c_{k-l,j-p} =
  \left(k|k|+\frac{2\pi}{T}j\right)c_{kj}, \qquad
  (k>0,\; j\in\mathbb{Z}),
\end{equation}
then $u(x,t)$ satisfies the Benjamin-Ono equation, $uu_x = Hu_{xx} -
u_t$.  The traveling waves at each end of the path have fewer non-zero
entries, namely
\begin{equation} \label{eqn:ckj:trav}
  \tilde{c}_{kj} = \left\{\begin{array}{ll}
    N\alpha + \frac{2\pi\nu}{NT} & \quad k=j=0, \\[4pt]
    2N\beta^{|k|/N} & \quad
    k\in N\mathbb{Z}\setminus\{0\},\; j=\frac{\nu k}{N} \\[2pt]
    0 & \quad \text{otherwise}.
  \end{array}\right\}, \quad\;
  \left(\alpha = \frac{1-3\beta^2}{1-\beta^2}\right).
\end{equation}
Here a tilde is used to indicate a solution about which we linearize.
Substitution of $c_{kj}=\tilde{c}_{kj} + \veps d_{kj}$ into
(\ref{eqn:lattice:formulation}) and matching terms of order $\veps$
leads to an eigenvalue problem with solution
\begin{equation}
  d_{kj} = \begin{cases}
    \hat{z}_{N,n}(k), & \quad k\in k_{N,n}+N\mathbb{Z}, \quad
    j = \frac{k\nu-m}{N}, \\
    \hat{z}_{N,n}(-k), & \quad k\in -k_{N,n}+N\mathbb{Z}, \quad
    j = \frac{k\nu+m}{N}, \\
    0 & \quad \text{otherwise},
  \end{cases}
\end{equation}
with $\hat{z}_{N,n}(k)$ as in (\ref{eqn:evec:formulas}).  The non-zero
coefficients $d_{kj}$ in this linearization are represented by open
squares in Figure~\ref{fig:lattice}.  Recall from
(\ref{eqn:evec:formulas}) that if $n\ge N$ and $k\le n-N$ then
$\hat{z}_{N,n}(k)=0$, but if $n<N$, the non-zero entries of
$\hat{z}_{N,n}(k)$ continue in both directions (with $k$ approaching
$+\infty$ or $-\infty$).  This is why the rows of open squares
terminate in the graphs in the top row of Figure~\ref{fig:lattice}
rather than continuing past the origin as in the graphs in the bottom
row.

\subsection{Elementary Symmetric Functions}

It is interesting that the lattice patterns that arise for the
exact solutions (beyond the linearization) contain only positive
integer combinations of the lattice points of the linearization and of
the traveling wave (treating the left and right half-planes
separately).  Somehow the double convolution in
(\ref{eqn:lattice:formulation}) leads to exact cancellation at all
other lattice sites!  This suggests that
the $c_{kj}$ have a highly regular structure that generalizes
the simple power law decay rate of the Fourier coefficients
$\hat{u}_\text{stat}(k;N,\beta)$ of the $N$-hump stationary
solution.

The first step to understand this is to realize that there is a close
connection between the trajectories of the Fourier coefficients and
the trajectories of the elementary symmetric functions of the
particles $\beta_1$, \dots, $\beta_N$ in (\ref{eqn:beta:rep}) above.
Specifically, because the Fourier coefficients of $\phi(x;\beta)$ in
(\ref{eqn:ubeta:hat}) are of the form $2\beta^k$ for $k\ge1$, we have
\begin{equation}
  \beta_1^k(t) + \cdots + \beta_N^k(t) = \frac{1}{2}c_k(t),
  \qquad \Bigl(k\ge1, \;
  c_k(t)=\frac{1}{2\pi}\int_0^{2\pi}u(x,t)e^{-ikx}\,dx\Bigr).
\end{equation}
Next we define the elementary symmetric functions $\sigma_j$ via
\begin{equation}
  \sigma_0 = 1, \qquad
  \sigma_j = \sum_{l_1<\cdots<l_j} \beta_{l_1}\cdots\beta_{l_j}, \qquad
  (j=1,\dots,N)
\end{equation}
so that
\begin{equation}
  P(z) := \prod_{l=1}^N (z-\beta_l) = \sum_{j=0}^N (-1)^j\sigma_jz^{N-j}.
\end{equation}
It is well known \cite{wilk228A} that
the companion matrix $\Sigma$ of $P$ has the Jordan
canonical form
\begin{equation*}
  \Sigma = \begin{pmatrix} 0 & 1 \\ & \ddots & \ddots \\
    0 & \cdots & 0 & 1 \\
    \pm\sigma_N & \cdots & -\sigma_2 & \sigma_1
\end{pmatrix},
\quad
V^{-1}\Sigma V=\begin{pmatrix} J_1 \\ & \ddots \\ & & J_m\end{pmatrix}, \quad
J_r = \begin{pmatrix}
\beta_{l(r)} & 1 & 0 \\ 0 & \ddots & 1 \\ 0 & 0 & \beta_{l(r)}
  \end{pmatrix},
\end{equation*}
where $l:\{1,\dots,m\}\rightarrow\{1,\dots,N\}$
is an enumeration of the
\emph{distinct} roots of $P(z)=0$ and the size of the Jordan block
$J_r$ is equal to the multiplicity of
$\beta_{l(r)}$.
As a result,
the trace of powers of $\Sigma$ will give the power sums of the
$\beta_l$, and hence the Fourier coefficients:
\begin{equation}\label{eqn:trace}
  c_k = 2\opn{tr}(\Sigma^k), \qquad\qquad (k\ge1).
\end{equation}
Thus, if the elementary symmetric functions are finite sums of
circular orbits, then the Fourier coefficients will be as well, and we
expect higher Fourier modes to involve more terms, in accordance with
our findings above.

Before presenting our main result, note that once the mapping
(\ref{eqn:conj1}) from $(N,\nu,n,m)$ to $(N',\nu',n',m')$ is known, we
can choose $N$, $\nu$, $N'$ and $\nu'$ independently, subject to the
conditions
\begin{equation}\label{eqn:restric}
  N'<N, \qquad\qquad \nu'>\frac{N'}{N}\nu.
\end{equation}
The first condition is merely a labeling convention while the second
is an actual restriction on which traveling waves are connected
together by a path of non-trivial solutions.  The formulas of
Conjecture~\ref{conj2} then imply that
\begin{equation}
  m = m' = N\nu' - N'\nu>0, \qquad n = N-N', \qquad n' = N-1.
\end{equation}
After extensive experimentation with data fitting on the numerical
simulations described in Section~\ref{sec:num}, we arrived at the form
(\ref{eqn:P:def3}) below for the polynomial $P$.  We then substituted
the ansatz (\ref{eqn:u:thm}) into (\ref{eqn:BO}) to obtain algebraic
relationships between $A$, $B$, $C$, $\alpha_0$, $\omega$, $N$, $N'$,
$\nu$ and $\nu'$, namely (\ref{eqn:alg1})--(\ref{eqn:alg3}) in
Appendix~\ref{sec:proof}.  We solved these using Mathematica to obtain
formulas for $A$, $B$ and $\omega$ in terms of $C$, $\alpha_0$, $N$,
$N'$, $\nu$ and $\nu'$.  We had to break the analysis into three cases
depending on whether $\nu$ is less than, equal to, or greater than
$\nu'$.  By comparing our exact solutions with previously known
representations of multi-periodic solutions \cite{matsuno:04}, we
found that all three cases could be unified by replacing $C$ and
$\alpha_0$ by two new parameters, $\rho$ and $\rho'$, related to
$C$ and $\alpha_0$ by (\ref{eqn:ABC}) below.  We give a
direct proof of the following theorem in Appendix~\ref{sec:proof}.

\begin{theorem} \label{thm:bif:trav} Let $N$, $N'$, $\nu$ and $\nu'$ be
  integers satisfying $N>N'>0$ and $N\nu'-N'\nu>0$.  There is a
  four-parameter family of time-periodic solutions connecting the
  traveling wave bifurcations $(N',\nu',N-1,m)$ and $(N,\nu,N-N',m)$,
  where $m=N\nu'-N'\nu$.  These solutions are of the form
\begin{equation}
  u(x,t) = \alpha_0 + \sum_{l=1}^N \phi(x;\beta_l(t)), \qquad
  \hat{\phi}(k;\beta) = \begin{cases}
    2\bar\beta^{|k|}, & \; k<0, \\
    0, & \; k=0, \\
    2\beta^k, & \; k>0,
    \end{cases} \label{eqn:u:thm}
\end{equation}
where $\beta_1(t)$, \dots, $\beta_N(t)$ are the roots of the
polynomial
\begin{equation}\label{eqn:P:def3}
  P(z) = z^{N} + A e^{-i\nu'\omega t}z^{N-N'} +
  Be^{-i(\nu-\nu')\omega t}z^{N'} + C e^{-i\nu\omega t}
\end{equation}
with
\begin{equation}
  \label{eqn:ABC}
\begin{aligned}
  A &= e^{i\nu'\omega t_0}e^{-iN'x_0}
  \sqrt{\frac{N-N'+\rho+\rho'}{N+\rho+\rho'}}
  \sqrt{\frac{(N+\rho')\rho'}{N'(N-N')+(N+\rho')\rho'}}, \\[2pt]
  B &= e^{i(\nu-\nu')\omega t_0}e^{-i(N-N')x_0}
  \sqrt{\frac{(N+\rho')\rho'}{N'(N-N')+(N+\rho')\rho'}}
  \sqrt{\frac{\rho}{N-N'+\rho}}, \qquad \\[2pt]
  C &= e^{i\nu\omega t_0}e^{-iNx_0}\sqrt{\frac{\rho}{N-N'+\rho}}
  \sqrt{\frac{N-N'+\rho+\rho'}{N+\rho+\rho'}}, \\[2pt]
  \alpha_0 &= \frac{N^2\nu'-(N')^2\nu}{m} -
  2\rho - \frac{2N'(\nu'-\nu)}{m}\rho',
  \qquad \omega = \frac{2\pi}{T} = \frac{N'(N-N')(N+2\rho')}{m}.
\end{aligned}
\end{equation}
The four parameters are $\rho\ge0$, $\rho'\ge0$,
$x_0\in\mathbb{R}$ and $t_0\in\mathbb{R}$.  The $N$- and $N'$-hump
traveling waves occur when $\rho'=0$ and $\rho=0$, respectively.  When
both are zero, we obtain the constant solution $u(x,t)\equiv
\frac{N^2\nu'-(N')^2\nu}{m}$.
\label{thm:4}
\end{theorem}

\begin{remark}\label{rk:phase}
  The parameters $x_0$ and $t_0$ are spatial and temporal phase
  shifts.  A straightforward calculation shows that if $u$ has
  parameters $\rho$, $\rho'$, $x_0$ and $t_0$ in
  Theorem~\ref{thm:bif:trav} while $\tilde{u}$ has parameters $\rho$,
  $\rho'$, 0 and 0, then $u(x,t) = \tilde{u}(x-x_0,t-t_0)$.
\end{remark}

There are two features of this theorem that are new.  First, it had
not previously been observed that the dynamics of the Fourier modes of
multiperiodic solutions was so simple.  And second, in our
representation, it is clear that these solutions reduce to traveling
waves in the limit as $\rho$ or $\rho'$ approaches zero. By contrast,
other representations become indeterminate in the equivalent limit,
and are missing a key degree of freedom (the mean) to allow
bifurcation between levels of the hierarchy of multi-periodic
solutions.

\subsection{Three Types of Reconnection}

We now wish to explain why following a path of non-trivial solutions
with the mean $\alpha_0$ held fixed sometimes leads to re-connection
with a different traveling wave and sometimes leads to blow-up of the
initial condition.  By Theorem~\ref{thm:bif:trav}, $\alpha_0$ depends
on the parameters $\rho$ and $\rho'$ via
\begin{equation}
  \alpha_0 = \alpha_0^* - 2\rho - \frac{2N'(\nu'-\nu)}{m}\rho',
  \qquad \alpha_0^* := \frac{N^2\nu' - (N')^2\nu}{m}.
\end{equation}
If we hold $\alpha_0$ fixed, then $\rho$ and $\rho'$ must satisfy
\begin{equation} \label{eqn:line}
  2\rho + \frac{2N'(\nu'-\nu)}{m}\rho' = (\alpha_0^* - \alpha_0).
\end{equation}
This is a line in the $\rho$-$\rho'$-plane whose intersection with
the first quadrant gives the set of legal parameters for a
time-periodic solution to exist.  We assume the mean is chosen so that
this intersection is non-empty.  If the $\rho$- or $\rho'$-intercept
of this line is positive, the corresponding traveling wave
bifurcation exists.  There are three cases to consider.

\underline{\emph{Case 1:}} ($\nu<\nu'$)  Both intercepts will be
positive as long as $\alpha_0<\alpha_0^*$.  Thus, a reconnection
occurs regardless of which side of the path we start on.

\underline{\emph{Case 2:}} ($\nu=\nu'$) The line (\ref{eqn:line}) is
vertical in this case, so $\rho=(\alpha_0^*-\alpha_0)/2$ remains constant
as we vary $\rho'$ from $0$ to $\infty$.  As $\rho'\rightarrow\infty$,
we see from (\ref{eqn:ABC}) that $T\rightarrow0$, $A\rightarrow1$, and
$B$ and $C$ both approach $\sqrt{\rho/(N-N'+\rho)}$.  In this limit,
$N'$ of the roots $\beta_l$ lie on the unit circle at $t=0$,
indicating that the norm of the initial condition blows up as
$\rho'\rightarrow\infty$.

\underline{\emph{Case 3:}} ($\nu>\nu'$) The line (\ref{eqn:line}) has
positive slope in this case.  If $\alpha_0<\alpha_0^*$, a bifurcation
from the $N'$-hump traveling wave exists.  If $\alpha_0>\alpha_0^*$, a
bifurcation from the $N$-hump traveling wave exists.  And if
$\alpha_0=\alpha_0^*$, a bifurcation directly from the constant
solution $u=\alpha_0^*$ to a non-trivial time periodic solution
exists.  In any of these cases, another traveling wave is not reached
as we increase $\rho$ and $\rho'$ to $\infty$.  Instead,
$T\rightarrow0$ and $A$, $B$ and $C$ all approach 1.  As a result, all
the roots $\beta_l$ approach the unit circle, indicating that the norm
of the initial condition blows up as $\rho,\rho'\rightarrow\infty$.

\begin{example}
  Consider the three-particle solutions on the path
  $e:(2,-1,2,3)\leftrightarrow(3,-3,1,3)$ in Figures~\ref{fig:bifur4}
  and \ref{fig:evol2}.  Since $-3=\nu<\nu'=-1$, we do not need to vary
  the mean in order to reconnect with a traveling wave on the other
  side of the path.  Suppose $\alpha_0<\alpha_0^*=1$ is held fixed.
  Then the parameters $\rho$ and $\rho'$ in Theorem~\ref{thm:bif:trav}
  satisfy
\begin{equation}\label{eqn:rho:rho:pr}
  \rho = \frac{1}{2}\left(1-\alpha_0-\frac{8}{3}\rho'\right), \qquad
  0\le\rho'\le\frac{3(1-\alpha_0)}{8}.
\end{equation}
  The solutions $u(x,t)$ on this path are of the form (\ref{eqn:u:thm})
  with particles $\beta_l(t)$ evolving as the roots of the polynomial
\begin{equation}\label{eqn:P:example}
  P(z) = z^3 + Ae^{i\omega t}z + Be^{2i\omega t}z^2 + C e^{3i\omega t},
\end{equation}
where
\begin{equation}
\begin{gathered}
  A = \sqrt{\frac{(9-3\alpha_0-2\rho')(3+\rho')\rho'}{
    (21-3\alpha_0-2\rho')(2+\rho')(1+\rho')}}, \qquad
  B = \sqrt{\frac{(3-3\alpha_0-8\rho')(3+\rho')\rho'}{
    (9-3\alpha_0-8\rho')(2+\rho')(1+\rho')}}, \\[4pt]
  C = \sqrt{\frac{(9-3\alpha_0-2\rho')(3-3\alpha_0-8\rho')}{
    (21-3\alpha_0-2\rho')(9-3\alpha_0-8\rho')}}, \qquad
  \omega = \frac{2\pi}{T} = \frac{2(3+2\rho')}{3}.
\end{gathered}
\label{eqn:A2:23}
\end{equation}
The transition from the two- to three-hump traveling wave occurs as we
decrease the bifurcation parameter $\rho'$ from $3(1-\alpha_0)/8$ to 0.
This causes $C$ to increase from 0 to
$\sqrt{\frac{1-\alpha_0}{7-\alpha_0}}$ and $A$ to decrease from
$\sqrt{\frac{3-3\alpha_0}{19-3\alpha_0}}$ to 0.  $B$ is zero at both
ends of the path.

\begin{figure}[t]
\begin{center}
\includegraphics[width=\linewidth,trim=0 0 0 0]{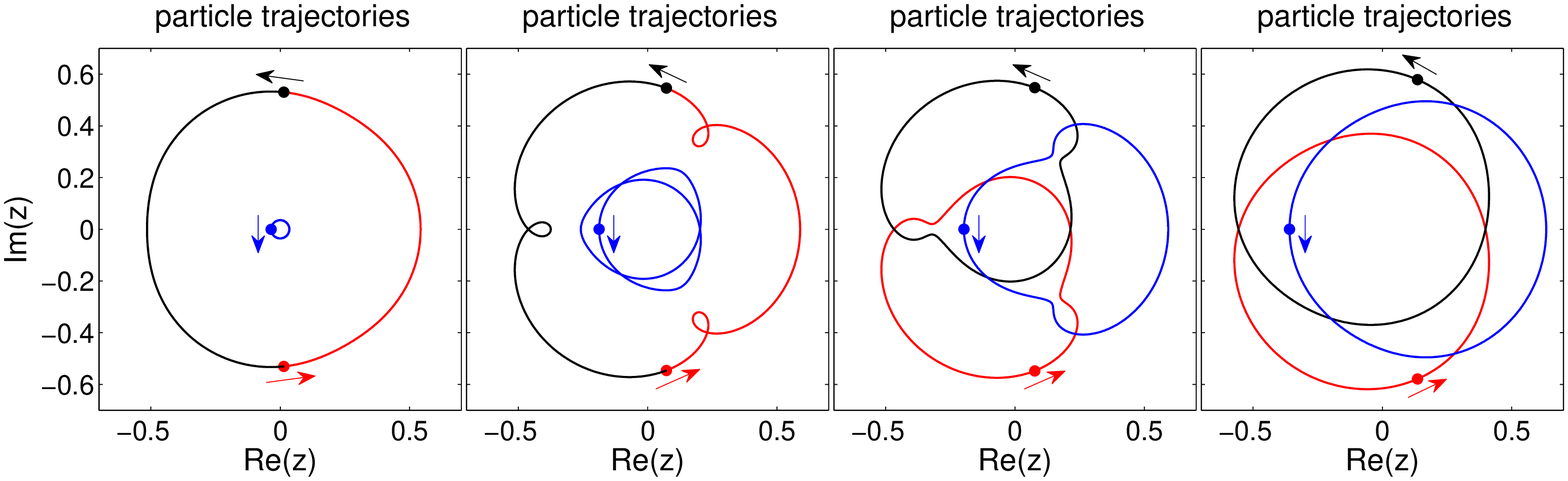}
\end{center}
\caption{Trajectories $\beta_l(t)$
  for four solutions on the path $(2,-1,2,3)\leftrightarrow(3,-3,1,3)$
  with mean $\alpha_0=0.544375$.  The
  markers give the position of the $\beta_l$ at $t=0$.  The value of
  $\rho'$ in (\ref{eqn:rho:rho:pr}) is, from left to right:
  $0.1707$, $0.1642$, $0.1634$ and $0.1369$.  In Figure~\ref{fig:evol2},
  $\rho'=0.0862$.
}\label{fig:beta3}
\end{figure}

The trajectories $\beta_1(t)$, $\beta_2(t)$ and $\beta_3(t)$ for
$\alpha_0=0.544375$ and four choices of $\rho'$ are shown in
Figure~\ref{fig:beta3}.  For this value of the mean, $\rho'$ varies
from $0.17086$ to $0$.
Note that the bifurcation from the two-hump
traveling wave causes a new particle to nucleate at the origin.  As
$\rho'$ decreases, the new particle's trajectory grows in amplitude
until it joins up with the orbits of the outer particles. There is a
critical value of $\rho'$ at which the particles collide and the
solution of the ODE (\ref{eqn:beta:ode}) ceases to exist for all time;
nevertheless, the representation of $u$ in terms of $P$ in
(\ref{eqn:u:from:P}) in Appendix~\ref{sec:proof} remains well-behaved
and does satisfy (\ref{eqn:BO}) for all time.  Thus, a change in
topology of the orbits does not manifest itself as a singularity in
the solution of the PDE.  As $\rho'$ decreases further, the three orbits
become nearly circular and eventually coalesce into a single circular
orbit (with $\nu=-3$) at the three-hump traveling wave.  The
``braided'' effect of the solution shown in Figure~\ref{fig:evol2} is
recognizable for $\rho'\le0.15$ or so for this value of the mean.
\end{example}

\section{Interior Bifurcations} \label{sec:interior}

\begin{figure}[p]
\begin{center}
\includegraphics[height=3in]{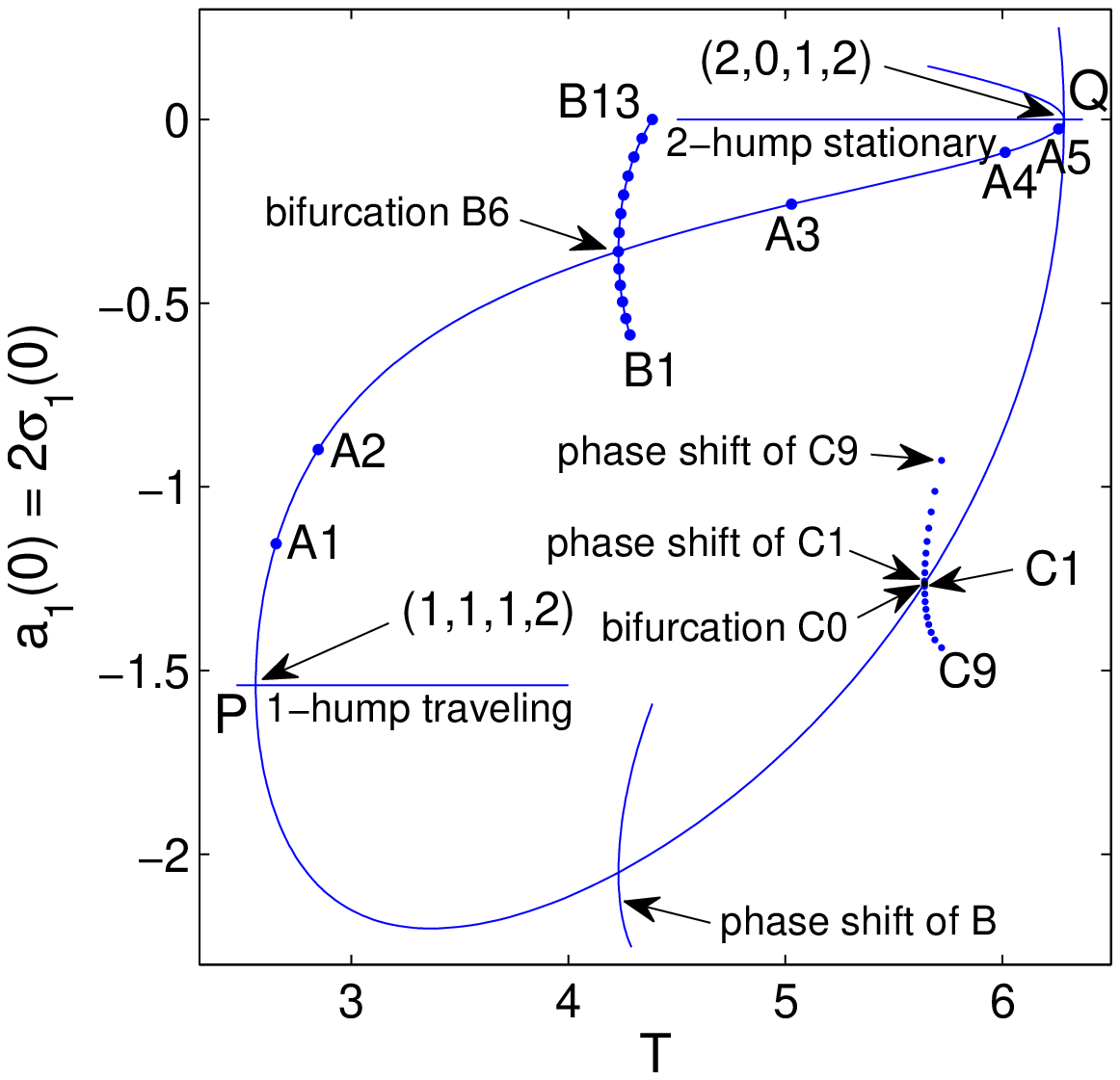}
\quad
\includegraphics[height=3in, trim=0 0 0 0]{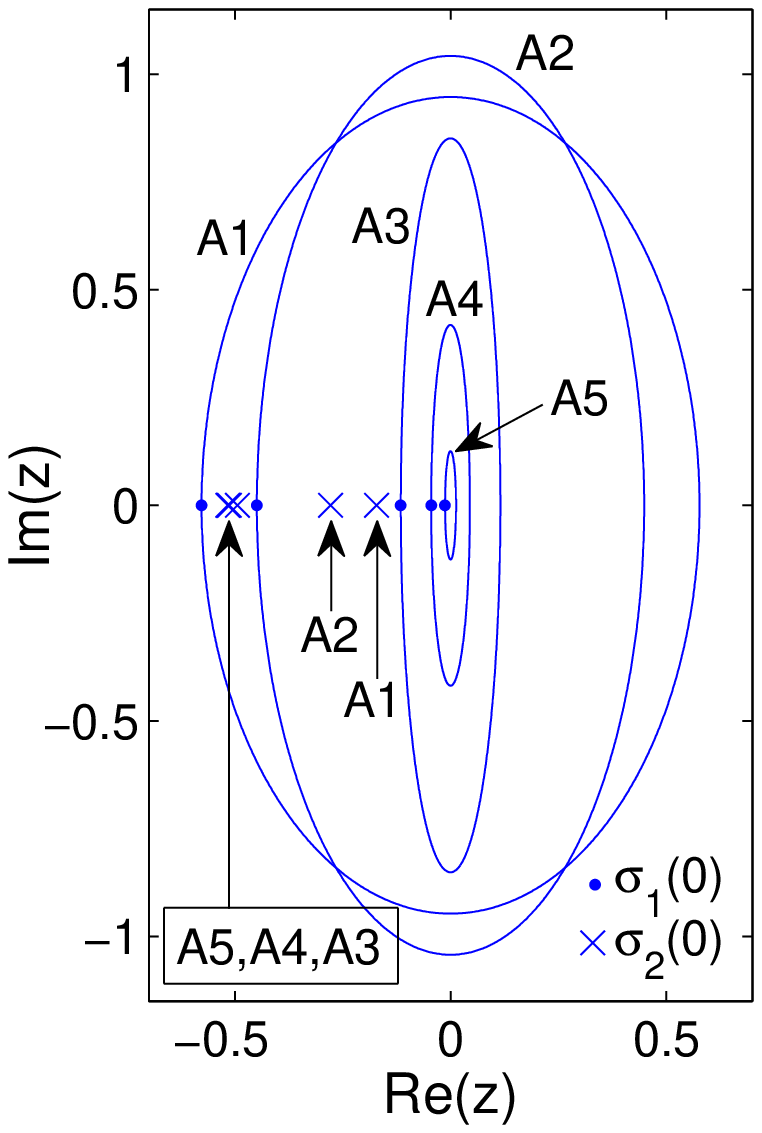}
\end{center}
\caption{\emph{Left:} Bifurcation diagram showing several interior
  bifurcations on the path $(1,1,1,2)\rightarrow(2,0,1,2)$.
  \emph{Right:} Trajectories of the elementary symmetric functions
  $\sigma_1(t)$, which have elliptical, clockwise orbits, and
  $\sigma_2(t)$, which remain stationary in time,
  for the solutions labeled A1-A5 in the bifurcation diagram.
}
\label{fig:bifurBd}
\end{figure}

We conclude this work by mentioning that our numerical method for
following paths of non-trivial solutions from one traveling wave
to another occasionally wanders off course, following an interior
bifurcation
rather than reaching the traveling wave on the other side
of the original path.  These interior bifurcations lead to new
paths of non-trivial solutions that are more complicated
than those on the original path.  For example, on the path
\begin{equation}
    (1,1,1,2) \; \longleftrightarrow \; (2,0,1,2),
\end{equation}
Theorem~\ref{thm:bif:trav} tells us that the exact solution is
a two-particle solution with elementary symmetric functions of
the form
\begin{equation}
  \sigma_1(t) = -(Ae^{-i\omega t}+Be^{i\omega t}), \qquad
  \sigma_2(t) = C.
\end{equation}
We freeze $\alpha_0<\alpha_0^*=2$, set
$\rho=\frac{1}{2}(2-\alpha_0-\rho')$, and determine that
\begin{gather}
\notag
A = e^{-i(x_0-\omega t_0)}
\sqrt{\frac{(4-\alpha_0+\rho')(2+\rho')\rho'}{(6-\alpha_0+\rho')
  (1+\rho')^2}}, \qquad
B = e^{-i(x_0+\omega t_0)}
\sqrt{\frac{(2-\alpha_0-\rho')(2+\rho')\rho'}{(4-\alpha_0-\rho')
  (1+\rho')^2}}, \\[4pt]
C = e^{-i(2x_0)}\sqrt{\frac{(4-\alpha_0+\rho')(2-\alpha_0-\rho')}{
  (6-\alpha_0+\rho')(4-\alpha_0-\rho')}}, \qquad
\omega = \frac{2\pi}{T} = 1 + \rho'.
\label{eqn:E:12}
\end{gather}
In Figure~\ref{fig:bifurBd}, we show the bifurcation diagram for the
transition from the one-hump right-traveling wave (labeled P) to the
two-hump stationary solution (labeled Q).  This diagram was computed
numerically before we had any idea that exact solutions for this
problem exist; therefore, we used the real part of the first Fourier
mode at $t=0$ for the bifurcation parameter rather than $\rho'$.  We
can obtain the same curves analytically as follows.  The upper curve
from P to Q (containing A1-A5) can be plotted parametrically by
setting $x_0=\pi/2$ and $t_0=\pi/2\omega$ in (\ref{eqn:E:12}), varying
$\rho'$ from $2-\alpha_0$ to $0$, holding $\alpha_0=0.544375$ fixed,
and plotting $-2(A+B)$ versus $T=\frac{2\pi}{1+\rho'}$.  The lower
curve from P to Q is obtained in the same fashion if we instead set
$x_0=t_0=0$.

As illustrated in Figure~\ref{fig:bifurBd},
solutions such as A1-A5 on the upper path have $\sigma_1(t)$ executing
elliptical, clockwise orbits that start out circular at the one-hump
traveling wave but become more eccentric and collapse to a point as we
progress toward the two-hump stationary solution Q.  Meanwhile,
$\sigma_2(t)$ remains constant in time, nucleating from the origin at
the one-hump traveling wave and terminating with
$\sigma_2\equiv-\sqrt{\frac{2-\alpha_0}{6-\alpha_0}}$ at the two-hump
stationary solution.  On the lower path, the major axis of
the orbit of $\sigma_1$ is horizontal rather than vertical and
$\sigma_2$ moves right rather than left as we move from P to Q.

\begin{figure}[p]
\begin{center}
\includegraphics[height=2.75in, trim=10 0 0 0, clip]{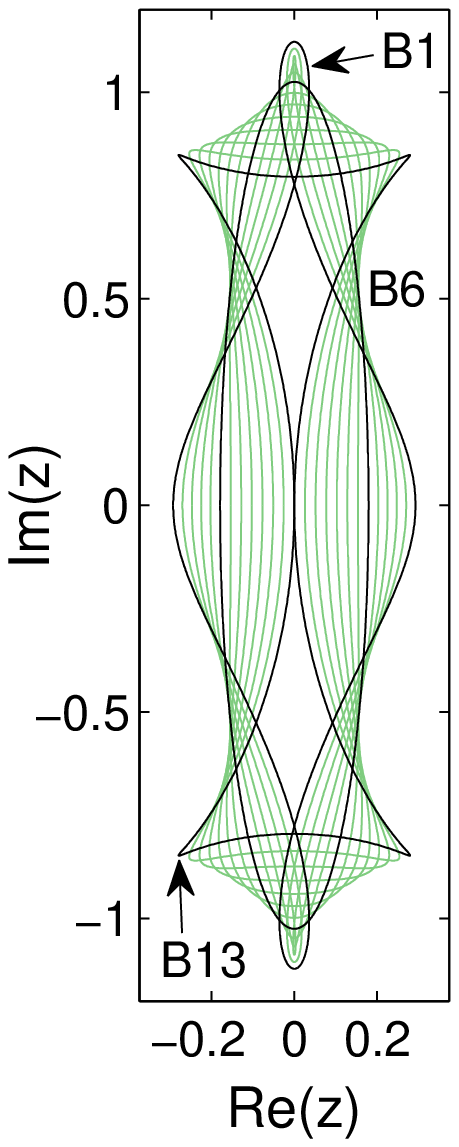}
\includegraphics[height=2.75in, trim=0 0 0 0, clip]{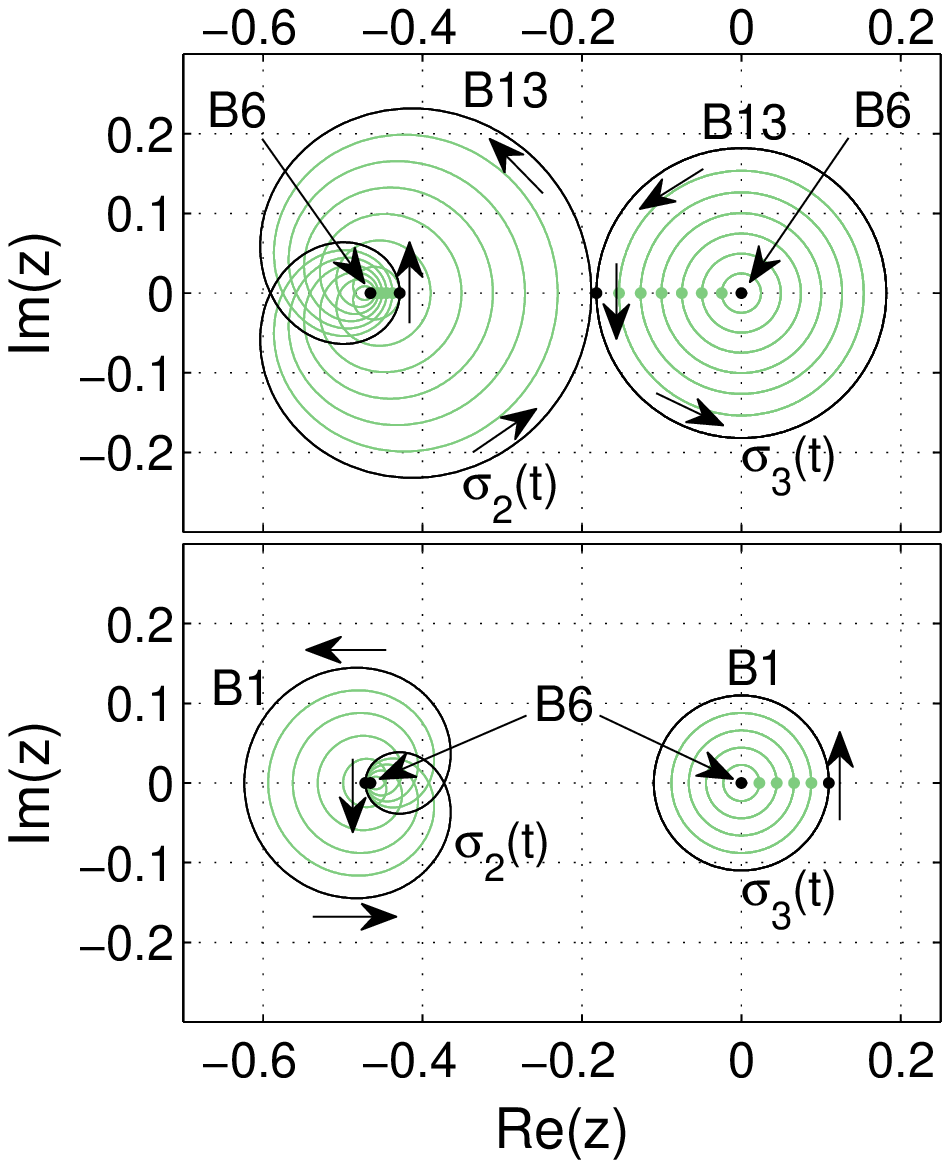}\quad
\includegraphics[height=2.75in, trim=0 -60 10 0]{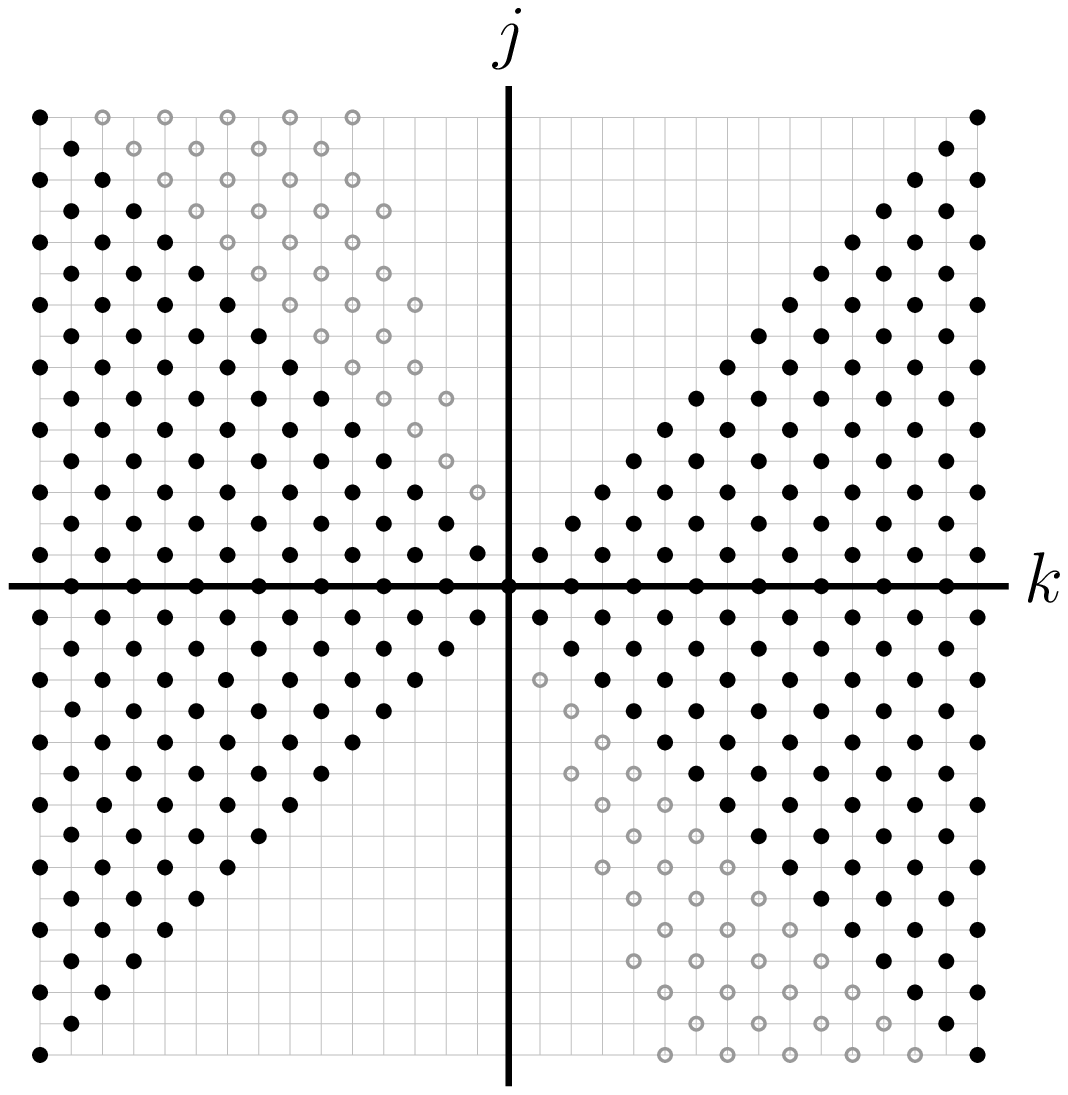}
\end{center}
\caption{ \emph{Left:} Trajectories of $\sigma_1(t)$ for solutions
  labeled B1-B13 in Figure~\ref{fig:bifurBd}.  \emph{Center:}
  Trajectories of $\sigma_2(t)$ and $\sigma_3(t)$.  Since B6 is on the
  original path from P to Q, $\sigma_2(t)$ is constant and
  $\sigma_3(t)\equiv0$ for this solution.  \emph{Right:} The interior
  bifurcation causes additional lattice coefficients $c_{kj}$ to
  become non-zero; grey circles represent the new terms.  }
\label{fig:spiroBd}
\end{figure}

When computing these paths from P to Q, we encountered two interior
bifurcations.  In the bifurcation labeled B6 in
Figure~\ref{fig:bifurBd}, an additional elementary symmetric function
nucleates at the origin and the trajectories of $\sigma_1$ and
$\sigma_2$ become more complicated.  Through data fitting, we find
that
\begin{align}
  \label{eqn:bifBd1}
  \sigma_1(t) &= -(Ae^{-i\omega t} + Be^{i\omega t} + C_1e^{3i\omega t}), \\
  \sigma_2(t) &= C + C_2e^{2i\omega t} + C_3e^{4i\omega t}, \\
  \label{eqn:bifBd3}
  \sigma_3(t) &= -C_4e^{3i\omega t},
\end{align}
where the new coefficients $C_j$ are all real parameters.
We have not attempted to derive algebraic relationships among these
parameters to obtain exact solutions.  These trajectories are shown in
Figure~\ref{fig:spiroBd} for the solutions labeled B1-B13 in the
bifurcation diagram.  The additional term in (\ref{eqn:bifBd1}) causes
the elliptical orbit of $\sigma_1(t)$ to deform by bulging out in the
vertical and horizontal directions while pulling in along the diagonal
directions (or vice versa, depending on which direction we follow the
bifurcation).  Meanwhile, $\sigma_2(t)$ ceases to be constant and
$\sigma_3(t)$ ceases to be zero.  To avoid clutter, we plotted the
trajectories $\sigma_2(t)$ and $\sigma_3(t)$ for B1-B6 separately from
B6-B13, illustrating the effect of following the bifurcation in one
direction or the other.  The additional terms in
(\ref{eqn:bifBd1})--(\ref{eqn:bifBd3}) cause the lattice pattern of
non-zero entries $c_{kj}=\frac{1}{T}\int_0^T c_k(t)e^{ij\omega t}\,dt$
to become more complicated, where we recall that in this case,
\begin{equation*}
  c_k(t) = \frac{1}{2\pi}\int_0^{2\pi}u(x,t)e^{-ikx}\,dx =
  2\opn{tr}\left[\begin{pmatrix} 0 & 1 & 0 \\ 0 & 0 & 1 \\
    \sigma_3(t) & -\sigma_2(t) & \sigma_1(t) \end{pmatrix}^k\,\,\right].
\end{equation*}
The solid dots in Figure~\ref{fig:spiroBd}
represent the non-zero entries of solutions on the original path
from P to Q while grey circles show the additional terms that
are non-zero after the bifurcation at B6.  Although this
bifurcation causes some of the unoccupied lattice sites
to be filled in, the new lattice pattern is rather
similar to the original pattern and maintains its checkerboard
structure.  Also, this bifurcation leads to symmetric perturbations of
the Fourier mode trajectories, and is also present (in a phase shifted
form) along the lower path from P to Q.

\begin{figure}[p]
\begin{center}
\includegraphics[height=2.4in, trim=0 -60 10 0]{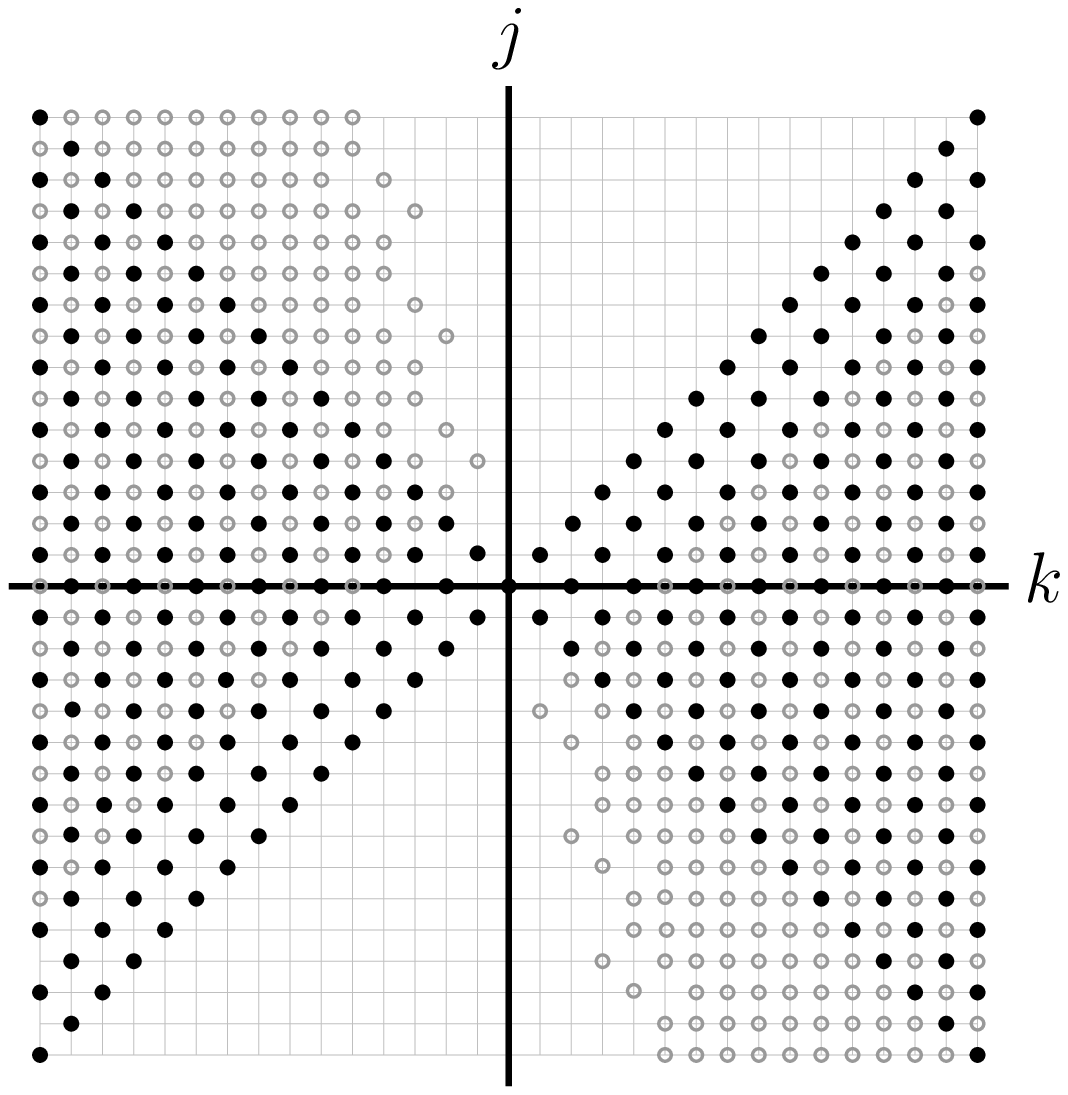}
\includegraphics[height=2.6in]{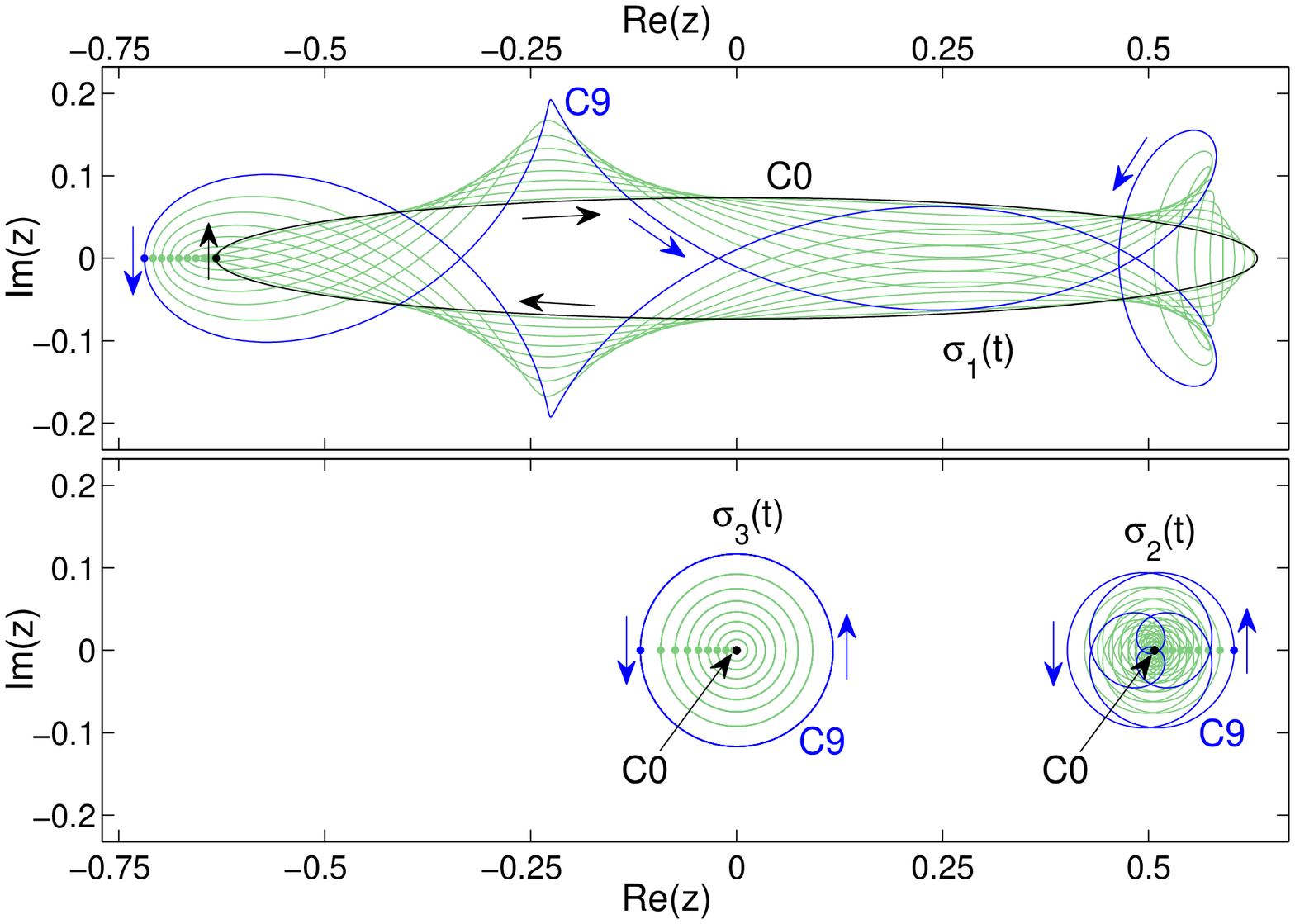}
\end{center}
\caption{\emph{Left:} This interior bifurcation causes more lattice
  coefficients to become non-zero than the interior bifurcation of
  Figure~\ref{fig:spiroBd}.  \emph{Right:} Trajectories of
  $\sigma_1(t)$, $\sigma_2(t)$, and $\sigma_3(t)$ for the solutions
  labeled C0-C9 in Figure~\ref{fig:bifurBd}. The long axis of the
  ellipse C0 is horizontal because we start from the bottom branch
  connecting P to Q in Figure~\ref{fig:bifurBd}.}
\label{fig:spiroBa}
\end{figure}

\begin{figure}[p]
\begin{center}
\includegraphics[height=2.75in]{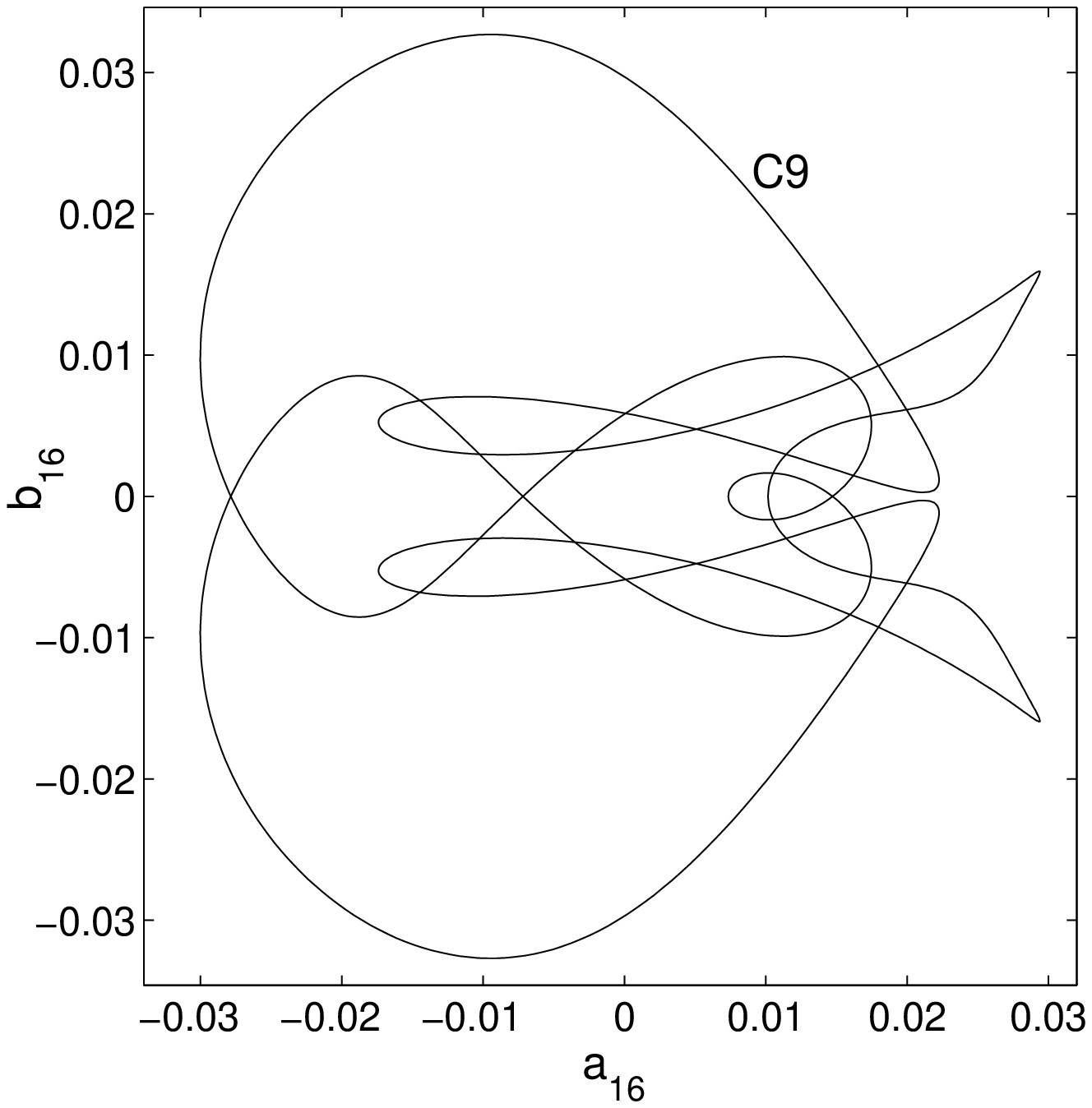}
\quad
\includegraphics[height=2.75in]{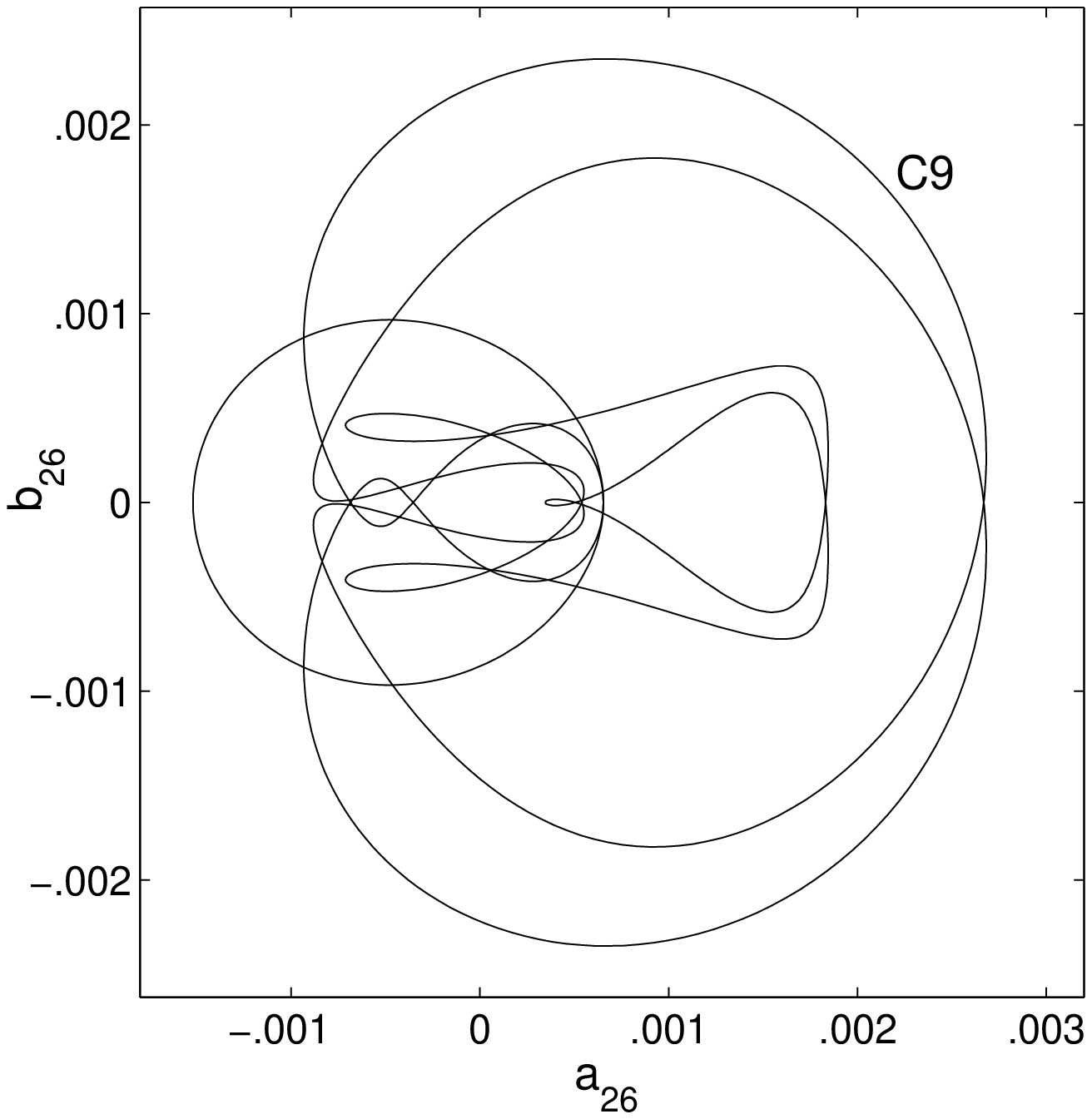}
\end{center}
\caption{The trajectories of the Fourier modes become very complicated
  after the interior bifurcation occurs.  Here we show the 16th and
  26th Fourier modes $c_k(t)=a_k(t)+ib_k(t)$ over one period.  It was
  clearly essential to use a high order (in fact spectrally accurate)
  numerical method to resolve these dynamics when computing
  time-periodic solutions.}
\label{fig:spiroBa16}
\end{figure}

In the bifurcation labeled C0 in Figure~\ref{fig:bifurBd}, the fill-in
pattern of the lattice representation is much more complicated, and in
fact the checkerboard structure of the non-zero coefficients $c_{kj}$
is destroyed; see Figure~\ref{fig:spiroBa}.  But actually, the
elementary symmetric functions behave similarly to the previous
case:  By fitting our numerical data, we find that
\begin{align}
    \label{eqn:bifBa1}
  \sigma_1(t) &= -(Ae^{-i\omega t} + Be^{i\omega t} + C_1e^{4i\omega t}), \\
  \sigma_2(t) &= C + C_2e^{3i\omega t} + C_3e^{5i\omega t}, \\
  \label{eqn:bifBa3}
  \sigma_3(t) &= -C_4e^{4i\omega t},
\end{align}
so each of the new terms executes one additional loop per cycle of
the periodic solution in comparison to the corresponding term in
(\ref{eqn:bifBd1})--(\ref{eqn:bifBd3}).  This extra loop causes a
star-shaped perturbation of the $\sigma_1$ ellipse instead of the
rectangular and diamond shaped perturbations seen previously in
Figure~\ref{fig:spiroBd}.  As a result, this bifurcation is not
present on the upper path from P to Q because the symmetry of the
perturbation does not respect the 90 degree rotation of the orbit
$\sigma_1(t)$ associated with the $\frac{\pi}{2}$-spatial and
$\frac{T}{4}$-temporal phase shifts that relate solutions on the upper
and lower paths from P to Q.
To follow the bifurcation at C0 in the other direction, we can use the
same numerical values for $A$, $B$, $C$, $C_1$, $C_2$, $C_3$, $C_4$ in
(\ref{eqn:bifBa1})--(\ref{eqn:bifBa3}) after changing the signs of the
latter four parameters.  This causes the trajectories of $\sigma_1$ in
Figure~\ref{fig:spiroBa} to be rotated $180^\circ$ with a
corresponding $\frac{T}{2}$ phase-shift in time so that the initial
position $\sigma_1(0)$ remains on the left side of the figure.
Meanwhile, the trajectory of $\sigma_2(t)$ experiences a $\frac{T}{2}$
phase-shift in time with no change in the location of the orbit, and
$\sigma_3(t)$ starts on the opposite side of its circular trajectory
about the origin.

In Figure~\ref{fig:spiroBa16}, we show the orbits of the 16th and
26th Fourier modes
for the solution labeled C9 in the bifurcation diagram of
Figure~\ref{fig:bifurBd}.  As the index of the Fourier mode increases,
these trajectories become increasingly complicated (involving more
non-zero terms $c_{kj}$ in the lattice representation), but also decay
exponentially so that the amplitude of the orbit is eventually smaller
than can be resolved using floating point arithmetic.  We emphasize
that these trajectories were resolved to full machine precision by
our general purpose numerical method for finding periodic solutions
of non-linear PDE (without any knowledge of the solitonic structure
of the solutions). Everything we learned about the form of the exact
solutions came about from studying these numerical solutions, which
was possible only because our numerical results are correct to
10-15 digits of accuracy.

\appendix
\section{Bifurcation formulas and rules}
\label{sec:T:a0:a}

In this section we collect formulas relating the period, mean and
decay parameter at a bifurcation.  We also identify bifurcation rules
governing the legal values of $\alpha_0$ for a given set of
bifurcation indices.

In computing the nullspace $\mc{N}=\ker DF(U_0,T)$ in
Section~\ref{sec:classify}, we considered $N$, $\nu$, $\beta$, $T$
(and hence $\alpha_0$) to be given and searched for compatible indices
$n$ and $m$.
The decay parameter $|\beta|$, the mean $\alpha_0$, and the
period $T$ cannot be specified independently; any two of them
determines the third.
We now derive formulas for the period
and mean in terms of $(N,\nu,n,m)$ and $\beta$.
To simplify the formulas, we work with
$\alpha=(1-3|\beta|^2)/(1-|\beta|^2)$ instead of $\beta$.  Note that
as we increase $|\beta|$ from $0$ to $1$, $\alpha$ decreases from $1$
to $-\infty$.  For the period, we have
\begin{equation}\label{eqn:T:alpha}
  T = \frac{2\pi m}{N\omega_{N,n}} =
  \begin{cases}
    \jd \frac{2\pi m}{Nn(N-n)} & \quad n<N, \\[5pt]
    \jd \frac{2\pi m}{N(n+1-N)(n+1+N(1-\alpha))} & \quad n\ge N,
  \end{cases}
\end{equation}
so the period is independent of $\beta$ when $n<N$, and otherwise
decreases to zero as $|\beta|$ varies from 0 to 1.
For the mean, $\alpha_0$, we note that
\begin{equation}
  cT=\frac{2\pi\nu}{N}, \qquad
  c=\alpha_0-N\alpha \qquad \Rightarrow \qquad
  \alpha_0 = N\alpha+\frac{2\pi\nu}{NT}.
\end{equation}
Hence, using $\frac{2\pi}{NT}=\frac{\omega_{N,n}}{m}$, we obtain
\begin{equation} \label{eqn:alpha0:alpha}
  \alpha_0 = \begin{cases}
    \jd N + \frac{n(N-n)}{m}\nu - (1-\alpha)N, & \quad n<N, \\[7pt]
    \jd N + \frac{(n+1-N)(n+1)}{m}\nu - \left(1 - \frac{n+1-N}{m}\nu\right)
    N(1-\alpha), & \quad n\ge N.
  \end{cases}
\end{equation}
Thus, as $|\beta|$ varies from 0 to 1, the mean $\alpha_0$ decreases
to $-\infty$ if $n<N$, and
otherwise either decreases to $-\infty$,
increases to $+\infty$, or is independent of $\beta$, depending on the
sign of $[m-(n+1-N)\nu]$.

In practice, we often wish to start with $N$, $\nu$, $n$, $m$ and
$\alpha_0$ and determine $T$ and $|\beta|$ from these.  However, not
all values of $\alpha_0$ are compatible with a given set of indices.
The bifurcation rules are summarized in Figure~\ref{fig:rules}.
\begin{figure}[t]
  \begin{center}
\fbox{\parbox{.9\linewidth}{
\begin{enumerate}
  \item $N\ge1$,\; $\nu\in\mathbb{Z}$,\; $n\ge1$,\; $m\ge1$
  \item if $n<N$ \;then\;
    \begin{itemize}
      \item $m\in n\nu + N\mathbb{Z}$
      \item $\alpha_0 \le \jd N + \frac{n(N-n)}{m}\nu$
    \end{itemize}
  \item if $n\ge N$ \;then\;
    \begin{itemize}
    \item $m\in (n+1)\nu + N\mathbb{Z}$
    \item if $m>(n+1-N)\nu$ \;then\;
      $\alpha_0\le N+\frac{(n+1-N)(n+1)}{m}\nu$
    \item if $m<(n+1-N)\nu$ \;then\;
      $\alpha_0\ge N+\frac{(n+1-N)(n+1)}{m}\nu$
    \item if $m=(n+1-N)\nu$ \;then\;
      $\alpha_0=n+1+N$
    \end{itemize}
  \end{enumerate}
  }}
  \end{center}
\caption{Bifurcation rules governing which values of $\alpha_0$
are compatible with the bifurcation indices $(N,\nu,n,m)$.
} \label{fig:rules}
\end{figure}
Solving (\ref{eqn:alpha0:alpha}) for $\alpha$ yields
\begin{equation}\label{eqn:alpha:alpha0}
  \alpha = \begin{cases}
    \jd 1 - \frac{(N-\alpha_0)m + n(N-n)\nu}{Nm}, & \quad
    n<N,
    \\[7pt]
    \jd 1 - \frac{(N-\alpha_0)m + (n+1-N)(n+1)\nu}{
      [m - (n+1-N)\nu]N}, & \quad n\ge N.
  \end{cases}
\end{equation}
The corresponding period is given by
\begin{equation} \label{eqn:T:N:nu:nm}
  T = \begin{cases}
    \jd \frac{2\pi m}{Nn(N-n)}, & \quad n<N,
    \\[7pt] \jd
    \frac{2\pi\left(\frac{m}{n+1-N} - \nu\right)}{N(n+1+N-\alpha_0)},
    & \quad n\ge N.
  \end{cases}
\end{equation}
In the indeterminate cases $\{n\ge N,\; m=(n+1-N)\nu,\; \alpha_0=n+1+N\}$, 
any $\alpha\le1$ is allowed and formula (\ref{eqn:T:alpha}) should be
used to determine $T$.

If we express $n$, $n'$, $m$ and $m'$ 
in terms of $N$, $\nu$, $N'$, $\nu'$, then
(\ref{eqn:T:alpha}) and (\ref{eqn:alpha0:alpha}) give
\begin{alignat}{2}
\notag
  T&=\frac{2\pi(N\nu'-N'\nu)}{N'(N-N')N}, &\qquad
  \alpha_0 &= \alpha_0^* - (1-\alpha)N,
  \qquad \alpha_0^* := \frac{N^2\nu'-(N')^2\nu}{N\nu'-N'\nu}\\
\label{eqn:trav:Ta0}
T'&=\frac{2\pi(N\nu'-N'\nu)}{N'(N-N')[N+(1-\alpha')N']}, &
  \alpha_0' &= \alpha_0^* -
  \frac{\nu'-\nu}{N\nu'-N'\nu}(N')^2(1-\alpha'),
\end{alignat}
where $\alpha=\frac{1-3|\beta|^2}{1-|\beta|^2}$ and
$\alpha'=\frac{1-3|\beta'|^2}{1-|\beta'|^2}$.
We note that the two traveling waves reduce to the same constant
function when $\beta\rightarrow0$ and $\beta'\rightarrow0$, which is
further evidence that a single sheet of non-trivial solutions connects
these two families of traveling waves.

\section{Proof of Theorem~\ref{thm:bif:trav}}\label{sec:proof}

As explained in Remark~\ref{rk:phase}, $x_0$ and $t_0$ are spatial and
temporal phase shifts, so we may set them to zero without loss of
generality.  We can express the solution directly in terms of the
elementary symmetric functions via
\begin{align}\label{eqn:u:from:P}
  u(x,t) &= \alpha_0 + \sum_{l=1}^N \phi(x;\beta_l(t)) =
  \alpha_0 + \sum_{l=1}^N 4\real\left\{\sum_{k=1}^\infty
    \beta_l(t)^k e^{ikx}\right\}  \\
  \notag
  &= \alpha_0 + \sum_{l=1}^N 4\real\left\{\frac{z}{z-\beta_l(t)}-1\right\}
  = \alpha_0 + 4\real\left\{\frac{z\pr_zP(z)}{P(z)}-N\right\},
  \quad (z = e^{-ix}).
\end{align}
Next we derive algebraic expressions relating $A$, $B$, $C$,
$\alpha_0$, $\omega$, $N$, $N'$, $\nu$ and $\nu'$ by substituting
(\ref{eqn:u:from:P}) into the Benjamin-Ono equation (\ref{eqn:BO}).
To this end, we include the time dependence of $P$ in the notation
and write (\ref{eqn:u:from:P}) in the form
\begin{equation} \label{eqn:u:from:gh}
  u(x,t) = \alpha_0 +
  2\left(\frac{i\partial_x g}{g}-N\right) +
  2\left(\frac{-i\partial_x h}{h}-N\right),
\end{equation}
where
\begin{align}
  \label{eqn:gh:def}
  g(x,t) &= P(e^{-ix},e^{-i\omega t}), \qquad
  h(x,t) = \overline{g(x,t)}, \\
 \label{eqn:P:def2}
  P(z,\lambda) &= z^N + A\lambda^{\nu'}z^{N-N'} + B\lambda^{\nu-\nu'}z^{N'} +
  C\lambda^{\nu}.
\end{align}
Note that $P$ is a polynomial in $z$ and a Laurent polynomial in
$\lambda$ (as $\nu$ and $\nu'$ may be negative).
We may assume $\omega>0$; if not, we can change the sign of $\omega$
without changing the solution by replacing $(A,B,\nu,\nu',N')$ by
$(B,A,-\nu,\nu'-\nu,N-N')$.  Assuming the roots $\beta_l(t)$ of
$z\mapsto P(z,e^{-i\omega t})$ remain inside the unit disk $\Delta$ for
all $t$, we have
\begin{equation}
  \left(\frac{i\partial_x g}{g}-N\right) = \sum_{l=1}^N \sum_{k=1}^\infty
  \beta_l(t)^ke^{ikx} \quad \Rightarrow \quad
  Hu = 2\left(\frac{\partial_x g}{g}+Ni\right) +
  2\left(\frac{\partial_x h}{h}-Ni\right).
\end{equation}
Using (\ref{eqn:u:from:gh}) and $\partial_t\Big(\frac{\partial_xg}{g}\Big)=
\partial_x\Big(\frac{\partial_tg}{g}\Big)$, (a technique we learned by
studying the bilinear formalism approach of \cite{satsuma:ishimori:79,
matsuno:04}), the equation
$\frac{1}{2}\left(u_t-Hu_{xx}+uu_x\right)=0$ becomes
\begin{equation} \label{eqn:BO:gh}
  \partial_x\left[
    i\left(\frac{\partial_tg}{g}-\frac{\partial_th}{h}\right) -
    \partial_x\left(\frac{\partial_xg}{g}+\frac{\partial_xh}{h}\right)
    +\frac{1}{4}\left((\alpha_0-4N)+2i\left(\frac{\partial_xg}{g}-
      \frac{\partial_xh}{h}\right)\right)^2\right]=0.
\end{equation}
The expression in brackets must be a constant, which we denote by
$\gamma$.  We now write
\begin{equation}
  P_{jk}=(z\partial_z)^j(\lambda\partial_\lambda)^kP(z,\lambda)\biggr
  \vert_{\parbox[b]{.45in}{$\js z=e^{-ix}$ \\[-5pt]
      $\js \lambda=e^{-i\omega t}$}}
\end{equation}
so that e.g.~$\partial_tg=-i\omega P_{01}$ and
$\partial_xh=i\bar{P}_{10}$.
Equation (\ref{eqn:BO:gh}) then becomes
\begin{equation} \label{eqn:P:alg}
  \begin{aligned}
  \gamma P_{00}\bar{P}_{00}
  & + \bar{P}_{00}\big[P_{20} + \omega P_{01} +
  (\alpha_0-4N)P_{10}\big] \\
  & + P_{00}\big[\bar{P}_{20} + \omega\bar{P}_{01} +
  (\alpha_0-4N)\bar{P}_{10}\big] +
  2P_{10}\bar{P}_{10}=0,
  \end{aligned}
\end{equation}
where we have absorbed
$\frac{1}{4}(\alpha_0-4N)^2$ into $\gamma$.
This equation may be written
\begin{equation*}
  e_1 \big\llbracket z^N\lambda^{-\nu}\big\rrbracket +
  e_2 \big\llbracket z^{N-2N'}\lambda^{2\nu'-\nu}\big\rrbracket +
  e_3 \big\llbracket z^{N-N'}\lambda^{\nu'-\nu}\big\rrbracket +
  e_4 \big\llbracket z^{N'}\lambda^{-\nu'}\big\rrbracket +
  e_5=0,
\end{equation*}
where $\llbracket a \rrbracket = a+\bar{a}=2\real\{a\}$,
\begin{equation*}
  e_1 = \big[\gamma+\nu\omega+N^2+(\alpha_0-4N)N\big]C, \qquad
  e_2 = \big[\gamma+\nu\omega+N^2+(\alpha_0-4N)N\big]AB,
\end{equation*}
and, after setting $\gamma=(3N-\alpha_0)N-\nu\omega$ to achieve
$e_1=e_2=0$,
\begin{align}
  \label{eqn:alg1}
&  e_3 = [(N')^2 - 2 N N' + N'\alpha_0 - \nu' \omega]B + 
  [(N')^2 + 2 N N' - N'\alpha_0 + \nu' \omega]AC = 0, \\
   \notag
& e_4 =
   \bigl[3N^2 - 4NN' + (N')^2 - (N-N')\alpha_0 + (\nu-\nu')\omega\bigr]BC
\\
& \hspace*{2in}
   \label{eqn:alg2}
   - \bigl[N^2 - (N')^2 - (N-N')\alpha_0 + (\nu-\nu')\omega\bigr]A = 0,
\\ \notag
& e_5 = (N\alpha_0 - \nu\omega - N^2) +
\big[(2N'-N)\alpha_0 + (\nu-2\nu')\omega + 3N^2 -8NN' + 4(N')^2\big]B^2 \\
\label{eqn:alg3}
&\quad + \big[(N-2N')\alpha_0 + 4(N')^2-N^2 +
(2\nu'-\nu)\omega\big]A^2 + \big[(3N-\alpha_0)N+\nu\omega\big]C^2 = 0.
\end{align}
Using a computer algebra system, it is easy to check that
(\ref{eqn:alg1})--(\ref{eqn:alg3}) hold when $A$, $B$, $C$, $\alpha_0$
and $\omega$ are defined as in (\ref{eqn:ABC}).  When $\rho'=0$, we have
$A=B=0$ and $\jd C=\sqrt{\frac{\rho}{N+\rho}}$\, so that
\begin{equation*}
  \beta_l(t) = \sqrt[N]{-C\lambda^{\nu}} = \sqrt[N]{-C}e^{-ict}, \qquad
  c = \frac{\omega\nu}{N} = \frac{N'(N-N')\nu}{m} = \alpha_0 - N
  \frac{1-3C^2}{1-C^2},
\end{equation*}
where each $\beta_l$ is assigned a distinct $N$th root of $-C$.  By
(\ref{eqn:N:trav}), this is an $N$-hump traveling wave with speed index
$\nu$ and period $T=\frac{2\pi}{\omega}$.  Similarly, when $\rho=0$, we
have $B=C=0$ and $\jd A=\sqrt{\frac{\rho'}{N'+\rho'}}$\, so that
\begin{equation*}
  \beta_l(t)=\left\{\begin{array}{cc}
      \sqrt[N']{-A}e^{-ict} & l\le N' \\
      0 & l>N' \end{array}\right\}, \qquad
  c = \frac{\omega\nu'}{N'} = \frac{(N-N')(N+2\rho')\nu'}{m} =
  \alpha_0 - N'\frac{1-3A^2}{1-A^2},
\end{equation*}
which is an $N'$-hump traveling wave with speed index $\nu'$ and
period $T=\frac{2\pi}{\omega}$.

Finally, we show that the roots of $P(\cdot,\lambda)$ are
inside the unit disk for any $\lambda$ on the unit circle, $S^1$.
We will use Rouch\'e's theorem \cite{ahlfors}.  Let
\begin{alignat*}{4}
    f_1(z) &= z^N & + \,&A\lambda^{\nu'}z^{N-N'} & &+B\lambda^{\nu-\nu'}z^{N'}
    & &+ C\lambda^\nu, \\
    f_2(z) &= z^N & + \,&A\lambda^{\nu'}z^{N-N'},\hspace*{-.1in} \\
    f_3(z) &= z^N & & & &+B\lambda^{\nu-\nu'}z^{N'}.\hspace*{-.1in}
\end{alignat*}
From (\ref{eqn:ABC}), we see that $\{A,B,C\}\subseteq[0,1)$,
$A\ge BC$, $B\ge CA$ and $C\ge AB$.  Thus,
\begin{equation}
\begin{aligned}
  d_2(z) := |f_2(z)|^2 &- |f_1(z)-f_2(z)|^2 = |\lambda^{-\nu'}z^{N'}+A|^2 -
  |B\lambda^{-\nu'}z^{N'}+C|^2 \\
  &= 1 + A^2 - B^2 - C^2 + 2(A-BC)\cos \theta \ge
  (1-A)^2 - (B-C)^2,
\end{aligned}
\end{equation}
where $\lambda^{-\nu'}z^{N'}=e^{i\theta}$.  Similarly,
\begin{equation}
  d_3 := |f_3(z)|^2 - |f_1(z)-f_3(z)|^2 \ge (1-B)^2 - (A-C)^2.
\end{equation}
Note that
\begin{alignat*}{3}
  &B\le A, \quad C\le B \quad & &\Rightarrow \quad
  B-C \le B-AB < 1-A \quad & &\Rightarrow \quad
  d_2(z)>0 \text{ for } z\in S^1, \\
  &B\le A, \quad C>B \quad & &\Rightarrow \quad
  |C-A|<1-B \quad & &\Rightarrow \quad
  d_3(z)>0 \text{ for } z\in S^1, \\
  &A\le B, \quad C\le A \quad & &\Rightarrow \quad
  A-C \le A-AB < 1-B \quad & &\Rightarrow \quad
  d_3(z)>0 \text{ for } z\in S^1, \\
  &A\le B, \quad C>A \quad & &\Rightarrow \quad
  |C-B|<1-A \quad & &\Rightarrow \quad
  d_2(z)>0 \text{ for } z\in S^1.
\end{alignat*}
Thus, in all cases, $f_1(z)=P(z,\lambda)$ has the same number of zeros
inside $S^1$ as $f_2(z)$ or $f_3(z)$, which each have $N$ roots inside
$S^1$.  Since $f_1(z)$ is a polynomial of degree $N$, all the roots are
inside $S^1$.

\bibliographystyle{plain}
\bibliography{refs}

\begin{thebibliography}{10}

\bibitem{ahlfors}
Lars Ahlfors.
\newblock {\em Complex Analysis}.
\newblock McGraw-Hill, New York, 1979.

\bibitem{ambrose:03}
D.~M. Ambrose.
\newblock Well-posedness of vortex sheets with surface tension.
\newblock {\em SIAM J. Math Anal.}, 35(1):211--244, 2003.

\bibitem{benj1}
D.~M. Ambrose and J.~Wilkening.
\newblock Time-periodic solutions of the {Benjamin}-{Ono} equation.
\newblock 2008.
\newblock (submitted), arXiv:0804:3623.

\bibitem{ambrose:wilkening:vtx}
D.~M. Ambrose and J.~Wilkening.
\newblock Computation of time-periodic solutions of the vortex sheet with
  surface tension.
\newblock 2009.
\newblock (in preparation).

\bibitem{amick:toland:BO}
C.~J. Amick and J.~F. Toland.
\newblock Uniqueness and related analytic properties for the {Benjamin}--{Ono}
  equation --- a nonlinear {Neumann} problem in the plane.
\newblock {\em Acta Math.}, 167:107--126, 1991.

\bibitem{benjamin:67}
T.~B. Benjamin.
\newblock Internal waves of permanent form in fluids of great depth.
\newblock {\em J. Fluid Mech.}, 29(3):559--592, 1967.

\bibitem{bock:kruskal:79}
T.~L. Bock and M.~D. Kruskal.
\newblock A two-parameter {Miura} transformation of the {Benjamin}--{Ono}
  equation.
\newblock {\em Phys. Letters}, 74A:173--176, 1979.

\bibitem{glowinski1}
M.~O. Bristeau, R.~Glowinski, and J.~P\'eriaux.
\newblock Controllability methods for the computation of time-periodic
  solutions; application to scattering.
\newblock {\em J. Comput. Phys.}, 147:265--292, 1998.

\bibitem{bfgs}
C.~G. Broyden.
\newblock The convergence of a class of double-rank minimization algorithms,
  {Parts I and II}.
\newblock {\em J. Inst Maths Applics}, 6:76--90, 222--231, 1970.

\bibitem{cabral:rosa}
M.~Cabral and R.~Rosa.
\newblock Chaos for a damped and forced {KdV} equation.
\newblock {\em Physica D}, 192:265--278, 2004.

\bibitem{camassa:lee:08}
R.~Camassa and L.~Lee.
\newblock Complete integrable particle methods and the recurrence of initial
  states for a nonlinear shallow-water wave equation.
\newblock {\em J. Comput. Phys.}, 227(15):7206--7221, 2008.

\bibitem{case:mero}
K.~M. Case.
\newblock Meromorphic solutions of the {Benjamin--Ono} equation.
\newblock {\em Physica}, 96A:173--182, 1979.

\bibitem{case:remarkable}
K.~M. Case.
\newblock The {Benjamin--Ono} equation: a remarkable dynamical system.
\newblock {\em Annals of Nuclear Energy}, 7:273--277, 1980.

\bibitem{cooper}
G.~J. Cooper and A.~Sayfy.
\newblock Additive {Runge}-{Kutta} methods for stiff ordinary differential
  equations.
\newblock {\em Math. Comp.}, 40(161):207--218, 1983.

\bibitem{davis:67}
R.~E. Davis and A.~Acrivos.
\newblock Solitary internal waves in deep water.
\newblock {\em J. Fluid Mech.}, 29(3):593--607, 1967.

\bibitem{dobro:91}
S.~Yu. Dobrokhotov and I.~M. Krichever.
\newblock Multi-phase solutions of the {Benjamin}-{Ono} equation and their
  averaging.
\newblock {\em Math. Notes}, 49:583--594, 1991.

\bibitem{doedel91b}
E.~J. Doedel, H.~B. Keller, and J.~P. Kern\'evez.
\newblock Numerical analysis and control of bifurcation problems: ({II})
  {Bifurcation} in infinite dimensions.
\newblock {\em Int. J. Bifurcation and Chaos}, 1(4):745--772, 1991.

\bibitem{fokas:ablowitz:83}
A.~S. Fokas and M.~J. Ablowitz.
\newblock The inverse scattering transform for the {Benjamin}--{Ono} equation
  --- a pivot to multidimensional problems.
\newblock {\em Stud. Appl. Math.}, 68:1--10, 1983.

\bibitem{tolandPlotnikovIooss}
G.~Iooss, P.I. Plotnikov, and J.F. Toland.
\newblock Standing waves on an infinitely deep perfect fluid under gravity.
\newblock {\em Arch. Rat. Mech. Anal.}, 177:367--478, 2005.

\bibitem{kaup:matsuno:98}
D.~J. Kaup and Y.~Matsuno.
\newblock The inverse scattering transform for the {Benjamin}--{Ono} equation.
\newblock {\em Stud. Appl. Math.}, 101:73--98, 1998.

\bibitem{carpenter}
C.~A. Kennedy and M.~H. Carpenter.
\newblock Additive {Runge}-{Kutta} schemes for convection-diffusion-reaction
  equations.
\newblock {\em Appl. Numer. Math.}, 44(1--2):139--181, 2003.

\bibitem{leveque}
R.~J. LeVeque.
\newblock On the interaction of nearly equal solitons in the {KdV} equation.
\newblock {\em SIAM J. Appl. Math.}, 47(2):254--262, 1987.

\bibitem{marsden:weinstein}
J.~Marsden and A.~Weinstein.
\newblock Reduction of symplectic manifolds with symmetries.
\newblock {\em Rep. Math. Phys.}, 5:121--130, 1974.

\bibitem{matsuno:BO:sol}
Y.~Matsuno.
\newblock Interaction of the {Benjamin}--{Ono} solitons.
\newblock {\em J. Phys. A}, 13:1519--1536, 1980.

\bibitem{matsuno:backland:85}
Y.~Matsuno.
\newblock Note on the {B\"acklund} transformation of the {Benjamin}--{Ono}
  equation.
\newblock {\em J. Phys. Soc. Jpn}, 54(1):45--50, 1985.

\bibitem{matsuno:04}
Y.~Matsuno.
\newblock New representations of multiperiodic and multisoliton solutions for a
  class of nonlocal soliton equations.
\newblock {\em J. Phys. Soc. Jpn.}, 73(12):3285--3293, 2004.

\bibitem{meyer}
K.~R. Meyer.
\newblock Symmetries and integrals in mechanics.
\newblock In M.~Peixoto, editor, {\em Dynamical Systems}, pages 259--72.
  Academic Press, New York, 1973.

\bibitem{nakamura:backlund:79}
A.~Nakamura.
\newblock B\"acklund transform and conservation laws of the {Benjamin}--{Ono}
  equation.
\newblock {\em J. Phys. Soc. Jpn}, 47(4):1335--1340, 1979.

\bibitem{nocedal}
Jorge Nocedal and Stephen~J. Wright.
\newblock {\em Numerical Optimization}.
\newblock Springer, New York, 1999.

\bibitem{ono:75}
H.~Ono.
\newblock Algebraic solitary waves in stratified fluids.
\newblock {\em J. Phys. Soc. Jpn.}, 39(4):1082--1091, 1975.

\bibitem{tolandPlotnikov}
P.I. Plotnikov and J.F. Toland.
\newblock {N}ash-{M}oser theory for standing water waves.
\newblock {\em Arch. Rat. Mech. Anal.}, 159:1--83, 2001.

\bibitem{satsuma:ishimori:79}
J.~Satsuma and Y.~Ishimori.
\newblock Periodic wave and rational soliton solutions of the {Benjamin}-{Ono}
  equation.
\newblock {\em J. Phys. Soc. Jpn.}, 46(2):681--687, 1979.

\bibitem{stoer:bulirsch}
J.~Stoer and R.~Bulirsch.
\newblock {\em Introduction to Numerical Analysis}.
\newblock Springer, New York, third edition, 2002.

\bibitem{DV:lorentz:04}
D.~Viswanath.
\newblock The fractal property of the {Lorenz} attractor.
\newblock {\em Physica D}, 190:115--128, 2004.

\bibitem{DV:couette:07}
D.~Viswanath.
\newblock Recurrent motions within plane {Couette} turbulence.
\newblock {\em J. Fluid Mech.}, 580:339--358, 2007.

\bibitem{Achain}
J.~Wilkening.
\newblock An algorithm for computing {Jordan} chains and inverting analytic
  matrix functions.
\newblock {\em Linear Algebra Appl.}, 427:6--25, 2007.

\bibitem{benj3}
J.~Wilkening.
\newblock An infinite branching hierarchy of time-periodic solutions of the
  {Benjamin}--{Ono} equation.
\newblock 2008.
\newblock (submitted).

\bibitem{wilk228A}
Jon Wilkening.
\newblock {\em Math 228A Lecture Notes: Numerical Solution of Differential
  Equations}.
\newblock UC Berkeley, 2007.
\newblock Available from author's webpage.

\bibitem{wulff:bif:rel}
C.~Wulff, J.~S.~W. Lamb, and I.~Melbourne.
\newblock Bifurcation from relative periodic solutions.
\newblock {\em Ergod. Th. Dynam. Sys.}, 21:605--635, 2001.

\end{thebibliography}

\end{document}